\RequirePackage{fix-cm}
\documentclass[twocolumn,epjc3]{svjour3}  
\smartqed  
\RequirePackage{graphicx}
\usepackage{color,amssymb}
\usepackage{slashed}
\usepackage{changes}
\usepackage{cancel}
\usepackage{hyphenat}
\def\be{\begin{eqnarray} &&}
\def\nonu{\nonumber \\ &&}
\def\nonub{\right. \nonumber \\ && \left.}
\def\nonubb{\right. \right.\nonumber \\ && \left.\left.}
\def\ee{\end{eqnarray}}
\def\psla{\slash \! \! \!}

\newcommand{\pv}[1]{\ensuremath{\textrm{PV}\left[#1\right]}}
\newcommand{\eqref}[1]{(\ref{#1})}
\newcommand{\pva}[0]{\ensuremath{\langle \rho_A \rangle}}
\newcommand{\pvb}[0]{\ensuremath{\langle \rho_B \rangle}}
\newcommand{\pvg}[0]{\ensuremath{\langle \rho_\gamma \rangle}}

\journalname{Eur. Phys. J. C}
\begin{document}
\title{  Fermion and Photon gap-equations
in Minkowski space within  the Nakanishi Integral Representation method}
\author{C\'edric Mezrag \thanksref{e0,addr1,addr2}
\and Giovanni Salm\`e \thanksref{e1,addr1}}
\thankstext{e0}{e-mail:cedric.mezrag@cea.fr}
\thankstext{e1}{e-mail:salmeg@roma1.infn.it}
\institute{Istituto  Nazionale di Fisica Nucleare, Sezione di Roma,
 P.le A. Moro 2, I-00185 Rome, Italy \label{addr1}\and IRFU, 
 CEA, Universit\'e Paris-Saclay, F-91191 Gif-sur-Yvette, 
  France\label{addr2}}

\date{Received: date }
\maketitle

\begin{abstract} The approach based on the Nakanishi integral representation of
n-leg transition amplitudes is extended to the treatment of the self-energies of
a fermion and an (IR-regulated) vector boson, in order to pave the way for constructing  
a comprehensive  application of the
technique to  
both gap- and Bethe-Salpeter equations, in  Minkowski space. The achieved result, namely a  6-channel coupled system of
integral equations, eventually allows one to determine the three 
K\"all\'en-Lehman weights for fully 
dressing the propagators of fermion and photon. A first consistency check is
also provided. The presented formal elaboration points to embed 
the characteristics 
of the non-perturbative regime at a more fundamental level. It yields  
a 
   viable  tool in  Minkowski space for the phenomenological investigation of  strongly interacting
    theories, 
    within a QFT framework  
   where the dynamical ingredients are made transparent and under control.

\end{abstract}


\section{Introduction}

The description of bound states, fully taking into account the general principles of the relativistic quantum field theory (QFT) 
and the needed non-perturbative regimes, is a  longstanding  and highly challenging problem.
As it is well-known, the formal solution of the problem can be traced back to the birth of
{relativistic} QFT, with the seminal paper by Salpeter and Bethe
 \cite{BS51}.
Starting from the analysis of the pole contributions  to the Green's function relevant for the bound state under scrutiny 
(e.g., the four-points Green's function for investigating two-body bound states),
 they introduced  an integral equation, known as the Bethe-Salpeter equation (BSE) for the bound-state amplitude, where the 
  kernel is obtained   from the two-particle irreducible diagrams,  describing  the dynamics inside the system.
   The systematic evaluation of the
  interaction kernel
  needs in turn the knowledge of other key ingredients: i) self-energies of both intermediate particles and quanta and ii) 
  vertex functions, as pointed out by Gell-Mann and Low \cite{GellMann:1951rw}. 
  Unfortunately, those $2$- and
  $3$-point functions are quantities to be determined through  the  infinite tower of Dyson-Schwinger equations  (DSEs)
  \cite{Dyson:1949ha,Schwinger:1951ex,Schwinger:1951hq} 
  (see for 
    introductory 
   reviews, e.g.,  Refs. 
   \cite{Roberts:1994dr,Alkofer:2000wg,Maris:2003vk,Bashir:2012fs,Eichmann:2016yit,Sanchis-Alepuz:2017jjd},
    and  references 
   quoted therein) that govern the whole set of $N$-point functions. 
   Therefore, in order to make feasible the construction of more and more realistic interaction kernels, model builders have to elaborate
  strategies for truncating  the DSEs infinite tower, as much self-consistently as possible, while
   retaining the dynamical effects, at the greatest extent {(see, e.g., 
    Ref. \cite{Qin:2020jig}
   for a closed-form of the BSE kernel, obtained by using 
     the acting
   symmetries).
   { Finally, it is worth mentioning that  within  the Hamiltonian framework 
   (suitable for the
   studies in Minkowski space), other relevant non-perturbative approaches,
    have been developed, like i) the discretized light-cone quantization
     \cite{Brodsky:1997de}, e.g.  recently applied  to positronium,  
     pion and kaon
    by using 
     the so-called Basis
    Light-front Quantization and a suitable truncation of  the Fock space 
    (see, e.g., Refs.\cite{Vary:2009gt,Lan:2019rba,Zhao:2020gtf}); 
    and ii) the Hamiltonian formulation of the
    lattice gauge theories that preserves the evolution of the states with a
    continuous real time
    \cite{Kogut:1974ag}, (see, e.g., Ref. \cite{Raychowdhury:2019iki} and
    references quoted therein for a recent QCD study and Ref. \cite{Banuls:2019bmf} for 
    interesting   quantum simulations of lattice gauge theories). }
  
 In the last decades, significant progresses have been done for implementing the 
 {approach based on  BSE plus truncated DSEs}\footnote{ We can roughly  group the truncation schemes in two sets: i) the ones exploiting a dressed vertex,  
     with different amount  of  complexity, and ii) the ones using a 
     simple bare vertex.},  and  a high degree of
 sophistication has been achieved, mainly in Euclidean space (see Refs.
 \cite{Roberts:1994dr,Alkofer:2000wg,Maris:2003vk,Bashir:2012fs,Eichmann:2016yit,Sanchis-Alepuz:2017jjd,Tandy:2003hn} and also 
 Refs. \cite{Chang:2009zb,Qin:2013mta,Qin:2014vya,Binosi:2016rxz,Binosi:2016wcx,Qin:2017lcd,Aguilar:2018epe}).
 {Moreover,  as it is well-known,
   DSEs can be obtained in a very efficient way by using the
    path-integral formalism, that in turn   acquires  a rigorous 
    mathematical meaning in the Euclidean space (see \emph{e.g.} Ref.
   \cite{Glimm:1987ng,Chatterjee:2018})}\footnote{
   It must recalled that 
   the complete equivalence between  Euclidean and relativistic QFT,
   under  a given set of necessary and sufficient conditions, was established by
   Osterwalder and Schrader, in the seventies 
   \cite{osterwalder1975}.}. 
   In particular,
     the most widely investigated field is the continuum QCD,  with an impressive  wealth of 
  applications in hadron physics, ranging from baryon and meson  spectra to elastic electromagnetic 
  (em) form factors and transition ones 
   (see, e.g.,  
  \cite{Roberts:1994dr,Alkofer:2000wg,Maris:2003vk,Bashir:2012fs,Eichmann:2016yit,Sanchis-Alepuz:2017jjd,Tandy:2003hn}).
    { Notice that also some timelike observables can be studied, e.g.
   by exploiting  i) analytic  models for the  running coupling (see the seminal Ref. 
   \cite{Shirkov:1997wi}); ii) the evaluation of a finite number of  moments of 
   proper Euclidean 
   correlation functions (see, e.g.,  Ref. \cite{Joo:2019jct} for a recent approach on lattice 
   QCD and Refs.  \cite{Chen:2016sno,Bednar:2018mtf} for continuum QCD results); iii)
   integral representations of the quantities primarily evaluated  in 
   Euclidean space or  
    algebraic Ans\"atze, properly tuned through spacelike data  
    (see, e.g.,
   \cite{Chang:2013pq,Gao:2016jka,Xu:2018eii,Ji:2001pj,Kekez:2020vfh}).}

   {Furthermore,}  also  applications to  QED 
    have been pursued at large extent 
   (see, e.g., Refs. \cite{Bashir:1997qt,Fischer:2004nq,Kizilersu:2009kg,Kizilersu:2013hea} and for a recent study in Minkowski space Ref. 
   \cite{Jia:2017niz}). The interest in  investigating the QED, in the whole dynamical range, may be 
   surprising, given the extraordinary accuracy 
   achieved in the comparison 
   with the data  by using perturbative tools (see, e.g., the case of the muon anomalous magnetic moment 
   \cite{PDG2018}). Indeed,  a close study of the 
   non-perturbative regime on the
   one side is  
   relevant  for shedding light {on
   QED at very short distances, as suggested, e.g., in Refs. 
   \cite{Bashir:1994az,Bashir:2011vg} for Euclidean studies of the 
   critical 
   coupling, below which chiral symmetry breaks down in quenched QED, and in 
   Ref. \cite{Reenders:1999bg} for an Euclidean investigation on how to escape the triviality fate of
   QED by adding a relevant four-fermion operator}.
   {On the other side, a non-perturbative  exploration of QED
   represents a needed  step for  approaches played in
   Minkowski space, and  eventually aiming to compare  the calculated outcomes with
    experimental  results for hadrons,  as  it has been  already done 
    {  by using  approaches} in Euclidean space.}

Our investigation will focus on the study of 
{  the charged fermion and photon 
gap-equations below the critical coupling, $\alpha_c \leq \pi/3$,
  (see, e.g., Refs.\cite{Bashir:1994az,Bashir:2011vg,Atkinson:1993mz})}, and 
 in order to help the reader to better appreciate the differences
  with other approaches  it is useful 
   to indicate, even in a simplistic way, the directions along 
   which we will move in what follows.
  For this reason, let us immediately 
  mention the two key ingredients
    we  will adopt: i) the Minkowski space, where the physical processes take place and  ii)  the 
   structure of
   the vertex function. 
   
   The vertex has a prominent role, and in our approach it is composed of 
   two contributions. The first term is the well-known part introduced by Ball and Chiu in Ref. \cite{Ball:1980ay}, that fulfills the 
   Ward-Takahashi
   identities (WTIs) (both the differential form and the finite-difference one). {The second term is}
   a transverse contribution (see, e.g.,  Refs
   \cite{Bashir:1997qt,Kizilersu:2009kg,Jia:2017niz,Kondo,Curtis:1990zs,Bashir:2011dp,Kizilersu:2014ela} 
   for a wide discussion), based on a minimal Ans\"atz proposed 
   in Ref. \cite{Qin:2013mta}. Such a transverse term is  able to  
   restore  the full multiplicative renormalization of both fermion and photon propagators (solutions of suitably truncated \linebreak DSEs), and in turn gracefully implements  a workable, 
   self\hyp consistent truncation scheme. Indeed, although the knowledge of the full content of the vertex requires the one of the full off-shell scattering matrix, the longitudinal
     WTI and its transverse counterparts relate respectively the divergence and
     the curl of the vertex in terms of the fermion 2-points function. These
     relations are exact for the divergence and truncated for the curl, allowing
     for a viable closed system \cite{Qin:2020jig,Qin:2013mta}.
     
     Another important ingredient, though  more technical, is represented by the so-called Nakanishi integral
     representation (NIR) of a generic $n$-leg transition amplitude  \cite{nak63,Nakarev,nakabook}. Indeed, such a
     tool has allowed one 
   to undertake new efforts for developing   methods for solving in Minkowski space both
   truncated DSEs  (see, e.g.,
  Refs. \cite{Jia:2017niz,Sauli:2001mb,Sauli:2001we,Sauli:2002tk,Sauli:2003zp,Sauli:2006ba,Sauli:2019hqd})
  and BSE
  (see, e.g., Refs. \cite{Kusaka,CK2006,CK2006b,CK2010,FSV2,FSV3,dePaula:2016oct,Gutierrez:2016ixt,dFSV2,AlvarengaNogueira:2019zcs,Sauli:2008bn,Sauli:2014uxa,Mello:2017mor,Frederico:2019noo,Ydrefors:2019jvu}, where systems  with and without spin degrees of freedom are
  investigated).
  
  The main motivation for adopting the NIR, closely related to the Stieltjes transform (see an application in Ref.
  \cite{Carbonell:2017kqa}),  is given by the possibility to express  the $n$-leg transition amplitudes
  through their all-order perturbative form.
  The  freedom needed for 
  exploring a non-perturbative regime is  assured via the  unknown Nakanishi weight 
  functions (NWFs), that are real
  functions fulfilling a uniqueness theorem, within the Feynman diagrammatic
  framework \cite{nakabook}.
   Such a freedom has shown all its relevance in the numerical studies of the bound states
    (by using both ladder and cross-ladder interaction kernels), that are the
   main instance where   the realistic description of the non-perturbative regime is necessary. Furthermore, 
   the NWFs to be used  for
   the self-energies  do not depend upon the external
   momenta,  greatly simplifying our formal elaboration, as shown in what follows.
   The  advantage of the NIR  is that the four-momentum dependence is 
   made explicit, allowing 
   direct algebraic
   manipulations, and eventually making affordable  
  analytic integrations. This is an important virtue of the NIR approach, 
  since
  it simplifies the treatment of the expected singularities.
  On the   phenomenology side, when  light-cone observables have to be evaluated, e.g. for describing the partonic
   structure of hadrons 
  \cite{Chang:2013pq,Chang:2014lva,Mezrag:2014jka,Gao:2014bca,Shi:2014uwa,Fanelli:2016aqc,Mezrag:2017znp,Chouika:2017rzs,Chouika:2017dhe,Shi:2018mcb},
   the explicit dependence upon the momenta facilitates the needed
  projection onto the light-cone. However, in the NIR context, the dynamical assumptions are still 
  much simpler than the one made in Euclidean calculations. 
  For instance, in the above mentioned works 
  when solving  Minkowskian BSE for mesons, the constituent
  fermions are most of the time considered perturbative-like, i.e. omitting the running 
 of the dressed quark mass (with the exception of Ref. \cite{Mello:2017mor} 
 where it has been proposed  to import the running mass of the quarks  from the Euclidean  lattice into the BSE framework).

 Our present effort aims at formally developing a method based on NIR for solving 
 a coupled system composed by the 
 gap-equations for both fermion and gauge boson, directly in Minkowski space. 
 It should be pointed out that the final goal (to be presented elsewhere) 
 of the program we are
 pursuing  is to provide both implementation and quantitative solutions of  the BSE with
 dressed propagators, in order to achieve a more and more realistic description
 of an  interacting system within the QFT framework, directly in the physical
 space.  
 The integral-equation system, we arrive at,  is   obtained by using
 a self-consistent truncation scheme of DSEs, valid
 in the whole dynamical range of  QED, and by adopting  both dimensional regularization  and 
   momentum-subtraction procedure
  for the renormalization. Moreover, to pragmatically   
  remove   
  the well-known IR divergences,  a tiny mass-regulator has been 
  introduced  
  for the gauge boson, (see, e.g.,
  Ref. \cite{Zuber} for a more general discussion and Ref. \cite{Kapec:2017tkm} for a recent
  analysis).
 In general, we  share the same spirit of  works {as}: i) Ref.
  \cite{Jia:2017niz} where, within a quenched approximation,  a spectral representation of the 
  fermion propagator was adopted, in combination
  with its K\"all\'en-Lehman (KL) representation, and, importantly,
   a vertex function was
  constructed by exploiting the form suggested by the Gauge Technique  \cite{Delbourgo:1977jc}  plus 
  transverse terms, added for matching
  the {\em perturbative expressions} of the
  renormalized fermion self-energy {(see also Refs. \cite{Frederico:2019noo,Jia:2019kbj} for further Minkowskian exploration of a
  massive QED, in quenched approximation)}; ii) Refs. 
   \cite{Sauli:2001mb,Sauli:2002tk,Sauli:2003zp} (see also 
  Ref. 
 \cite{Sauli:2019jkb} for a first study of the transverse vertex contribution), where a more direct link to the NIR technique 
 (with different sets of approximations) can be found. 
 Simplifying,  the main difference with the previous works is a fully dressing of  fermion and photon self-energies, by 
  introducing  a vertex function 
 composed by the standard
  Ball-Chiu component \cite{Ball:1980ay} and a minimal Ans\"atz for the purely transverse contribution \cite{Qin:2013mta}, able to
  ensure the multiplicative renormalizability of the whole approach.

The paper is organized as follows. In Sect. \ref{sect_GF}, the general formalism is introduced for fermion and photon 
propagators  and self-energies, in terms of
the KL representations and  the NIR, respectively.
In Sect. \ref{sect_vert}, the adopted vertex function  is discussed. In Sect. \ref{sect_GEQ},
the gap-equations are introduced and the main result of our formal analysis, i.e. the coupled system 
of   integral equations for determining the NWFs,
of both  electron and photon  self-energies, is illustrated.
In Sect. \ref{sect_res} an initial  application  of the coupled system, based on its first
 iteration,
 is shown.
In Sect. \ref{sect_CP}, the conclusions of our  analysis of the truncated DSEs within the NIR framework 
are drawn and the 
perspectives of the future numerical
studies are presented. Finally, it has to be emphasized that the Appendices have been written in a detailed form 
for making as simple as
possible a check of the whole formalism, and therefore they have to be considered an essential part 
of the work.

\section{General formalism}
\label{sect_GF}
 In this Section, we summarize the general formalism that will be used in our investigation of   QED in Minkowski space
(see the review in Ref. \cite{Roberts:1994dr} for the Euclidean version).
We introduce first  the expression of the self-energy ($2$-leg
transition amplitude in the Nakanishi language \cite{nakabook}, that emphasizes the set of external momenta) 
in terms of NIR, for both fermion and photon. Then, the KL representations of the corresponding propagators are 
given. The
main goal of this initial step is the relations between
 KL weights and  NWFs (see  Refs. \cite{Sauli:2002tk,Sauli:2003zp}, for an analogous approach, but with 
  renormalization constants $Z_1=Z_2=1$ and with a bare vertex  function or the Ball-Chiu one, respectively).

 The suitable renormalization scheme we adopt is the momentum subtraction one, applied on the mass-shell (MOM), 
 as discussed in what follows. {This scheme is suitable when asymptotic states exist, and this
 property  is actually needed for the KL representation we have adopted.}
{ Clearly, when  a confined phase of the QED 
 establishes, likely for large values of photon running mass, 
   the KL representation becomes unproved (see, e.g., Ref. \cite{Bashir:2011vg} and references quoted therein for   the analysis of the confining phase in QED within the gap-equation formalism in
 Euclidean space, and  Ref. \cite{Sauli:2020dmx} for a more recent investigation
in Minkowski space)}.
It should be anticipated that both electron and photon self-energies can be
 nicely renormalized  by applying such a scheme, given the benefit from the presence of the  transverse component 
 of the vertex
  function.

\subsection{The renormalized propagator of a fermion}
\label{subsect_selff}
By adapting  the notations  in Ref. \cite{Roberts:1994dr},
 one can write the following relations involving the renormalized propagator 
 of a fermion and the  regularized self-energy.
 
The renormalized fermion propagator is given  by
\be
S_R(\zeta,p)= {i\over \psla p-m(\zeta)
-\Sigma_R(\zeta;p)+i\epsilon}
\label{propR}\ee
with $\zeta$ the renormalization point and $\Sigma_R(\zeta;p)$ the 
renormalized
 self-energy. From Lorentz invariance, one can write
 \be
 \Sigma_R(\zeta;p)= \psla p~{\cal A}_R(\zeta;p)  +  {\cal B}_R(\zeta;p)~~,
 \label{self_R}
 \ee
 with ${\cal A}_R(\zeta;p)$ and ${\cal B}_R(\zeta;p)$ suitable scalar functions. In terms of
 the expression in Eq. \eqref{self_R}, the renormalized propagator reads
 \be
S_R(\zeta,p)
\nonu
=i~{\psla p~
\Bigl(1-{\cal A}_R(\zeta;p)\Bigr)+m(\zeta)+{\cal B}_R(\zeta;p)\over p^2 \Bigl(1-{\cal A}_R(\zeta;p)\Bigr)^2-\Bigl(m(\zeta)+{\cal
B}_R(\zeta;p)\Bigr)^2+i\epsilon}
\label{SR2}\ee
 Noteworthy, by requesting that the renormalized propagator for 
 $p^2\to\zeta^2$ has a  pole at the  mass $m(\zeta)=m_{phys}$ \footnote{For the sake of generality, we will leave the notation $m(\zeta)$ 
in the following expressions, though an on-mass-shell renormalization is adopted.}
and the 
same residue of the free propagator, one finds the constraints to be fulfilled by
 the two scalar functions, at the renormalization point. Needless to say,    those
 constraints are crucial for establishing the relations between the {\em regularized
 self-energy} and the  two renormalization constants $\delta m$
 and $Z_2(\zeta,\Lambda)$ (see what follows). As a matter of fact, from the well-known 
 general approach illustrated, e.g., in  Ref. \cite{Zuber} 
 (or adopting Eq. \eqref{SR2} and imposing 
$-i\Bigl(p^2-m^2(\zeta)\Bigr)~S_R(\zeta;p) \to \psla p_{on}+m(\zeta)$ for $p\to p_{on}$, with
$p^2_{on}=m^2(\zeta)$) one gets
\be
m(\zeta)~{\cal A}_R(\zeta;\zeta)+{\cal B}_R(\zeta;\zeta)= 0 ~~,
\nonu
{\cal A}_R(\zeta;\zeta)+2m(\zeta) \nonu \times
\left[m(\zeta)~{\partial {\cal A}_R(\zeta;p)\over \partial p^2}+
{\partial {\cal B}_R(\zeta;p)\over \partial p^2}\right]_{p^2=\zeta^2}=0~~.
\label{SR_cond}\ee
{These two equations define the standard on-shell QED 
renormalization scheme.}
Bringing in mind that the natural outcome of our formal elaboration will be 
a system of
integral equations, needed for determining ${\cal A}_R$ and ${\cal B}_R$, we adopt the
following renormalization conditions {defining the RI'/MOM scheme}
 (see Ref.\cite{Kizilersu:2001pd}, for the renormalization
independent method in the unquenched QED)
\be
\label{cond1}
{\cal A}_R(\zeta;\zeta)=0 ~, ~~~  {\cal B}_R(\zeta;\zeta)=0~ ~~~.
\ee
It is worth noticing  the following remarks about this choice: i) 
{it preserves the pole at the physical mass of the fermion}; ii) it allows a numerical
 simplification, avoiding to implement boundary conditions where there is an interplay between
${\cal A}_R$ and ${\cal B}_R$; and last but not least iii) {exchanging the physical mass for the current mass evaluated at a space-like momentum, it is formally similar to the RI'/MOM scheme exploited} 
in the literature
devoted to the non-perturbative studies of QFT, e.g. in the context of the investigation on the lattice (see the {discussions on the}  
RI'/MOM scheme 
e.g., in Refs. \cite{Sturm:2009kb,Gracey:2013sca}) as well as   in continuous approaches (see, e.g., Refs.
\cite{Roberts:1994dr,Kizilersu:2001pd}). 
Since at the present stage of the novel approach we are exploring, the two 
boundary conditions in Eq. \eqref{cond1} turn out to 
simplify the determination of the two renormalization constants,
we will leave the 
study of QED in the standard renormalization scheme, 
Eq. \eqref{SR_cond}, for further investigation.

  The propagator $S_R$ can be expressed  in terms of the regularized quantity, $\Sigma(\zeta,\Lambda;p)$,
  where $\Lambda$ stands for a Poincar\'e invariant regulator, e.g. $\Lambda=1/\epsilon$ within a dimensional 
  regularization framework  with $d=4-\epsilon$. To make the mathematical notation less heavy, in what follows
   it is understood that the relations involving renormalized quantities
  hold only in the limit   $\Lambda\to\infty$ {(notice that above the critical coupling, one expects to
  meet well-known difficulties for QED, as illustrated e.g., in Refs. \cite{Reenders:1999bg,Kizilersu:2001pd} )}. Hence, one writes 
 \be
S_R(\zeta,p)={ 1\over Z_2(\zeta,\Lambda)}~S(\zeta,\Lambda;k)
\nonu
=
{ 1\over Z_2(\zeta,\Lambda)}~{i \over \psla p- m(\zeta)+\delta m -\Sigma(\zeta,\Lambda;p)+ i \epsilon}~,
\label{propR1}\ee
where $Z_2(\zeta,\Lambda)$ is the renormalization factor affecting the
fermionic field and    $\delta m=  m(\zeta)- m_0$,  with $m_0$ the  bare mass.
  The analogous form of Eq. \eqref{self_R}, for the {\em regularized self-energy} reads (it is useful to include the
  renormalization constant $Z_2$ in the definition)
\be
\Sigma_Z(\zeta,\Lambda;p)=Z_2(\zeta,\Lambda)~\Sigma(\zeta,\Lambda;p)
\nonu
=\psla p~
{\cal A}_Z(\zeta,\Lambda;p)  + 
{\cal B}_Z(\zeta,\Lambda;p)~,
\label{self_reg}\ee
with   ${\cal A}_Z(\zeta,\Lambda;p)$ and  ${\cal B}_Z(\zeta,\Lambda;p)$
  suitable scalar functions.
 In particular, comparing Eq. \eqref{propR} and Eq.
\eqref{propR1}, one obtains 
\be
{\cal A}_R(\zeta;p)={\cal A}_Z(\zeta,\Lambda;p) - \Bigl(Z_2(\zeta,\Lambda)-1\Bigr)~~,
\nonu
{\cal B}_R(\zeta;p)={\cal B}_Z(\zeta,\Lambda;p) - 
\Bigl[ m(\zeta)~ (1-Z_2(\zeta,\Lambda))
\nonu +Z_2(\zeta,\Lambda) ~\delta m\Bigr]~~.
\label{calAB}\ee
Indeed,  those relations amount to   the outcomes of the  subtraction scheme for the renormalization of 
each scalar function.
Moreover, by taking into account Eq. \eqref{cond1}, one has
\be
{\cal A}_Z(\zeta,\Lambda;\zeta)= Z_2(\zeta,\Lambda)-1~~,
\nonu
{\cal B}_Z(\zeta,\Lambda;\zeta)= 
m(\zeta) (1-Z_2(\zeta,\Lambda))+Z_2(\zeta,\Lambda) ~\delta m~~,
\label{renorc}\ee
and therefore in the limit $\Lambda \to \infty$:
\be
\Sigma_R(\zeta;p)=\Sigma_Z(\zeta,\Lambda;p)-\left.\Sigma_Z(\zeta,\Lambda;p)\right|_{p^2=\zeta^2}
\nonu
=\psla p \Bigl[{\cal A}_Z(\zeta,\Lambda;p)-\left.{\cal A}_Z(\zeta,\Lambda;p)\right|_{p^2=\zeta^2}\Bigr]
\nonu+\Bigl[ {\cal B}_Z(\zeta,\Lambda;p)-\left.{\cal B}_Z(\zeta,\Lambda;p)\right|_{p^2=\zeta^2}\Bigr]~~.
\label{self_R1}\ee
 
Pursuing our goal of establishing a formal framework where one can get actual solutions of the gap equation, and eventually 
 describe the renormalized propagator, we usefully introduce
 the NIR for the fermionic self-energy. This can be achieved
by  starting from  the
approach proposed for a scalar case by Nakanishi (see Ref. \cite{nakabook}), for  summing up the infinite
contributions to a given n-leg amplitude, and generalizing in two respects. One is the transition from scalars to fermions, and the second
one, more important, from a perturbative to a non-perturbative regime. Those steps have been explored for the
BSEs in Refs. \cite{Kusaka,CK2006,CK2006b,CK2010,FSV2,FSV3,dePaula:2016oct,Gutierrez:2016ixt,dFSV2,AlvarengaNogueira:2019zcs,Sauli:2008bn,Sauli:2014uxa,Mello:2017mor,Frederico:2019noo,Ydrefors:2019jvu}.
For the fermion self-energy,  it is necessary to introduce two NWFs, since one has to deal with two scalar functions. 
Hence, 
  the 
{\em regularized
self-energy}  can be written in terms of the following scalar functions
\be
{\cal A}_Z(\zeta,\Lambda;p)=
   \int_{s_{th}}^\infty ds 
   ~\frac{\rho_A(s,\zeta,\Lambda)}{p^2-s+i\epsilon} ~~,
  \nonu 
  {\cal B}_Z(\zeta,\Lambda;p) =  \int_{s_{th}}^\infty ds~ 
   \frac{\rho_B(s,\zeta,\Lambda)}{p^2-s+i\epsilon}~~,
\label{NIR_f}\ee
with $s_{th}$  the multiparticle  threshold and $\rho_{A(B)}$ the NWFs. It should be recalled that the NWFs 
are real
functions, and do not depend upon the external momenta. This last remark will be useful for simplifying the formal
elaboration aiming to get the suitable integral equations for $\rho_{A(B)}$.

 Moreover, the NWFs  have  to fulfill the relation entailed by
 Eq. \eqref{renorc}, i.e.
\be
 Z_2(\zeta,\Lambda)= 1 +\int_{s_{th}}^\infty ds 
   ~\frac{\rho_A(s,\zeta,\Lambda)}{\zeta^2-s+i\epsilon} ~~,
\nonu 
Z_2(\zeta,\Lambda) ~\delta m= \int_{s_{th}}^\infty ds 
   ~\frac{m(\zeta)~\rho_A(s,\zeta,\Lambda)+ \rho_B(s,\zeta,\Lambda)}
   {\zeta^2-s+i\epsilon}~~.
\nonu  \label{Z2_NIR}  \ee
It is easily seen that NWFs with a  constant behavior for  $s \to \infty$ generate an expected logarithmic divergence. 
  
   By using Eqs. \eqref{calAB},  \eqref{renorc} and \eqref{NIR_f} one can write
 \be
 {\cal A}_R(\zeta;p)=\lim_{\Lambda \to \infty}\left[{\cal A}_Z(\zeta,\Lambda;p)-\left. {\cal
 A}_Z(\zeta,\Lambda;p)\right|_{p^2=\zeta^2}\right]
 \nonu
 =(\zeta^2-p^2) ~\int_{s_{th}}^\infty ds 
   ~\frac{\rho_A(s,\zeta)}{(p^2-s+i\epsilon)~(\zeta^2-s+i\epsilon)} ~~,
  \label{AR_NIR}
   \ee
 \be
 {\cal B}_R(\zeta;p)=\lim_{\Lambda \to \infty}\left[{\cal B}_Z(\zeta,\Lambda;p)-
 \left.{\cal B}_Z(\zeta,\Lambda;p)\right|_{p^2=\zeta^2}\right]
\nonu
= 
 (\zeta^2-p^2)~\int_{s_{th}}^\infty ds 
   ~\frac{\rho_B(s,\zeta)}{(p^2-s+i\epsilon)~(\zeta^2-s+i\epsilon)} ~~,
    \label{BR_NIR}
    \ee
where the notation $\rho_{A(B)}(s,\zeta) = \rho_{A(B)}(s,\zeta,\Lambda\to \infty)$ is adopted from now on.
   
It should be pointed out that {the actual form of the NFWs will follow from the solution of the
coupled system of integral equations we are going to elaborate.} {Anticipating on the next sections,} a possible constant behavior  
of the NWFs $\rho_{A(B)}$ for 
 $s \to \infty$
would be regularized by the  quadratic dependence  upon $s$ in the 
denominator, allowing to safely take  $\Lambda\to \infty$. {A situation in which $\rho_{A(B)}$ is not bounded at infinity would create regulator dependent results, and may appear above the critical coupling in 
QED as already mentioned.}

Dealing with the gap-equations, it is fruitful to use  the KL  representation 
of  the 
renormalized propagators, and therefore one has to establish
the relation between KL weights   and  NWFs  of the corresponding self-energy (see also Ref. \cite{Sauli:2001mb} for the scalar case 
 and  Refs. \cite{Sauli:2002tk,Sauli:2003zp} for  $QED_{3+1}$).
Recalling the following 
KL representation
\be
S_R(\zeta,p)=
 i\mathcal{R}_S{ \psla p +m(\zeta)\over  p^2 -m^2(\zeta)  +i\epsilon} 
 \nonu + 
  i \int_{s_{th}}^\infty ds \frac{\slashed{p} \sigma_V(s,\zeta) +
  \sigma_S(s,\zeta)}{p^2-s +i\epsilon}~~,
\label{lehmR}\ee 
where $\mathcal{R}_S$ is the fermion propagator 
residue,  controlled by the choice of the 
renormalization scheme (here we have adopted RI'/MOM). Using $ \Sigma_R(\zeta,p)$ from Eq. \eqref{self_R}, one gets
\be
i \int_{s_{th}}^\infty ds \frac{\slashed{p} \sigma_V(s,\zeta) +
  \sigma_S(s,\zeta)}{p^2-s +i\epsilon}
  \nonu =i~{\psla p~
\Bigl(1-{\cal A}_R(\zeta;p)\Bigr)+m(\zeta)+{\cal B}_R(\zeta;p)\over {\cal D}(p;\zeta)+i\epsilon}
\nonu -i{\cal R}_S~{ \psla p +m(\zeta)\over  p^2 -m^2(\zeta)  +i\epsilon}~~,
\ee
with 
\be {\cal D}(p;\zeta)=p^2 \Bigl(1-{\cal A}_R(\zeta;p)\Bigr)^2-\Bigl(m(\zeta)+{\cal
B}_R(\zeta;p)\Bigr)^2
\nonu\ee
By evaluating the needed  traces, one can obtain the  following relations 
\be
\int_{s_{th}}^\infty ds \frac{\sigma_V(s,\zeta) }{p^2-s +i\epsilon}=
{
1-{\cal A}_R(\zeta;p)
\over  {\cal D}(p;\zeta)
+i\epsilon}-{{\cal R}_S\over p^2 -m^2(\zeta)  +i\epsilon}
  \nonu
  \int_{s_{th}}^\infty ds \frac{
  \sigma_S(s,\zeta)}{p^2-s +i\epsilon}=
  { m(\zeta)+{\cal B}_R(\zeta;p)   \over  {\cal D}(p;\zeta)
 +i\epsilon}\nonu-{{\cal R}_S~m(\zeta)\over p^2 -m^2(\zeta)  +i\epsilon}
 \label{klw}\ee
If one assumes that both  KL  weights and NWFs 
 match the hypotheses for applying the 
  {\em Sokhotski-Plemelj formula}, that reads
  \be
 \int_{-\infty}^\infty ds \frac{ f(s) }{\omega-s +i\epsilon}
= \pv{{f(s)\over \omega-s}} -i \pi f(\omega)~~,
 \label{a1_pv}\ee
with an  understood $ \theta(s-s_{th})$ inside $f(s)$, then 
 one can manipulate the singular integrals in the lhs of Eq. \eqref{klw} and the rhs of
  Eqs. \eqref{AR_NIR} and \eqref{BR_NIR} as follows 
 \be
 \int_{s_{th}}^\infty ds~{\sigma_{V(S)}(s',\zeta)\over (\omega-s' +i\epsilon)}
 \nonu
 = \pv{{\sigma_{V(S)}(s',\zeta)\over (\omega-s')}}
 -i\pi \sigma_{V(S)}(\omega,\zeta)~~,
 \label{app1_KL2} \ee
 \be
\int_{s_{th}}^\infty ds ~
\frac{\rho_{A(B)}(s,\zeta)}
 {(\omega-s+i\epsilon)~(\zeta^2-s+i\epsilon)}
 \nonu= \pv{\frac{\rho_{A(B)}(s,\zeta)}
 {(\omega-s)~(\zeta^2-s)}} -i\pi~{\rho_{A(B)}(\omega,\zeta)\over (\zeta^2-\omega)}~.
 \label{app1_AB}\ee
 Let us recall that $\rho_{A(B)}(s=\zeta^2,\zeta)=0$ and values $\omega \ge s_{th}$ are relevant in what
 follows. By {inserting}   Eqs. \eqref{app1_AB} in  \eqref{AR_NIR} and
 \eqref{BR_NIR}, the real and the imaginary parts of ${\cal A}_R(\zeta;\omega)$  become
  \be
  \Re e \Bigl\{{\cal A}_R(\zeta;\omega)\Bigr\}= (\zeta^2-\omega)~ \pva~~,\nonu  \Im m \Bigl\{{\cal
  A}_R(\zeta;\omega)\Bigr\}=-\pi\rho_{A}(\omega,\zeta)~~,
\ee 
with the notation $\pva$ indicating the principal value in Eq. \eqref{app1_AB}. 
Analogous expressions hold for ${\cal B}_R(\zeta;\omega)$.
Hence,  one can formally gets the following relations 
 between 
  KL weights  and   NWFs for $\omega>\omega_{th}=s_{th}$
 \be
   \sigma_V(\omega,\zeta) =  \frac{D_I\Bigl[1-(\zeta^2-\omega)\pva\Bigr]-\rho_A(\omega,\zeta)D_R}{D_R^2+\pi ^2 D_I^2} 
  \nonu
   \sigma_S(\omega,\zeta) =  \frac{D_I\Bigl[m(\zeta)+(\zeta^2-\omega)\pvb\Bigr]+\rho_B(\omega,\zeta)D_R}
   {D_R^2+\pi ^2 D_I^2}~~,
  \nonu\label{klwf}\ee
  where 
  \be
  D_R=\omega\Bigl[(1-(\zeta^2-\omega)\pva)^2-\pi^2 \rho^2_A(\omega,\zeta) \Bigr] \nonu- 
  \Bigl[(m(\zeta)+(\zeta^2-\omega)\pvb )^2-\pi^2 \rho_B^2(\omega,\zeta)\Bigr] ~~,
  \nonu
  D_I=2 \omega\rho_A(\omega,\zeta)\Bigl[1-(\zeta^2-\omega)\pva\Bigr]
  \nonu + 2\rho_B(\omega,\zeta)\Bigl[m(\zeta)+
  (\zeta^2-\omega)\pvb\Bigr]~~.
  \ee
It has to be pointed out that the knowledge of the  KL weights
  $\sigma_{S(V)}(\omega,\zeta)$  for $\omega>\omega_{th}$  is enough for determining the fermion propagator 
for all the possible values of $p^2$.
  
 \subsection{The renormalized propagator of a photon}
 \label{subsect_selfp}
 In the {\em Landau gauge}, the 
 free propagator of the photon reads
\be
D^{\mu\nu}(q)=-i~{T^{\mu\nu}(q)\over q^2 -\zeta^2_p+i\epsilon}~~,
\ee
where $T^{\mu\nu}(q)$  is the standard transverse projector 
\be
T^{\mu\nu}(q) = g^{\mu\nu}-{q^\mu q^\nu\over q^2}~~,
\label{tranT}\ee
 with its useful properties, 
\be
T_{\mu\nu}(q) = g_{\mu\nu}-{q_\mu q_\nu\over q^2}~~,\quad \quad \quad 
T_{\mu\alpha}(q)~T^\alpha_{~\nu}(q)=T_{\mu\nu}(q)
\nonu
~T_{\mu\nu}(q)~ g^{\nu\mu}= T^\mu_{~\mu}(q)=~3
\label{proj}\ee
and $\zeta_p$ is a IR-regulator, (see, e.g.,
Ref. \cite{Zuber}). For the sake of  light notation,  the dependence upon $\zeta_p$  will be understood  in the
renormalized quantities.
Hence, the renormalized photon propagator reads
\be
D_R^{\mu\nu}(\zeta,q)=-i{T^{\mu\nu}\over (q^2-\zeta^2_p+i\epsilon)~\Bigl[ 1 + \Pi_R(\zeta;q)\Bigr]}~~,
\label{propRph}\ee
where $\Pi_R(\zeta;q)$ can be called the photon self-energy, fulfilling the following condition,
able to lead to  the correct
residue at photon pole
\be
\Pi_R(\zeta;\zeta_p)=0~~.
\label{condph}
\ee
Notice that the photon propagator would present a problematic pole  if there exists a critical value,
 $q_{sing}$, such that $1 + \Pi_R(\zeta;q_{sing}) = 0$. {Interestingly, 
 the IR pole in \eqref{propRph}
  could be removed if  also $\Pi_R(q^2)$ develops an IR pole, i.e. the 
  so-called Schwinger mechanism \cite{Schwinger:1962tp},
 that leads to a massive photon (see, e.g., Ref. \cite{Burden:1991uh} for the absence of this
 phenomenon in QED$_3$ and Ref. \cite{Aguilar:2015bud}  for a comprehensive review in the case
 of QCD).}

The relation between $D_R^{\mu\nu}(\zeta;q)$ and  both the regularized self-energy  and the  renormalization constant $Z_3(\zeta,\Lambda)$ is
\be
D_R^{\mu\nu}(\zeta;q)
\nonu =-i{T^{\mu\nu}\over Z_3(\zeta,\Lambda)~(q^2-\zeta^2_p+i\epsilon)~
\Bigl[ 1 + \Pi(\zeta,\Lambda,q)\Bigr]}~~.
\label{propreph}\ee
Comparing the denominators in Eqs. \eqref{propRph} and \eqref{propreph} 
one has for $\Lambda \to \infty$
\be
\Pi_R(\zeta,q)=  \Pi_Z(\zeta,\Lambda;q)+Z_3(\zeta,\Lambda)-1~~,
\label{self_Rph}\ee
with $$\Pi_Z(\zeta,\Lambda;q)=Z_3(\zeta,\Lambda)\Pi(\zeta,\Lambda;q)~~.$$ By imposing the condition in Eq.
\eqref{condph}, one gets the following normalization
\be
\Pi_Z(\zeta,\Lambda;\zeta_p)=1-Z_3(\zeta,\Lambda)~~,
\label{normph}\ee
and writes for $\Lambda\to\infty$
\be
\Pi_R(\zeta;q)=~
\Pi_Z(\zeta,\Lambda;q)-\Pi_Z(\zeta,\Lambda;\zeta_p)~~,
\label{self_Rph1}\ee

It is also useful to recall that the renormalized propagator  fulfills the  well-known integral equation, given by 
\be
D_R^{\mu\nu}(\zeta,q)=D^{\mu\nu}(q)+D^{\mu\alpha}(q)
\nonu \times ~\Bigl[i\Pi_{\alpha\beta}^R(\zeta,q)\Bigr]
D_R^{\beta\nu}(\zeta,q)~~,
\label{inteph}
\ee
where 
 $\Pi^{\mu\nu}_R(\zeta,q^2)$ is the renormalized
vacuum polarization tensor, defined by
\be
\Pi^{\mu\nu}_R(\zeta,q)= -q^2~T^{\mu\nu}(q)~\Pi_R(\zeta;q)~~.
\ee
This quantity is involved in the gap-equation for the photon (see Sect. \ref{sect_GEQ} for more details).

Analogously to the fermion case, one introduces   the following NIR for
$\Pi_Z(\zeta,\Lambda;q)$
\be 
\Pi_Z(\zeta,\Lambda;q)=\int_{s^p_{th}}^\infty ds ~
  {\rho_\gamma(s,\zeta,\Lambda)\over (q^2-s+i\epsilon)}~~,
\label{self_nir}\ee 
where the real function $ \rho_\gamma(s,\zeta,\Lambda)$ is the NWF for the regularized photon self-energy, 
and $s^p_{th}$ the
multiparticle threshold, i.e. $s^p_{th}=4m^2(\zeta)$.

Using Eqs. \eqref{normph} and \eqref{self_nir}, one gets the following expression for $Z_3(\zeta,\Lambda)$
\be
Z_3(\zeta,\Lambda)= 1-\int_{s^p_{th}}^\infty ds ~
  {\rho_\gamma(s,\zeta,\Lambda)\over (\zeta^2_p-s+i\epsilon)}~~.
\label{Z3_NIR}\ee
The same observation below Eq. \eqref{Z2_NIR} is relevant also for Eq. \eqref{Z3_NIR}.

{By exploiting}  Eqs. \eqref{normph} and \eqref{self_nir} in \eqref{self_Rph1},  
 $\Pi_R(\zeta;q)$ can be written 
in terms of NWFs, viz
\be
  \label{NIRph}
  \Pi_R(\zeta;q) =(\zeta^2_p-q^2) \nonu \times ~
  \int_{s^p_{th}}^\infty ds ~
  \frac{\rho_\gamma(s,\zeta)}{(\zeta^2_p-s+i\epsilon)~(q^2-s+i\epsilon)}~~,
\ee
where $\rho_\gamma(s,\zeta)=\rho_\gamma(s,\zeta,\Lambda \to
\infty)$.

The KL representation of $ D^R_{\mu\nu}(\zeta,q)$ reads
\be
D^R_{\mu\nu}(\zeta,q) = 
  -i T_{\mu\nu}(q)
  \nonu \times ~\left(\frac{1}{q^2-\zeta^2_p+i\epsilon}+ \int_{s^p_{th}}^\infty 
  ds \frac{\sigma_\gamma(\omega,\zeta)}{q^2-s+i\epsilon}\right) ~~,
\label{lehmRph}
\ee
and has to be compared 
with the following expression obtained from Eq. \eqref{propRph}
\be
D^{\mu\nu}_R(\zeta,q) =-iT_{\mu\nu}(q)
~\Biggl[{1\over q^2-\zeta^2_p+i\epsilon} \nonu - 
{\Pi_R(\zeta;q)\over (q^2-\zeta^2_p+i\epsilon )~\Bigl (1 + \Pi_R(\zeta;q)\Bigr)}\Biggr]~~.
\ee
Hence one gets   
\be
\int_{s^p_{th}}^\infty ds \frac{\sigma_\gamma(s,\zeta)}
{q^2-s+i\epsilon}
=~-{\Pi_R(\zeta;q)\over (q^2-\zeta^2_p+i\epsilon )~\Bigl (1 + \Pi_R(\zeta;q)\Bigr)}~~.
\nonu\ee
By using Eqs. \eqref{a1_pv} and \eqref{NIRph}, the real and imaginary parts of $\Pi_R(\zeta;q)$ can be easily written in terms of the NWF 
$\rho_\gamma(\omega,\zeta)$ as follows (recall that $q^2\ge s^p_{th}$)
\be
 \Re e \Bigl\{\Pi_R(\zeta;q^2)\Bigr\}=(q^2 -\zeta^2_p)~\pv{\frac{\rho_\gamma(s,\zeta)} {(q^2-s)~(\zeta^2_p-s)}}~~,
 \nonu \Im m \Bigl\{\Pi_R(\zeta;q^2)\Bigr\}= -\pi \rho_\gamma(q^2,\zeta)~.
 \label{ReIMpi}\ee
Finally, by using {once} more Eq. \eqref{a1_pv}, one obtains   the desired  relation 
between  
$\rho_\gamma$  and  $\sigma_\gamma$, given by 
\be
 \sigma_\gamma(\omega,\zeta)=-~{1\over (\omega-\zeta^2_p)}
 \nonu \times~{\rho_\gamma(\omega,\zeta)\over \Bigl[(1+(\zeta^2_p-\omega)\pvg)^2+\pi^2
 \rho^2_\gamma(\omega,\zeta)\Bigr]}~~,
\label{sigmag}\ee
with $\omega\geq s^p_{th}$ and $\pvg$ the principal value in Eq. \eqref{ReIMpi}.

\section{ The renormalized vertex function}
\label{sect_vert}
The amputated three-leg transition amplitude, or vertex function, is the basic ingredient
for any dynamical approach that aims at determining the self-energies of 
particle and quanta, involved in a given theory. Unfortunately, the fully dressed
vertex function can be formally obtained only through the proper DSE where, 
in turn,  the four-leg
transition amplitude (i.e. the fully  off-shell fermion-antifermion scattering kernel in the case of QED) is present. This fact
makes clear the structure of the infinite tower of DSEs, where each n-leg
transition amplitude fulfills an integral equation containing {transition amplitudes with a
number of legs greater than $n$}. In spite of this, by using  general principles,  
one can  devise an   overall form of the vertex, in terms of the Dirac structures allowed by both the
 Lorentz
covariance, the parity conservation  and time reversal (see, e.g., Refs \cite{Ball:1980ay,Ball:1980ax}), when QED is investigated. Following well-known steps, one
decomposes   the vertex into two parts:
i) the standard 
component
introduced in the early eighties by Ball and  Chiu \cite{Ball:1980ay}, in order to   fulfill WTIs
and to avoid any kinematical singularity, and ii) 
a contribution purely transverse, i.e.
containing  the possible Dirac structures orthogonal to the momentum transfer
$q=p_f-p_i$ (see Fig. \ref{fig_vert}. for the pictorial representation and the kinematics). As a matter of
fact, one  writes the renormalized vertex (or the regularized one, with the proper modification in the notations)
as follows
\be
\Gamma^\mu_R(\zeta,p_f,p_i)=\Gamma^\mu_{R,BC}(\zeta,p_f,p_i)+\Gamma^\mu_{R,T}(\zeta,p_f,p_i)
\ee
where $q\cdot \Gamma_{R,T}(\zeta,p_f,p_i)=0$  and  $\Gamma^\mu_{R,BC}(\zeta,p_f,p_i)$ is the Ball-Chiu vertex
dictated by the WTI i.e.
\be
 q\cdot \Gamma_{R}(\zeta,p_f,p_i)= q\cdot \Gamma_{R,BC}(\zeta,p_f,p_i)\nonu
 =
iS^{-1}_R(\zeta,p_f)-iS^{-1}_R(\zeta,p_i)
\nonu
=\psla p_f -m(\zeta)-\Sigma_R(\zeta,p_f) -\Bigl[\psla p_i -m(\zeta)-\Sigma_R(\zeta,p_i)\Bigr]
\nonu=
\psla p_f ~\Bigl[1 -{\cal A}_R(\zeta;p_f)\Bigr] -{\cal B}_R(\zeta;p_f)
\nonu-
\psla p_i ~\Bigl[1 -{\cal A}_R(\zeta;p_i)\Bigr] +
{\cal B}_R(\zeta;p_i)
\ee
The actual expression of $\Gamma^\mu_{R;BC}$
 \cite{Ball:1980ay}, is given by
\be
  \label{eq:BallChiu}
  \Gamma^\mu_{R;BC} (\zeta, p_f,k_f) = \frac{\gamma^\mu}{2}~
  F_{{\cal A}_+}(p_f,p_i,\zeta)
  \nonu-\frac{(\psla p_f+\psla p_i) (p_f+p_i)^\mu}{2}~ F_{{\cal A}_-}(p_f,p_i,\zeta)
 \nonu - (p_f+p_i)^\mu~ F_{\cal B}(p_f,p_i,\zeta)  ~~,
\ee
where
\be
F_{{\cal A}_+}(p_f,p_i,\zeta)=2-{\cal A}_R(\zeta;p_f)-{\cal A}_R(\zeta;p_i)
\nonu =2  +
\int_{s_{th}}^\infty ds 
  ~{\rho_A(s,\zeta)\over(\zeta^2-s+i\epsilon)}
   \nonu \times  ~\left[{(p^2_f-\zeta^2)\over(p^2_f-s+i\epsilon)}
   +{(p^2_i-\zeta^2)  
   \over (p^2_i-s+i\epsilon)}\right]
 ~~ ,
  \nonu
F_{{\cal A}_-}(p_f,p_i,\zeta)=~{{\cal A}_R(\zeta;p_f)-{\cal A}_R(\zeta;p_i)\over (p^2_f-p^2_i)}
\nonu-\int_{s_{th}}^\infty ds 
   ~{\rho_A(s,\zeta)\over(p^2_f-s+i\epsilon)~(p^2_i-s+i\epsilon)}~~,
\nonu
F_{\cal B}(p_f,p_i,\zeta)=~{{\cal B}_R(\zeta;p_f)-{\cal B}_R(\zeta;p_i)\over (p^2_f-p^2_i)}=
  \nonu- \int_{s_{th}}^\infty ds 
   ~{\rho_B(s,\zeta)\over(p^2_f-s+i\epsilon)~(p^2_i-s+i\epsilon)}~~.
  \label{FAi} 
\ee
While $\Gamma^\mu_{R;BC}$ is elaborated starting from 
 WTIs and the crucial request
 of avoiding kinematical singularities, the transverse part $\Gamma^\mu_{R;T}$
  has to fulfill the constraint imposed by  the curl
of the current, $q^\mu \Gamma^\nu_{R}- q^\nu \Gamma^\mu_{R}$ \cite{Kondo} (see also the analysis 
in Ref. \cite{Qin:2013mta}), 
and it can be expressed in terms of {\em eight}
Dirac structures, $T^\mu_i$,  such that $q\cdot T_i=0$ (see Ref. \cite{Ball:1980ay} for the complete list) 
and eight scalar functions, ${\cal F}_i$, viz
\be
\Gamma^\mu_{R;T} (\zeta, p_f,p_i) = \sum_{i=1,8} {\cal F}_i(p_f,p_i,\zeta)~T^\mu_i(p_f,p_i)
\ee
In general the  functions ${\cal F}_i(p_f,p_i,\zeta)$
 cannot be  written only in terms of
${\cal A}_R$ and ${\cal B}_R$ 
 \cite{Qin:2013mta}, but the whole set of functions has to cooperate for ensuring another 
 fundamental property: the multiplicative renormalizability of both self-energies (see Eqs. \eqref{propR1} and
 \eqref{propreph})
 and vertex, viz
 \be
\Gamma^\mu_R(\zeta,p_f,p_i)=
Z_1(\zeta,\Lambda)~\Gamma^\mu(\zeta, \Lambda;p_f,p_i)
\label{Gamma_R}\ee
with the constraint $Z_1(\zeta,\Lambda)=Z_2(\zeta,\Lambda)$. 
  It is fundamental to
 notice that, given the DSEs, the multiplicative renormalizability of both  two-leg and three-leg functions are
 intimately  related. This has been elucidated by 
 a vast literature, in different frameworks. In particular, in Refs.
 \cite{Bashir:1997qt,Kizilersu:2009kg,Kizilersu:2013hea,Jia:2017niz,Curtis:1990zs,Kizilersu:2014ela} (and references
 quoted therein)
a close analysis, ranging  from a  first perturbative study to  non-perturbative ones, was carried out,  
pointing to the  role played 
by leading logarithms in determining the aforementioned property, through an
unavoidable  cooperation between  the scalar functions present in $\Gamma^\mu_{R;BC}$ and $\Gamma^\mu_{R;T}$.
Differently, in Refs. \cite{Bashir:1997qt,Bashir:2011dp,Brown:1989hy}, within a quenched approximation,
 the requirement of multiplicative renormalization is implemented by 
 looking for   solutions  of the fermion gap-equation with a power-law 
behavior.

 In our unquenched approach, we take into account the transverse vertex, retaining 
 only some contributions, as it
 will be explained in what follows. Indeed, this is a distinctive  feature of
 our work, in comparison with approaches sharing the same spirit, i.e.
 exploiting  spectral representations of both  propagators and self-energies 
 (see Refs. \cite{Jia:2017niz,Sauli:2001mb,Sauli:2002tk,Sauli:2003zp,Sauli:2006ba,Sauli:2019hqd}).
 In particular we consider the following two Dirac structures,  of the eight 
 identified in Ref. \cite{Ball:1980ax},
 \be
T^\mu_3(p_f,p_i)= q^2 \gamma^\mu- q^\mu \psla q
\nonu
T^\mu_8(p_f,p_i)= p^\mu_f\psla p_i -p^\mu_i \psla p_f -i \gamma^\mu
\sigma_{\nu \rho} p^\nu_i p^\rho_f
 \nonu
 =-i  \gamma_5\epsilon^{\mu}_{ ~\alpha\nu\rho} \gamma^{\alpha}p^\nu_i q^\rho
\ee
with $\sigma_{\nu \rho}=i[\gamma_\nu,\gamma_\rho]/2$ and $\epsilon^{0123}=+1$. Notice 
that   there is an
overall different sign with respect to Ref. \cite{Ball:1980ay}. 

It is worth mentioning that in the fermion massless  case (relevant for studying the dynamical generation 
of the mass) only the Dirac structures with $i=2,~3,~5,~8$ 
contribute \cite{Kizilersu:2009kg}, and moreover, in the same limiting case,  $T^\mu_3$ and $T^\mu_8$ allow 
one to implement 
the  gauge covariance of the fermion propagator \cite{Dong:1994jr}. Finally, as pointed out in Ref.
 \cite{Qin:2013mta}
  the
contribution $T^\mu_8$ is able to generate an anomalous magnetic moment term, within a perturbative framework
\cite{Chang:2009zb}.
 
\begin{figure}
\centerline{\includegraphics[width=4.0cm]{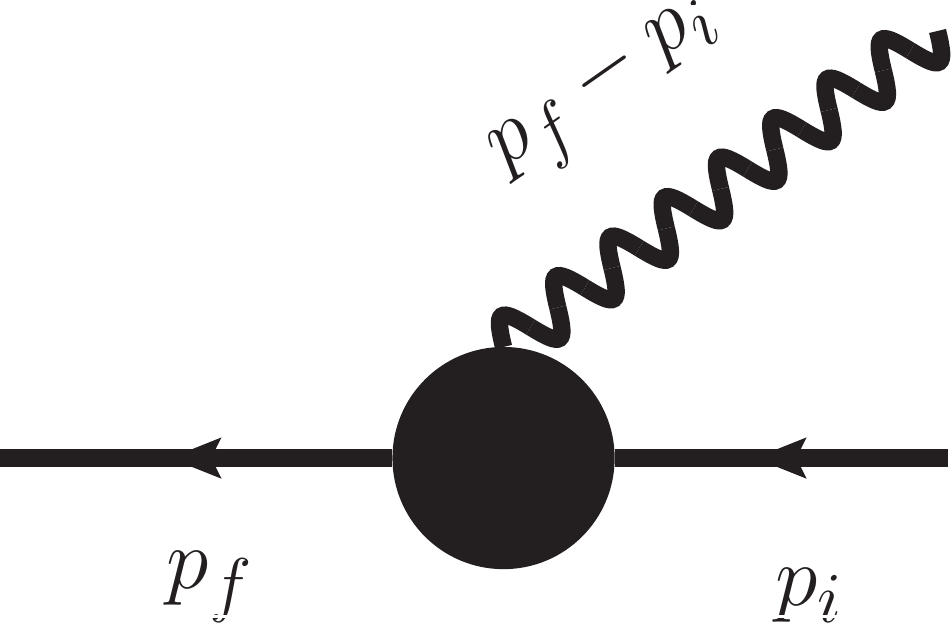}}
\caption{ The pictorial representation of the regularized 
fermion-photon vertex, with a fermion absorbing a photon.}
\label{fig_vert}
\end{figure}

The adopted  expressions of  ${\cal F}_3$ and ${\cal F}_8$ are the ones given in 
  \cite{Qin:2013mta} (see also \cite{Bashir:2011dp}),  where 
  a very detailed formal analysis of $\Gamma^\mu_T$
was carried out  (adopting   a Euclidean metric) and general expression were found.
In particular, due to the curl of the current, it turns out that  ${\cal F}_3$ and ${\cal F}_8$ can be {\em minimally} chosen as linear combinations of 
  ${\cal A}$ and ${\cal B}$  
(see subsect. \ref{subsect_selff}). In this way, 
one has  a workable Ans\"atz for $\Gamma^\mu_{R;T} $, that has the virtue
of closing the equations involving the fermion and photon self-energies.
The actual ${\cal F}_3$ and ${\cal F}_8$ are given by (see also Ref. \cite{Bashir:2011dp})
\be
{\cal F}_3( p_f,p_i,\zeta)= -{1\over 2} 
 F_{{\cal A}_-}(p_f,p_i,\zeta)
\nonu
{\cal F}_8( p_f,p_i,\zeta)=   F_{{\cal A}_-}(p_f,p_i,\zeta)
\ee
that match the expected perturbative behavior for $p^2_f>>p^2_i$ 
(see the discussion in Ref. \cite{Bashir:2011dp}).
Hence, one can write
\be
\Gamma^\mu_{R;T} (\zeta, p_f,p_i) 
=- {1 \over 2}
\nonu \times
\Bigl[q^2 \gamma^\mu- q^\mu \psla q+2i  \gamma_5\epsilon^{\mu\alpha}_{~~ \nu\rho} 
\gamma_{\alpha}p_i^\nu q_\rho\Bigr]~F_{{\cal A}_-}(p_f,p_i,\zeta)~~.
\label{vtren_qin}\ee

In conclusion, we use  $\Gamma^\mu_{R}$ given by the sum of  
the  Ball-Chiu vertex \cite{Ball:1980ay} and one of the minimal Ans\"atze for 
$\Gamma^\mu_{R;T}$,
proposed in Ref. \cite{Qin:2013mta}. In particular,  by  using Eq. \eqref{vtren_qin}
one can i)  fulfill the multiplicative renormalizability,
 ii)  establish  a non-perturbative  framework, where    a closed
 coupled system of  integral equations allows one to investigate the self-energies of
both fermion and photon.


\section {Coupled gap equations}
\label{sect_GEQ}
This Section, in particular subsections \ref{subsect_gapF} and  \ref{subsect_gapPh}, contains the 
main outcomes of our formal elaboration that aims to get a 
mathematical tool for
determining the fermion and photon NWFs and eventually yield 
 the fermion and photon
 self-energies. In order to accomplish such a task,
it is necessary to proceed by writing down the DSEs for the  
 self-energies (see, e.g., Ref. \cite{Roberts:1994dr} 
 for a general introduction), and   insert the results obtained in Sect.
\ref{sect_GF} (see details
 in  \ref{app_3} 
and \ref{app_4}). 

The DSE  for the regularized  fermion self-energy, defined in Eq.
\eqref{self_reg},  is given by 
\be
\Sigma_Z(\zeta,\Lambda;p)= -iZ_1(\zeta,\Lambda)~e^2_R \int_{\Lambda} {d^4k\over
(2\pi)^4}
\nonu \times ~~ \gamma^\beta~ D^R_{\beta\alpha}(\zeta, p-k)~  S_R(\zeta,k)~ 
 \Gamma^\alpha_R(\zeta;k,p)~~, 
\label{self_regF}\ee
where it is important to emphasize that the dependence upon $\Lambda$ means that the rhs can have singular contributions (indeed this is
the case). But such terms 
  become
finite after introducing a suitable regularization procedure, that in our case it turns out to be the dimensional one
 with $d=4-\epsilon$ and $\Lambda=1/\epsilon$ (see the details in
   \ref{app_3}).
Then,  the  renormalized self-energy fulfills
\be
\Sigma_R(\zeta;p)= -iZ_1(\zeta,\Lambda)~e^2_R ~\int_{\Lambda} {d^4k\over
(2\pi)^4}~ \gamma^\beta~ S_R(\zeta,k)
\nonu \times ~\Bigl \{D^R_{\beta\alpha}(\zeta, p-k)~    \Gamma^\alpha_R(\zeta;k,p)
\nonu
-\left[ D^R_{\beta\alpha}(\zeta, p-k)~    \Gamma^\alpha_R(\zeta;k,p)
\right]_{p^2=\zeta^2}\Bigr\}~~.
\label{dse_F}\ee
The  two scalar functions  describing $\Sigma_R(\zeta;p)$ (see Eq. \eqref{self_R}) can be obtained by 
evaluating  the suitable traces, i.e. 
\be
{\cal A}_R(\zeta;p)=
{1 \over 4p^2} ~\textrm{Tr} \Bigl[\psla p~\Sigma_R(\zeta;p)\Bigr]~~, \nonu
{\cal B}_R(\zeta;p)= {1 \over 4} ~\textrm{Tr} \Bigl[\Sigma_R(\zeta;p)\Bigr]
\label{dse_F1}\ee
In the next subsect.
 \ref{subsect_gapF} the results of the traces will be presented and the
  relation with the NWFs established.
 
In the Landau gauge we are adopting (recall that the polarization 
tensor is transverse in this gauge),
 one can  start from the following expression of the
regularized polarization tensor (see  \ref{app_4} for details) in terms of the renormalized quantities 
\be
  \Pi^{\mu\nu} (\zeta,\Lambda;q)= -q^2~T^{\mu\nu}~\Pi(\zeta,\Lambda;q)
  \nonu
   -i  {Z_1(\zeta,\Lambda)\over Z_3(\zeta,\Lambda)}~e_R^2
   \int_{\Lambda}  \frac{d^4k}{(2\pi)^4} 
   \nonu \times \textrm{Tr}\Bigl\{\gamma^\mu
    S_R(\zeta,k)~\Gamma^\nu_{R}(\zeta;k,k-q)~S_R(\zeta,k-q)\Bigr\}
~~.
\label{ten_pol}\ee
Notice that $T^{\mu\nu}$ is a symmetric tensor, 
and therefore also the rightmost term it has to be. One can convince himself by 
recalling that one has at disposal only one four-vector, $q^\mu$ for
constructing  antisymmetric contributions.
 It is understood that $\Pi^{\mu\nu}$, microscopically described by the second line   in Eq. \eqref{ten_pol}, 
  must satisfy the transversity property,
 i.e $ q_\mu\Pi^{\mu\nu}=0,~~ \Pi^{\mu\nu} q_\nu=0$. Hence, one has to verify that  
  \be
\int_{\Lambda}  \frac{d^4k}{(2\pi)^4} \textrm{Tr}\Bigl\{\psla q
    S_R(\zeta,k)\nonu \times\Gamma^\nu_{R}(\zeta;k,k-q)~S_R(\zeta,k-q)\Bigr\}
   =0~~,
   \nonu
   \int_{\Lambda}  \frac{d^4k}{(2\pi)^4}  \textrm{Tr}\Bigl\{\gamma^\mu
    S_R(\zeta,k)\nonu \times ~\Gamma_{R}(\zeta;k,k-q)\cdot q~S_R(\zeta,k-q)\Bigr\}=0~~.
\label{ten_tran1}   \ee
Since, we are  adopting a vertex that automatically fulfills the WTI, 
the second line in Eq. \eqref{ten_pol} can be easily demonstrated by using the 
WTI itself and the
dimensional regularization,  in order to make formally allowed a  shift in the integrand. As a matter of fact, one
gets
 \be  
\int  \frac{d^dk}{(2\pi)^4} \textrm{Tr} \Bigl\{\gamma^\mu
    S_R(\zeta,k)\nonu \times ~\Gamma_{R}(\zeta;k,k-q)\cdot q~S_R(\zeta,k-q)\Bigr\}
    \nonu
    = \int  \frac{d^dk}{(2\pi)^4} \textrm{Tr}\Bigl\{\gamma^\mu
     \left[S_R(\zeta,k-q)-~S_R(\zeta,k)\right]\Bigr\}
   =0 
   \label{ten_tran2}
 \ee 
The equality in the first line of Eq. \eqref{ten_tran1} is  more involved,
 but for ensuring that the microscopic calculation of $\Pi^{\mu\nu}$ be proportional to $T^{\mu\nu}$
 one should recover the structure 
 $$g^{\mu\nu} A+ {q^\mu q^\nu\over q^2} B$$
with the needed relations $A=-B$. This is guaranteed by Eq. \eqref{ten_tran2}, that follow from the fulfillment of
WTI. 
 
 In order to single out the photon self-energy, $\Pi(\zeta,\Lambda;q)$, one can
 proceed by saturating the polarization tensor with any combination of $g^{\mu\nu}$ and
 $q^\mu q^\nu/q^2$, but  it is extremely useful
 to take full advantage
and guidance from the analyses carried out in perturbative regime, (see, e.g., 
  Refs. \cite{Roberts:1994dr,Kizilersu:2009kg}).  Hence, 
 one  can saturate both sides in Eq.
\eqref{ten_pol} with the  tensor $\mathcal{P}^{\mu\nu}$ given by
\be
{\cal P}^{\mu\nu}=g^{\mu\nu} - 4 {q^\mu q^\nu\over q^2}~~.
\label{projN}\ee
This tensor has been introduced in previous works 
(see Refs.   \cite{Kizilersu:2009kg,Brown:1988bm,Brown:1988bn}) in order to project 
$T^{\mu\nu}$ on its $q^\mu q^\nu$ part, avoiding to  deal with quadratic 
singularities proportional to $g^{\mu\nu}$ present in $\Pi^{\mu\nu}$.
We emphasize that  such a projector is adopted for convenience reasons.
As a matter of fact, apparent quadratic singularities 
are met in the following elaboration,
  but the choice of the vertex presented in Sect. \ref{sect_vert}   (see also  
   \ref{app_4}) ensures 
  their cancellations. These apparent singularities are easily 
  bypassed by exploiting ${\cal P}^{\mu\nu}$, without carrying out a lengthy 
  algebra
(see also  Refs.  \cite{Brown:1988bm,Brown:1988bn}). As a final remark, it should be pointed out that the formal manipulation shown 
in  \ref{app_4} 
 needs a dimensional regularization of some terms and therefore 
one should  substitute $4$ with a generic dimension $d$ in the expression of  ${\cal P}^{\mu\nu}$.

In conclusion, one gets
the following expressions for 
 the regularized self-energy 
\be
q^2~\Pi_Z(\zeta,\Lambda;q)=-i
     {Z_1(\zeta,\Lambda)\over 3}~e_R^2
    \int_{\Lambda}  \frac{d^4k}{(2\pi)^4} ~{\cal P}_{\mu\nu}
    \nonu \times \textrm{Tr}
    \left[\gamma^\mu
    S_R(\zeta,k)\Gamma_R^\nu(\zeta,k,q)~S_R(k-q)\right]~~,
 \label{self_regPh}   \ee
    where $\Pi_Z(\zeta,\Lambda;q)=Z_3(\zeta,\Lambda)~\Pi(\zeta,\Lambda;q)$. This entails for the 
     renormalized self-energy, Eq. \eqref{self_Rph1}, 
    \be
    \Pi_R(\zeta;q)=-i
     Z_1(\zeta,\Lambda)~{4\over 3}~e_R^2
    \int_{\Lambda}  \frac{d^4k}{(2\pi)^4} ~{\cal P}_{\mu\nu}
    \nonu \times 
    \Bigl\{{1\over 4q^2}\textrm{Tr}\left[\gamma^\mu
    S_R(\zeta,k)\Gamma_R^\nu(\zeta,k,q)~S_R(\zeta,k-q)\right]
    \nonu
    -{1\over 4\zeta^2_p}\textrm{Tr}\left[\gamma^\mu
    S_R(\zeta,k)\Gamma_R^\nu(\zeta,k,q)~S_R(\zeta,k-q)\right]_{q^2=\zeta^2_p}\Bigr\}~~.
  \nonu  \label{self_renPh} \ee
 \subsection{ The fermion gap equation and the NWFs}
 \label{subsect_gapF}
As it is shown in details in  \ref{app_3},
 one can exploit the NIR of   ${\cal A}_R(\zeta;p)$ and  ${\cal B}_R(\zeta;p)$,
 Eq. \eqref{dse_F1}, and  the KL representations of both  fermion and photon
 propagators, 
 Eqs.   
 \eqref{lehmR} and \eqref{lehmRph} respectively, for obtaining the following relations
 \be
 {\cal A}_R(\zeta;p)=\int_{s_{th}}^\infty ds 
   ~\frac{(\zeta^2-p^2)~\rho_A(s,\zeta)}{(p^2-s+i\epsilon)~(\zeta^2-s+i\epsilon)}
  \nonu =
   {\cal T}_A(\zeta,\Lambda;p) -
\left. {\cal T}_A(\zeta,\Lambda;p)\right|_{p^2=\zeta^2}
\nonu\label{dse_Ar}
\ee
and 
\be 
{\cal B}_R(\zeta;p)
=\int_{s_{th}}^\infty ds 
   ~\frac{(\zeta^2-p^2)~\rho_B(s,\zeta)}{(p^2-s+i\epsilon)~(\zeta^2-s+i\epsilon)}
   \nonu =
    {\cal T}_B(\zeta,\Lambda;p) -
\left. {\cal T}_B(\zeta,\Lambda;p)\right|_{p^2=\zeta^2}\nonu
 \label{dse_Br}\ee
  where
  \be
{\cal T}_A(\zeta,\Lambda;p)=
-iZ_1(\zeta,\Lambda)~e^2_R ~\int_{0}^\infty 
  d\omega ~\bar\sigma_\gamma(\omega,\zeta,\zeta^2_p)
  \nonu\times\int_{0}^\infty ds'
~\int_{\Lambda} {d^4k\over(2\pi)^4}~\frac{1}{(p-k)^2-\omega+i\epsilon}
~{1\over  k^2 -s'  +i\epsilon}
\nonu \times ~{1\over 4 p^2}\textrm{Tr}\left\{ \left[
   \slashed{k} \bar \sigma_V(s',\zeta,s'_{th}) +
  \bar\sigma_S(s',\zeta,s'_{th})\right]
 \bar\Gamma^\beta_T~\psla p\gamma_\beta\right\}~,
\nonu\label{eq:TA}\ee
and
\be 
{\cal T}_B(\zeta,\Lambda;p)=
-iZ_1(\zeta,\Lambda)~e^2_R ~\int_{0}^\infty 
  d\omega ~\bar\sigma_\gamma(\omega,\zeta,\zeta^2_p)
  \nonu\times~\int_{0}^\infty ds'
~\int_{\Lambda} {d^4k\over
(2\pi)^4}~\frac{1}{(p-k)^2-\omega+i\epsilon}
{1\over  k^2 -s'  +i\epsilon}
\nonu \times~{1\over 4 } \textrm{Tr}\left\{ \left[
    \slashed{k} ~\bar\sigma_V(s',\zeta,s^\prime_{th}) +
 \bar \sigma_S(s',\zeta,s^\prime_{th})\right]
~\bar\Gamma^\beta_T
~\gamma_\beta\right\}~~.
\label{eq:TB}\ee
In Eqs. \eqref{eq:TA} and \eqref{eq:TB}, we have
\be
 \bar\sigma_\gamma(\omega,\zeta,\zeta_p)=\delta\Bigl(\omega-\zeta^2_p\Bigr)
 +\sigma_{\gamma} (\omega,\zeta)\Theta(\omega-\zeta^2_p)~~,
\nonu
    \bar{\sigma}_{S(V)}(s',\zeta,s'_{th}) = \delta\Bigl(s'-m^2(\zeta)\Bigr)
   \nonu +\sigma_{S(V)} (s',\zeta)\Theta\Bigl(s'-s'_{th}\Bigr)~,
    \nonu
    \bar\Gamma^\beta_T=\frac{\gamma^\beta_T}{2}~F_{{\cal A}_+}(k,p,\zeta)  
  -\Bigl((\psla p+\psla k)~p^\beta_T~
  +(p-k)^2 {\gamma^\beta_T\over 2}
  \nonu+i  \gamma_5\epsilon^{\beta\alpha \nu\rho} \gamma_{\alpha}p_\nu k_\rho
\Bigr)~F_{{\cal A}_-}(k,p,\zeta)   
   - 2p^\beta_T ~F_{\cal B}(k,p,\zeta) 
\label{def_sig} \ee
with $\gamma_T^\nu = \gamma^\nu -  q^\nu q\cdot \gamma /q^2$ and
 $p^\mu_T = p^\mu - q^\mu q\cdot p /q^2$.
 
In both  $F_{{\cal A}_+}$ and $F_{{\cal A}_-}$,  
a term $\mathcal{A}_R(\zeta;p)$ is present. The one in $F_{{\cal A}_+}$ generates
 a severe
divergent behavior in $k$ (see Eqs. \eqref{eq:TA} and \eqref{eq:TB}) that cannot be regularized by subtraction, since
the corresponding term in ${\cal T}_{A(B)}$, being evaluated at $p^2=\zeta^2$,
yields
$\mathcal{A}_R(\zeta;\zeta)=0$, by definition.
 A simple power counting in $k^2$ reveals that 
in ${\cal T}_A$ and 
${\cal T}_B$, only the combination  proportional to 
$ F_{{\cal A}_+}-(k^2-p^2)F_{{\cal A}_-}$ allows one to mitigate the
divergent
behavior due to $\mathcal{A}_R(\zeta;p)$ in $F_{{\cal A}_+}$, leading to 
a logarithmic divergence 
that can be regularized by the  subtraction in Eqs.\eqref{dse_Ar} and \eqref{dse_Br} (see details in
  \ref{app_3}). 
In fact, one has 
\be
  \label{eq:Unfolding}
  F_{{\cal A}_+}(k,p,\zeta)-(k^2-p^2)F_{{\cal A}_-}(k,p,\zeta) 
  \nonu =2\left(1-\mathcal{A}_R(\zeta;k) \right)~~.
\ee
 This cancellation highlights the intrinsic limitation of 
 the BC vertex, since it is necessary to go beyond such a  contribution 
 for restoring the  multiplicative renormalizability. This has been
 known from a long time (see, e.g. Ref. \cite{Curtis:1990zs}), but it is 
 relatively more recent the suggestion
 that the constraints coming from the curl of the vertex allow one to elaborate
  transverse contributions suitable for ensuring 
   the multiplicative renormalizability  (see, e.g.,  Ref.
   \cite{Qin:2013mta}). In  \ref{app_3} it is explicitly shown 
    how non-multiplicatively renormalizable contributions, from the BC
    term, Eq. \eqref{eq:BallChiu},  and the transverse ones, Eq. \eqref{vtren_qin},
   cancel  each other.

   Once the explicit expressions of the relations in Eqs \eqref{dse_Ar} and \eqref{dse_Br},  are obtained 
   as  in Eq \eqref{app_3_ARf} and \eqref{app_3_BRf}, respectively, 
   by using a spacelike external momentum 
    $p$ in order to avoid unnecessary formal complexities (recall that the NWFs are real functions that do not depend upon the
    external momenta as one can  also assess a posteriori), 
    one can extract the integral equations fulfilled by the 
   corresponding NWFs
   $\rho_A$ and $\rho_B$, 
 after  
assuming that the uniqueness theorem by Nakanishi \cite{nakabook} 
 can be applicable to the non-perturbative regime.

In particular comparing Eq \eqref{app_3_ARf} and the lhs of Eq. \eqref{dse_Ar}, one gets the desired relation 
for $\rho_A$ (see  \ref{app_3})
\be
\Theta\Bigl(y-s_{th}\Bigr)~ \rho_A(y,\zeta)
  = 
 {3\over (4\pi)^2}~e^2_R \lim_{\Lambda \to \infty}~Z_1(\zeta,\Lambda)
\nonu \times 
 ~\int_{0}^\infty~d\omega~\bar \sigma_\gamma(\omega,\zeta,\zeta_p,\Lambda)
~\int_0^1 d\xi~\int_{0}^\infty~ds'
\nonu~\Biggl\{ 
\bar\sigma_V(s',\zeta,s'_{th},\Lambda) ~
 \Biggl[\xi~
 \Theta\Bigl(y\xi (1-\xi) - \xi \omega -(1-\xi) s'\Bigr)
 \nonu- \int_0^{1-\xi}dt
\Theta\Bigl( yt (1-t) - \xi \omega -t s'\Bigr) \Biggr]
+  \bar\sigma_V(s',\zeta,s'_{th},\Lambda)
\nonu \times
\Biggl[~\int_{s_{th}}^\infty~ds~
 ~\rho_A(s,\zeta,\Lambda)~{\cal C}^{(0)}_{AV}(\zeta,\omega,s,s',\xi,y)
\nonu 
+  y~\int_{s_{th}}^\infty~ds~
 ~\rho_A(s,\zeta,\Lambda)~{\cal C}^{(1)}_{AV}(\zeta,\omega,s,s',\xi,y)\Biggr]
\nonu-  y\bar \sigma_S(s',\zeta,s'_{th},\Lambda)
 \int_0^{1-\xi}dt\int^{1-\xi-t}_0 {dw} 
 \nonu \times~\int_{s_{th}}^\infty ds
~\rho_B(s,\zeta,\Lambda)  
\nonu \times ~  \Delta' \Bigl[y  -s+
  {s {\cal A}_4(t,w)-\xi \omega -t s' -ws\over{\cal A}_4(t,w)}\Bigr]\Biggr\}
~~,
\label{rhoA}
\ee
with  ${\cal A}_4(t,w)=(t+w)~(1-t-w)$, 
\be
 \Delta'\Bigl[y-s+(sA-B)/A\Bigr]
 \nonu
 =
 {\delta[y-s + (sA-B)/A] - \delta(y-s)
\over  (sA-B)}~~,
\label{DeltapText}
\ee  
\be
 {\cal C}^{(0)}_{AV}(\zeta,\omega,s,s',\xi,y)= 
  ~{1\over (\zeta^2-s+i\epsilon)}~
 \nonu
 \times~\Biggl\{\xi~
 \Theta\Bigl[y\xi (1-\xi) - \xi \omega -(1-\xi) s'\Bigr]
 \nonu- \int_0^{1-\xi}dt~
\Theta\Bigl[ yt (1-t) - \xi \omega -t s'\Bigr] \Biggr\}
 \nonu + \int_0^{1-\xi}dt
 \Biggl\{{1\over (1-\xi)}
 ~\delta\Bigl[y -{\xi \omega +t s' +(1-\xi-t) s\over \xi(1-\xi)}\Bigr] 
\nonu -\int^{1-\xi-t}_0 {dw\over {\cal A}_4(t,w)}
 \delta\Bigl[ y -{\xi \omega +t s' +w s\over{\cal A}_4(t,w)}\Bigr]
 \Biggr\} 
~~,\nonu
\label{cal_AV0}
\ee
and
\be
{\cal C}^{(1)}_{AV}(\zeta,\omega,s,s',\xi,y)= 
   \int_0^{1-\xi}dt
 ~\Biggl\{{ (1+\xi)\xi}
 \nonu \times~
 \Delta'\Bigl[y -s+{s \xi(1-\xi)-\xi \omega -t s' -(1-\xi-t) s\over 
 \xi(1-\xi)}\Bigr]
 \nonu-\int^{1-\xi-t}_0 \hspace{-0.4 cm}{dw}
 ~ \Delta'\Bigl[y -s+{s{\cal A}_4(t,w) -\xi \omega -t s' -ws 
 \over{\cal A}_4(t,w)}\Bigr]  
\Biggr\} 
~.
\nonu\label{cal_AV1}
\ee
It has to be pointed out that the presence of the function $\Delta'$
 does not prevent a quantitative investigation (to be
presented elsewhere)  once an integration with a 
function smooth enough is carried out. Indeed, the existence of NWFs fulfilling a suitable smoothness
property will be the target of future numerical investigations.
 In the meanwhile, we should consider as an encouraging hint the results of
 the first quantitative check discussed in Sect. \ref{sect_res} since the  obtained first-order NWFs lead to well-defined integrals in Eq. \eqref{rhoA} when 
 iterating further.

As to $\rho_B$, after comparing Eq \eqref{app_3_BRf} and the lhs of Eq. \eqref{dse_Br}, one extracts
\be
\Theta\Bigl(y-s_{th}\Bigr)~\rho_B(y,\zeta)
 =
 -~{3 \over  (4\pi)^2}~e^2_R~\lim_{\Lambda \to \infty}~
 Z_1(\zeta,\Lambda )
\nonu \times~\int_{0}^\infty~d\omega~\bar \sigma_\gamma(\omega,\zeta,\zeta_p,\Lambda)
~\int_0^1 d\xi \int_{0}^\infty~ds'
\nonu \times~\Bigg\{  
\bar \sigma_S(s',\zeta,s'_{th},\Lambda)~ 
 ~
 \Theta\Bigl[ y\xi (1-\xi) - \xi \omega -(1-\xi) s'\Bigr]
\nonu+
\bar\sigma_S(s',\zeta,s'_{th},\Lambda)~
 \nonu \times \Biggl[~\int_{s_{th}}^\infty~ds~
{\rho_A(s,\zeta,\Lambda)}~{\cal C}^{(0)}_{AS}(\zeta,\omega,s,s',\xi,y)
\nonu+y~
 ~\int_{s_{th}}^\infty~ds~
{\rho_A(s,\zeta,\Lambda)}~{\cal C}^{(1)}_{AS}(\zeta,\omega,s,s',\xi,y)\Biggr]
\nonu+y~
\bar \sigma_V(s',\zeta,s'_{th},\Lambda)~
 \int_0^{1-\xi}dt\int^{1-\xi-t}_0 {dw}
\nonu \times ~ \int_{s_{th}}^\infty~ds~
\rho_B(s,\zeta,\Lambda) 
~\nonu \times 
  ~\Delta' \Bigl[y  -s
  +{s {\cal A}_4(t,w) -\xi \omega -t s' -ws\over{\cal A}_4(t,w)}\Bigr]
 \Biggr\}~~,
 \label{rhoB}
\ee 
  with 
\be
{\cal C}^{(0)}_{AS}(\zeta,\omega,s,s',\xi,y)=
{1\over \zeta^2-s+i\epsilon}
\nonu \times~ 
 \Theta\Bigl[ y \xi (1-\xi)- \xi \omega -(1-\xi) s'
\Bigr]
 + {1\over \xi (1-\xi)}
 \nonu \times~\int_0^{1-\xi}dz
 ~\delta\Bigl[ y -{\xi \omega +z s' +(1-\xi-z)
 s\over \xi(1-\xi)}\Bigr]  
~~,
\label{cal_AS0}\ee 
and
\be
{\cal C}^{(1)}_{AS}(\zeta,\omega,s,s',\xi,y)=
{1\over \zeta^2-s+i\epsilon}~ 
  \int_0^{1-\xi}dt
   ~ \Biggl\{ (1-\xi)
   \nonu \times~
  \Delta' \Bigl[y  -s + {s \xi (1-\xi) -\xi \omega -t s' -(1-\xi-t) s
 \over \xi (1-\xi)}\Bigr]
 \nonu+
 \int^{1-\xi-t}_0 {dw}
 \nonu \times ~
~\Delta' \Bigl[y 
 -s+{s {\cal A}_4(t,w) -\xi \omega -t s' -ws\over{\cal A}_4(t,w)}\Bigr]
 \Biggr\}~~.
\label{cal_AS1}\ee
 
 \subsection{The photon gap equation and the NWF}
 \label{subsect_gapPh}
 In  \ref{app_4},  the details are
 given for obtaining the integral equation fulfilled by the NWF 
 $\rho_\gamma$  (see  Eq. \eqref{NIRph}), exploiting both the integral equation  that determines the 
 renormalized photon self-energy, Eq. \eqref{self_renPh} and the uniqueness theorem \cite{nakabook}.  
 One can write 
 \be
 \Pi_R(\zeta;q)=~ \int_{\zeta^2_p}^\infty ds ~
  \frac{(\zeta^2_p-q^2)~\rho_\gamma(s,\zeta)}{(\zeta^2_p-s+i\epsilon)~(q^2-s+i\epsilon)}
  \nonu =
  \Bigl[{\cal T}_P(\zeta,\Lambda;q) -
\left. {\cal T}_P(\zeta,\Lambda;q)\right|_{q^2=\zeta^2_p}\Bigr]~~,
 \ee
 where
 \be
 {\cal T}_P(\zeta,\Lambda;q)=-i
     Z_1(\zeta,\Lambda)~{4\over 3}~e_R^2
    \int_{\Lambda}  \frac{d^4k}{(2\pi)^4} ~{\cal P}_{\mu\nu}
    {1\over 4q^2}
    \nonu \times ~\textrm{Tr}\left[\gamma^\mu
    S_R(\zeta,k)\Gamma_R^\nu(\zeta,k,q)~S_R(\zeta,k-q)\right]~~.
 \nonu\ee
Then, following the formal steps in  \ref{app_4}, where a spacelike $q^2$ has been adopted 
for a straightforward elaboration without  loss of generality (as in the fermionic case), one gets
\be
{ \Theta(y-s^p_{th})~\rho_\gamma(y,\zeta)=} -{e_R^2\over (2\pi)^2}~
\lim_{\Lambda\to\infty}~Z_1(\zeta,\Lambda)\int_{0}^\infty ds 
\nonu\times ~
     \int_{0}^\infty ds'~ \int_0^1 d\xi
~  \Biggl\{ \bar\sigma_V(s',\zeta,s'_{th},\Lambda) ~ 
\bar\sigma_V(s,\zeta,s_{th},\Lambda)
\nonu \times ~
 2 
 \xi (1-\xi)  
 \Theta\Bigl[ y \xi(1-\xi)- \xi s'-(1-\xi)s) \Bigr] 
 \nonu\times ~\left( 1
 + \int_{s_{th}}^\infty d\omega ~
{\rho_A(\omega,\zeta ,\Lambda)\over 
(\zeta^2-\omega +i\epsilon)}  \right) 
 \nonu +\int_{s_{th}}^\infty d\omega 
   ~\rho_A(\omega,\zeta,\Lambda)~{\cal C}_\gamma(s,s',\xi,\omega) 
  \nonu+
  2\bar\sigma_S(s',\zeta,s'_{th},\Lambda) ~ \bar\sigma_V(s,\zeta,s_{th},\Lambda)~
  \nonu \times~\int_{s_{th}}^\infty d\omega 
   ~\rho_B(\omega,\zeta,\Lambda)~
  ~ \int_0^{1-\xi} dv\int_0^{1-\xi-v} dw ~(v+w)\nonu \times
  ~{1 -2 (v+w)\over {\cal A}_4^2(v,w)}
   ~{\partial\over \partial y}\delta \Bigl[y -{\cal A}_7(s,s',\omega,v,\xi,w) \Bigr]   
 \Biggr\}~~,
 \label{rhog}
\ee 
with 
\be
{\cal A}_7(s,s',\omega,v,\xi,w)
\nonu =
{vs'+(\xi+w)\omega+(1-\xi-v-w)s\over (v+w)~(1-v-w)}~~,
\nonu
{\cal C}_\gamma(s,s',\xi,\omega)=\int_0^{1-\xi} dv\int_0^{1-\xi-v} {dw\over {\cal A}_4(v,w)}
 \nonu \times ~ \Biggl\{ 2~\bar\sigma_V(s',\zeta,s'_{th}) ~ \bar\sigma_V(s,\zeta,s_{th}) 
  \nonu \times~\Biggl(\omega~ 
~{\partial\over \partial y}\delta \Bigl[ y-{\cal A}_7(s,s',\omega,v,\xi,w)\Bigr]   
 \nonu +
   \delta\Bigl[y -{\cal A}_7(s,s',\omega,v,\xi,w)\Bigr]\Biggr) 
\nonu+~\bar\sigma_S(s',\zeta,s'_{th}) ~ \bar\sigma_S(s,s'_{th},\zeta)  
\nonu \times 
~{\partial\over \partial y}\delta \Bigl[ y-{\cal A}_7(s,s',\omega,v,\xi,w)\Bigr] \Biggr\}~~,
\label{cal_gamma}\ee
and ${\cal A}_4(v,w)=(t+w)~(1-t-w)$. In order to obtain numerical 
solutions, the derivatives of the Dirac $\delta$ can be traded for the 
derivative of the $\rho$ functions, which, contrary to the $\bar{\sigma}$, 
does not contain itself any Dirac $\delta$. Just like in the fermion case, 
the first check performed in Sec. \ref{sect_res} yields results compatible 
with well defined integrals in \eqref{rhog}. One can also note that in the right-hand side
 of Eq. \eqref{rhog}, the dependence upon  $\rho_\gamma$ is not explicit, but buried into  the fermionic quantities.

\subsection{Some remarks}
 Concluding this Section, that presents our formal results, some remarks are in order.
 The  
coupled systems for determining $\rho_A$, $\rho_B$ and $\rho_\gamma$ is composed by the set 
of Eqs.
\eqref{rhoA}, \eqref{rhoB} and  \eqref{rhog}, supplemented with Eqs. \eqref{klwf} and
 \eqref{sigmag}. The fact that the NWFs do not depend on external momenta
  has been extensively used to build the system of equations. Indeed, Eqs. \eqref{klwf} and 
  \eqref{sigmag} are obtained from timelike (above threshold) momenta, 
  taking advantage of the real and imaginary part decomposition, while Eqs. \eqref{rhoA}, 
  \eqref{rhoB} and \eqref{rhog} are derived for spacelike external momenta to avoid 
  complications coming from singularities.
  Beside the above feature, the uniqueness theorem allows one to
  finalize the  formal steps.
  Interestingly, derivatives of Dirac delta distribution naturally appear in our derivations.
In summary, i) the NWFs are real functions that do not depend upon the external momentum,
 as {\em a posteriori}  can be checked  by  a direct inspection of
  the coupled system;  ii) once the NWFs are numerically  evaluated, 
   the scalar functions
${\cal A}_R(\zeta;p)$, ${\cal B}_R(\zeta;p)$ and $\Pi_R(\zeta;q)$ 
are known for {\em any value of any momenta}; iii) the
presence of the derivative of the delta-function is not an issue
 from the numerical point of view, as already observed
when the NIR approach  has been applied to the numerical solution of BSE 
(see, e.g., Refs. \cite{dePaula:2016oct,dFSV2}).

\section{A first application}
\label{sect_res}

After establishing the formal results, i.e. the system of integral equations that 
 the NWFs $\rho_A$, $\rho_B$ and $\rho_\gamma$ have to fulfill, it is important to test the consistency.
 Following the same spirit of the first applications of the NIR approach to the two-scalar system, where the 
 Wick-Cutokski model had to stem from the formal elaboration 
 (see, e.g., Refs. \cite{Hwang:2004mr,FSV1}), 
 we have performed  the  first 
iteration of the 
coupled system, as an initial
step, in view of the quantitative investigation in full, 
to be presented elsewhere. It should be pointed out that the first iteration
 offers the possibility to partly 
check our results, as it corresponds to the  standard one-loop computation.    Once the analytical expressions of $\rho^{(1)}_A$,  $\rho^{(1)}_B$ and   $\rho^{(1)}_\gamma$ 
have been obtained, we have carried out  the evaluations of i)  the KL weights for both fermion 
and photon propagators, ii)  the fermionic running mass and iii)  the charge
 renormalization function, and eventually
compared with one-loop results (see, e.g., Ref. \cite{Zuber}, but noting
the different renormalization scheme).

The  numerical investigation, aimed at establishing the validity of our approach by assessing 
 the convergence
of the iterative method,
will be presented  elsewhere. 
We stress that, generally speaking, the result of the
 iterative procedure may differ significantly from the first iteration 
 one {(see, e.g., Refs. \cite{Sasagawa:2016xol,Tanaka:2017sdd} and
  Ref. \cite{Dudal:2012zx} for  
 QCD studies)}, and that we perform here a basic test of consistency.
In  \ref{app_5} all the  details 
for obtaining the aforementioned first iteration,  Eqs.
\eqref{app5_rhoA}, \eqref{app5_rhoB} and \eqref{app5_rhog} respectively, 
 are illustrated with also some of their features, while in 
  \ref{app_6},  the    full 
 expressions of the coupled system  is
summarized for the convenience of the interested reader.

In Fig. \ref{fig_sigmaF}, the first-order K\"all\'en-Lehman weights for the fermionic propagator, Eq. \eqref{lehmR}, 
 are presented for different values of the  IR regulator $\zeta_p$. 
They can be easily obtained from Eq. \eqref{klwf} after inserting  $\rho^{(1)}_A$ and  $\rho^{(1)}_B$
given by (see Eqs.
\eqref{app5_rhoA}, with its careful discussion,  and  \eqref{app5_rhoB})
\be
\Theta\Bigl(y-s_{th}\Bigr)~ \rho^{(1)}_A(y,\zeta)
  = -{e^2_R\over 2(4\pi)^2}~{1\over \zeta^2_py^2}
  \nonu \times ~\Theta(y-m^2(\zeta))\Biggr\{\Theta\Bigl[[m(\zeta)+ \zeta_p]^2- y\Bigr]~\Bigl(y-m^2(\zeta)\Bigr)^3
     \nonu
   + \Theta\Bigl[y-[m(\zeta)+ \zeta_p]^2\Bigr]~\Bigl(y-m^2(\zeta)\Bigr)^3
  \nonu \times ~ 
  \Bigl[1- f(y,\zeta,\zeta_p^2)\Bigr]
  \Biggr\}~,
 \nonu \label{rhoA1}
\ee
with
\be
  \label{eq:fdef}
  f(y,\zeta^2,\zeta_p^2) =
  \sqrt{1 -\zeta^2_p~{2y+2m^2(\zeta)-\zeta^2_p\over (y-m^2(\zeta))^2}}~
 \nonu \times \Biggl[ 1+\zeta^2_p~{y+m^2(\zeta)-2 \zeta^2_p\over \Bigl(y-m^2(\zeta)\Bigr)^2}
   \Biggr]~~,
\ee
and 
\be
\Theta\Bigl(y-s_{th}\Bigr)~\rho^{(1)}_B(y,\zeta)
\nonu = -~ \frac{3e_R^2}{(4\pi)^2} ~ m(\zeta) ~  
\Theta\Bigl[y-[m(\zeta)+\zeta_p]^2\Bigr]
    \nonu
   \times~ 
\frac{1}{y}\sqrt{[y-m^2(\zeta)-\zeta_p^2]^2-4m^2(\zeta)\zeta_p^2}
  \label{rhoB1}
  \ee
Qualitatively
$\sigma_V$ and $\sigma_S$ are quite similar, though  width and tail of 
$\sigma_S$ are slightly larger than 
the $\sigma_V$ ones.
The common features are i) the negative values 
(in Ref. \cite{Jia:2017niz} the scalar weight is also negative), that is a consequence of our 
choice of the
renormalization scheme, and ii) the sharp peaks for $\zeta_p\to 0$.
The latter are very close to the threshold and  
  clearly depend upon 
  the IR-regulator $\zeta_p$, while the tails are unaffected, as expected.
To complete the discussion it is useful to {recall} that the residue at
the pole of the renormalized
fermion propagator, ${\cal R}_S$, is not equal to $1$,  as expected  in the
  RI'/MOM scheme  
  \cite{Sturm:2009kb,Gracey:2013sca}. Table \ref{tab:residue} summarizes 
  the different values of the residues associated with our curves in Fig. \ref{fig_sigmaF}.
 \begin{table}[t]
   \centering
   \begin{tabular}[h]{|c|cccc|}
     \hline
     $100~ \zeta_p/m$ & $4$ & $3$ & $2$ & $1$ \\
     \hline
     ${\cal R}_S$ & 1.10 & 1.11 & 1.13 & 1.16 \\
     \hline
   \end{tabular}
   \caption{Values of the fermion propagator residue ${\cal R}_S$ for different $\zeta_p$.}
   \label{tab:residue}
 \end{table}
 
   Another simple check is the formal comparison between the
   expression of $\rho^{(1)}_B(y,\zeta)$ for vanishing values of $\zeta_p$, with
   the expression one can extract from standard one-loop computations of
    ${\cal B}_R(\zeta;p)$ (see \emph{e.g.} \cite{Zuber}),
     but within the RI'/MOM scheme. In the limit $\zeta_p\to 0$,   one has from Eq. \eqref{rhoB1}
   \be
   \lim_{\zeta_p\to 0}\Theta\Bigl(y-s_{th}\Bigr)~~\rho^{(1)}_B(y,\zeta)=-3 ~
   {\alpha_{em}\over 4\pi}~\Theta\Bigl(y-m^2(\zeta)\Bigr)~
   \nonu \times ~m(\zeta)~\left[1 -{m^2(\zeta)\over y}\right]   
   \ee
   with $\alpha_{em}=e^2_R/4\pi$. Moreover from the definition in 
   Eq. \eqref{BR_NIR} one
   writes
   \be
   \Im m\Bigl\{{\cal B}_R(\zeta,p)\Bigr\} = 
   -\pi \Theta\Bigl(p^2-s_{th}\Bigr)~~\rho^{(1)}_B(p^2,\zeta).
  \label{imBR} \ee
   After imposing ${\cal B}_R(\zeta,\zeta)=0$, perturbative computations 
   yield (see \emph{e.g.} \cite{Zuber})
   \be
   {\cal B}^{1\textrm{-loop}}_R(\zeta,p)=m(\zeta)~{\alpha_{em}\over 4\pi}~3~{m^2-p^2\over p^2}
\nonu \times ~\ln\Bigl(1 -{p^2\over m^2(\zeta)}\Bigr)~~,
\ee
 Hence, for $p^2>m^2(\zeta)$ the logarithm becomes complex, and adopting the  same
 analytic continuation as in Ref. \cite{Zuber} (i.e. $\ln(-\rho)=\ln \rho -i\pi$),
 eventually one has
 \be
   \Im m\Bigl\{{\cal B}^{IZ}_R(\zeta,p\Bigr\} = \pi ~m(\zeta)~3~{\alpha_{em}\over 4\pi}~
   \left[1-{m^2(\zeta)\over p^2}\right]~~.
  \label{IBR_IZ} \ee
  that coincides with the result one gets from Eqs. \eqref{rhoB1} and \eqref{imBR}.
  As a final remark, one should point out that in the same limit ${\cal
  A}_R(\zeta,p)$ vanishes both in our case (see Eq. \eqref{app5_rhoAl}) as well as in Ref. \cite{Zuber}.  

Figure \ref{fig_sigmaG} shows the K\"all\'en-Lehman weight for the photon  propagator, Eq. \eqref{sigmag},
 obtained from $\rho^{(1)}_\gamma$ given by (see also Eq.
\eqref{app5_rhog}) 
\be
\Theta(y-s^p_{th})~\rho^{(1)}_\gamma(y,\zeta)
 =  ~- \frac{e_R^2}{3(2\pi)^2}\Theta(y)\Theta(y-4m^2(\zeta))
 \nonu \times ~\Bigl(1+2{m^2(\zeta) \over y}\Bigr)~\sqrt{1-4\frac{m^2(\zeta)}{y}}
~~.\label{rhoG1}
\ee
The independence from  the IR-regulator $\zeta_p$, as shown in the expression of $\rho^{(1)}_\gamma$, is the standard feature of the one-loop calculation, and only the
higher-order contributions will make apparent such a dependence. Differently from the fermion
case, the KL weight of the photon is positive, as expected. This bosonic result points to the highly
non trivial interplay of the two scalar functions ${\cal A}_R$ and ${\cal B}_R$, in order to obtain
positive KL weights for the fermionic source.
\begin{figure}
\includegraphics[width=8.5cm]{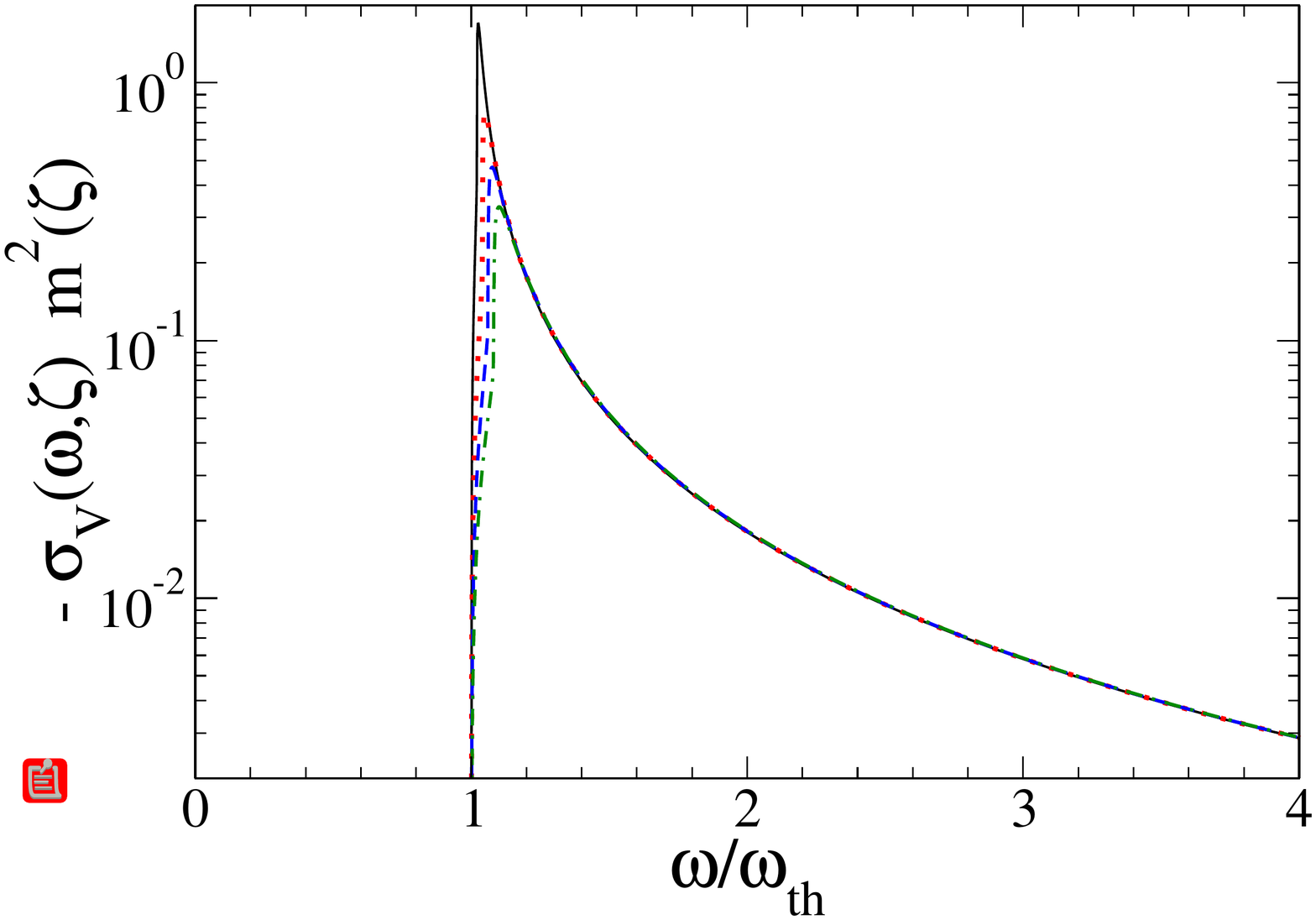}
\hspace{-1.0cm}\includegraphics[width=8.5cm]{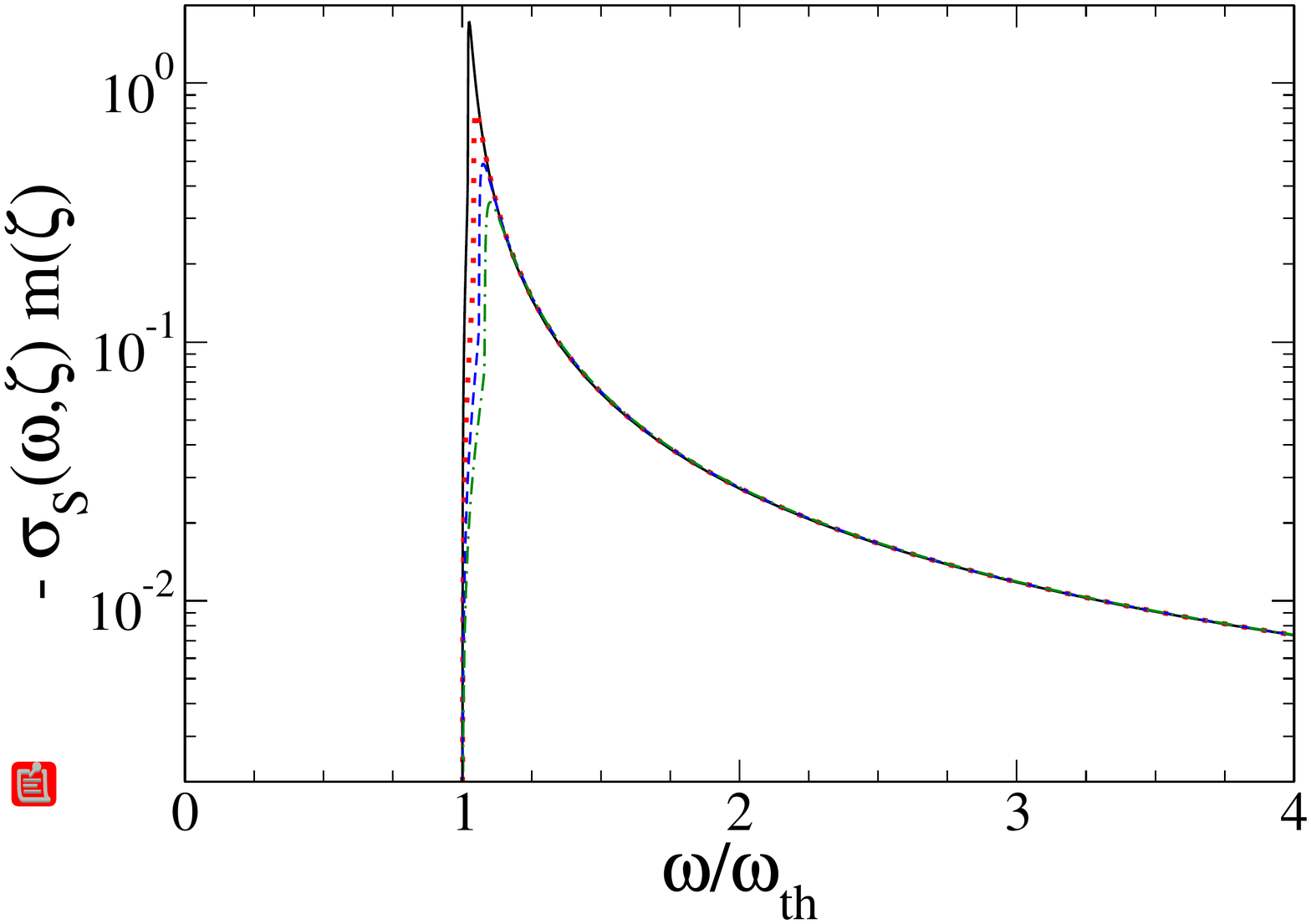}
\vspace*{-0cm}
\caption{The first iteration of the K\"all\'en-Lehman weights for the fermionic propagator, Eq. \eqref{lehmR}, 
for different values of the 
  IR-regulator, $\zeta_p$. The threshold is given by
   $\omega_{th}=m^2(\zeta)$.
Left panel the vector weight, $\sigma_V(\omega,\zeta)$. Right panel: scalar weight $\sigma_S(\omega,\zeta)$. 
 Solid line: $\zeta_p=0.01~m(\zeta)$.
 Dotted line:  $\zeta_p=0.02~m(\zeta)$.  Dashed line:  $\zeta_p=0.03~m(\zeta)$. Dash-dotted line:  
 $\zeta_p=0.04~m(\zeta)$.}
\label{fig_sigmaF}
\end{figure}
\begin{figure}
\begin{center}
\includegraphics[width=8.5cm]{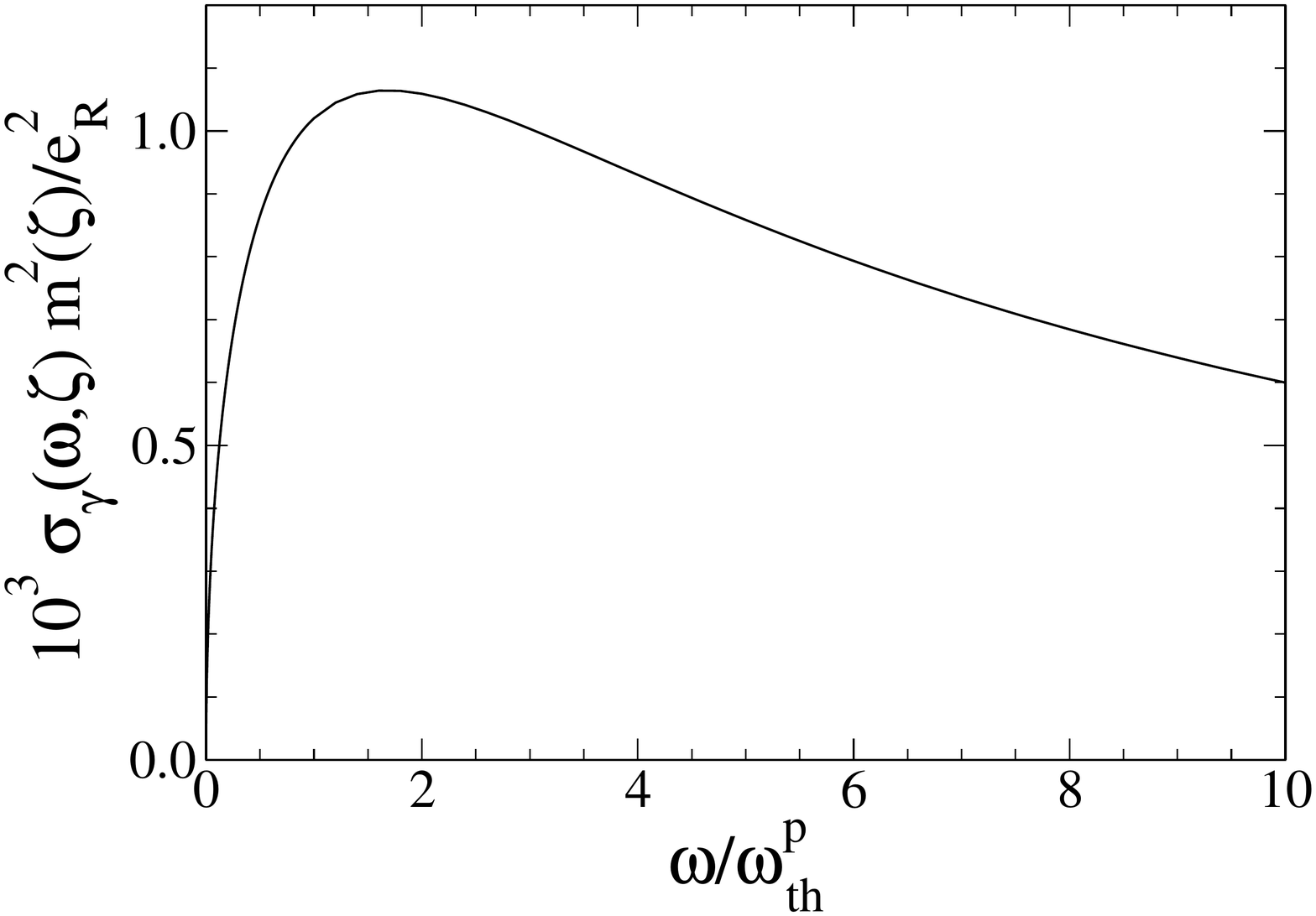}
\vspace*{-0cm}
\caption{ The first iteration of the K\"all\'en-Lehman weight for the photon
propagator, Eq. \eqref{lehmRph}. N.B. in this case 
there is no dependence upon  the 
  IR-regulator, $\zeta_p$, as shown in Eq. \eqref{rhoG1}. The threshold is given by $\omega_{th}^p=[2m(\zeta)]^2$.}
\label{fig_sigmaG}
\end{center}
\end{figure}
We have also calculated the running mass (see  Eq. \eqref{SR_cond} for the value at the
 renormalization point) 
 \be
 M(\zeta;p)={m(\zeta)+{\cal B}_R(\zeta;p)\over 1-{\cal A}_R(\zeta;p)} \nonu
 = m(\zeta) +{m(\zeta){\cal A}_R(\zeta;p)+{\cal B}_R(\zeta;p)\over 1-{\cal A}_R(\zeta;p)} ~~,
\label{run_mass}
\ee
and the charge renormalization function (.  Eq. \eqref{condph} for the value at the renormalization point)
\be G(\zeta;q^2)={\alpha_R(\zeta;q^2)\over \alpha_{em}}= {1\over 1+\Pi_R(\zeta;q)}~~.
\label{run_ch}\ee

In Fig. \ref{fig_RGIMbt}, the running mass is shown for values of the four-momentum below
the threshold, $s_{th}=m^2(\zeta)$, {adopting a tiny $\zeta_p$ up to
$\zeta_p/m(\zeta)=10^{-4}$}, while in Fig. \ref{fig_RGIMtime} both real and  imaginary terms,
generated
for $p^2\ge s_{th}$, are presented. Notice that  the positive sign of the
imaginary part is a consequence of the first-order calculation (see.  Eq. \eqref{IBR_IZ} and the
vanishing of ${\cal A}_R$ for $\zeta_p\to 0$).

\begin{figure}
\begin{center}
\includegraphics[width=8.5cm]{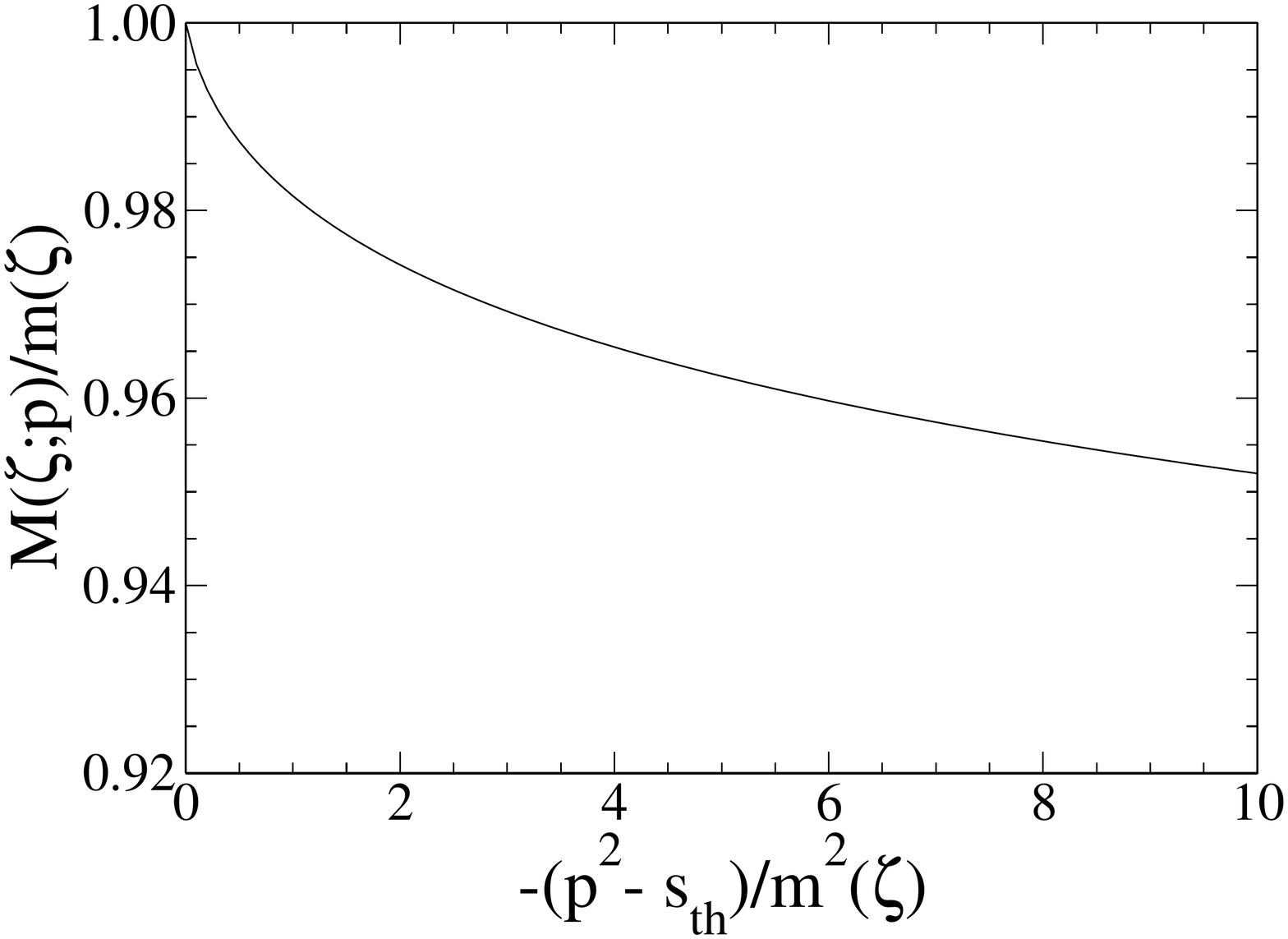}
\caption{ The  invariant mass  $M(\zeta;p)$, Eq. \eqref{run_mass}, below the threshold 
$s_{th}=m(\zeta)^2$, vs $p^2$.}
\label{fig_RGIMbt}\end{center}
\end{figure}
\begin{figure}
\includegraphics[width=8.5cm]{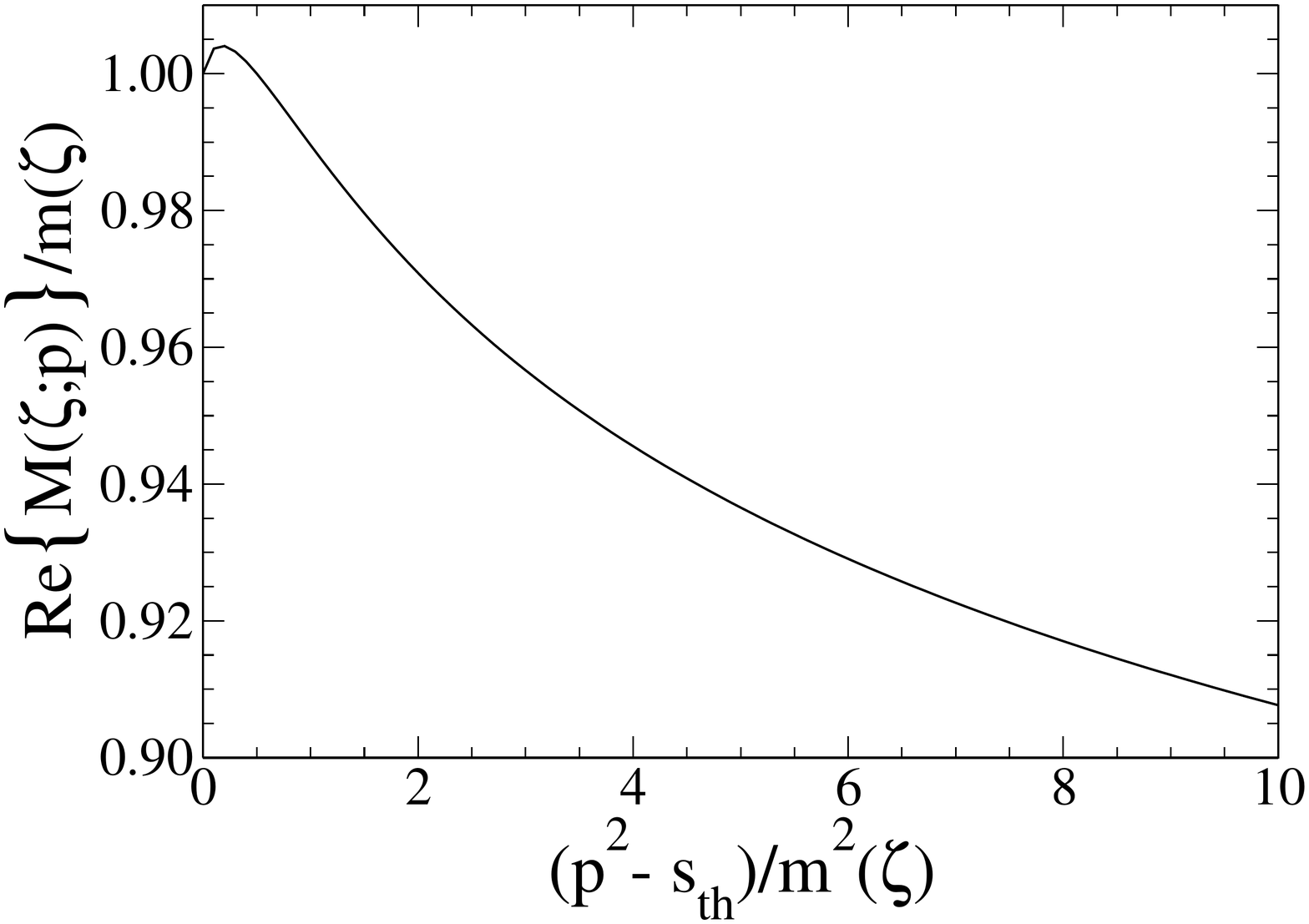}
\hspace{-1.0cm}\includegraphics[width=8.5cm]{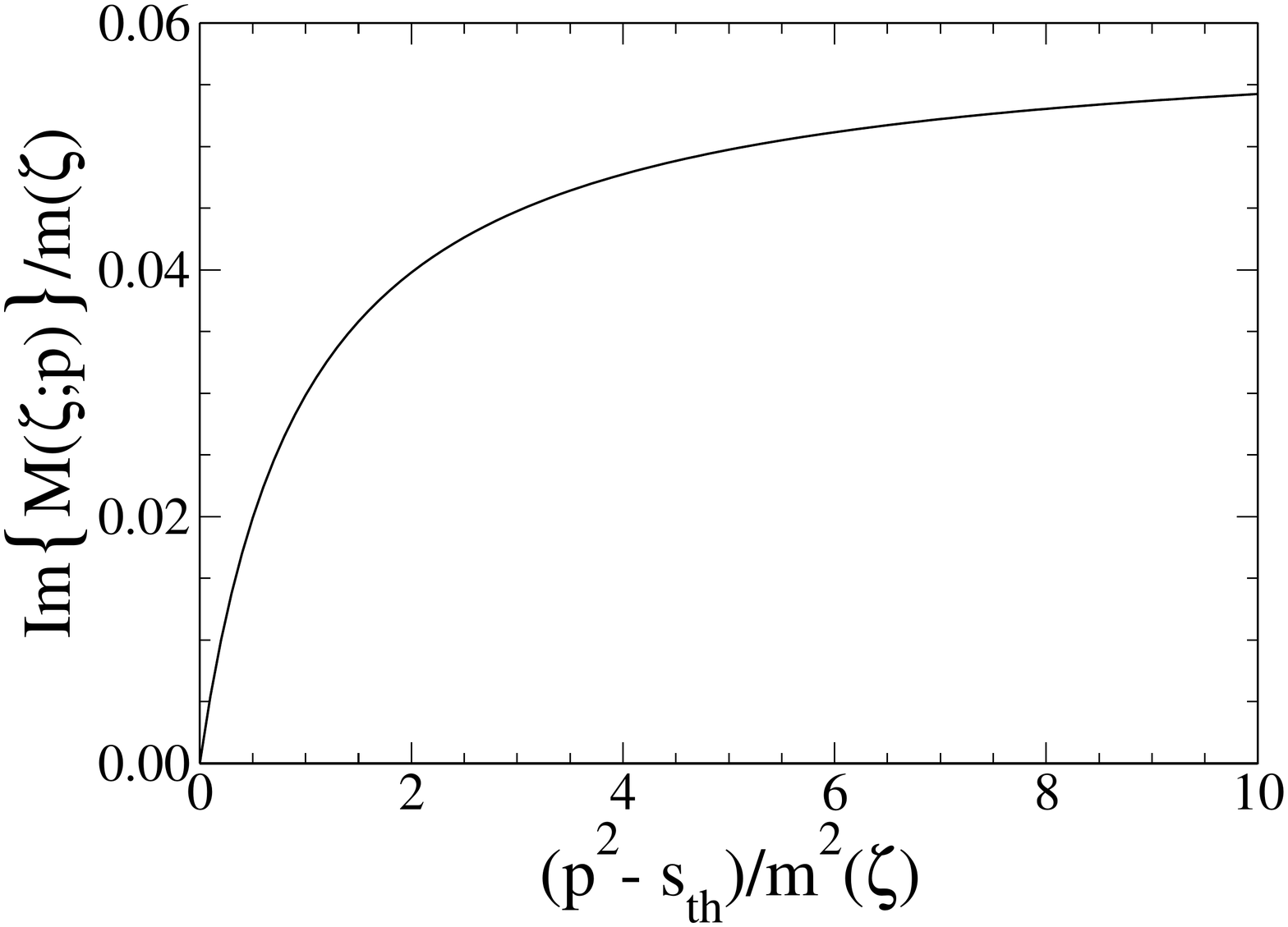}
\caption{ The same as in Fig. \ref{fig_RGIMbt}, but in the timelike region.
Left panel:
real part of the running mass. Right panel: imaginary part of the running mass.}
\label{fig_RGIMtime}
\end{figure}
\begin{figure}
\begin{center}
\includegraphics[width=8.5cm]{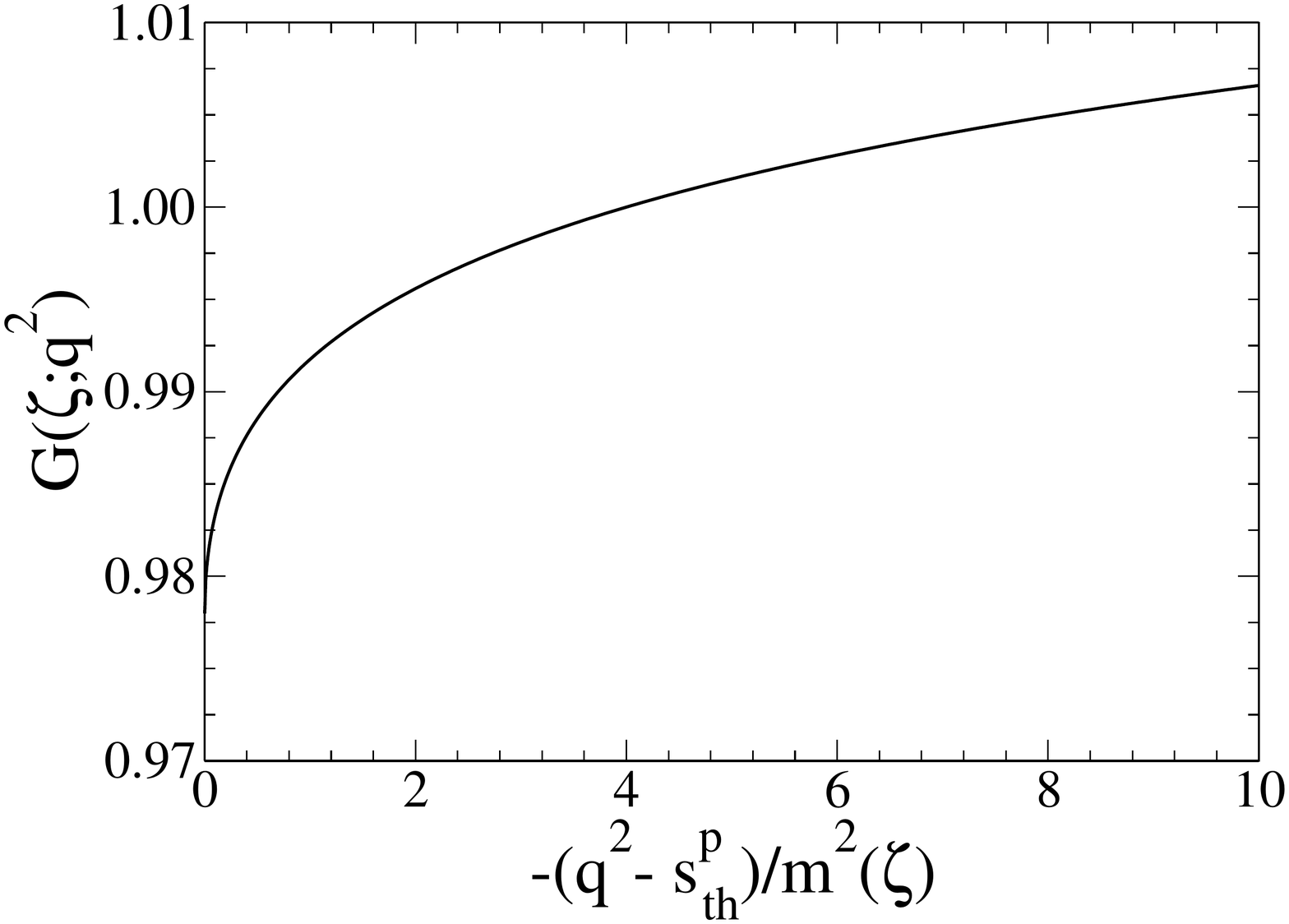}
\vspace*{-0cm}
\caption{ The running charge $G(\zeta;q^2)$,  Eq. \eqref{run_ch},  below the threshold vs $q^2$,
with $s^p_{th}=4m^2(\zeta)$.}
\label{fig_Rchbt}
\end{center}
\end{figure}
\begin{figure}
\includegraphics[width=8.5cm]{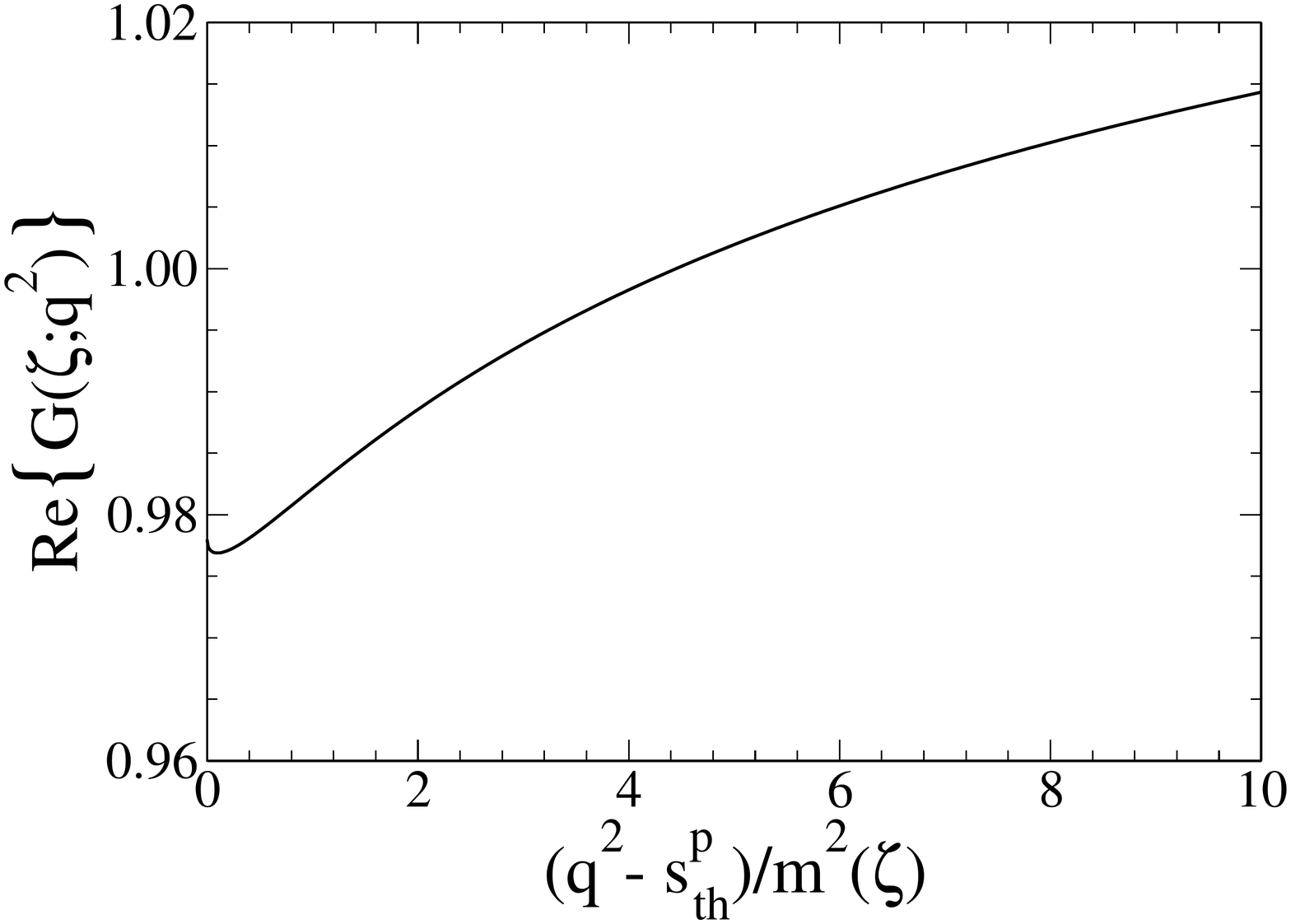} 
\hspace{-1.0cm}\includegraphics[width=8.5cm]{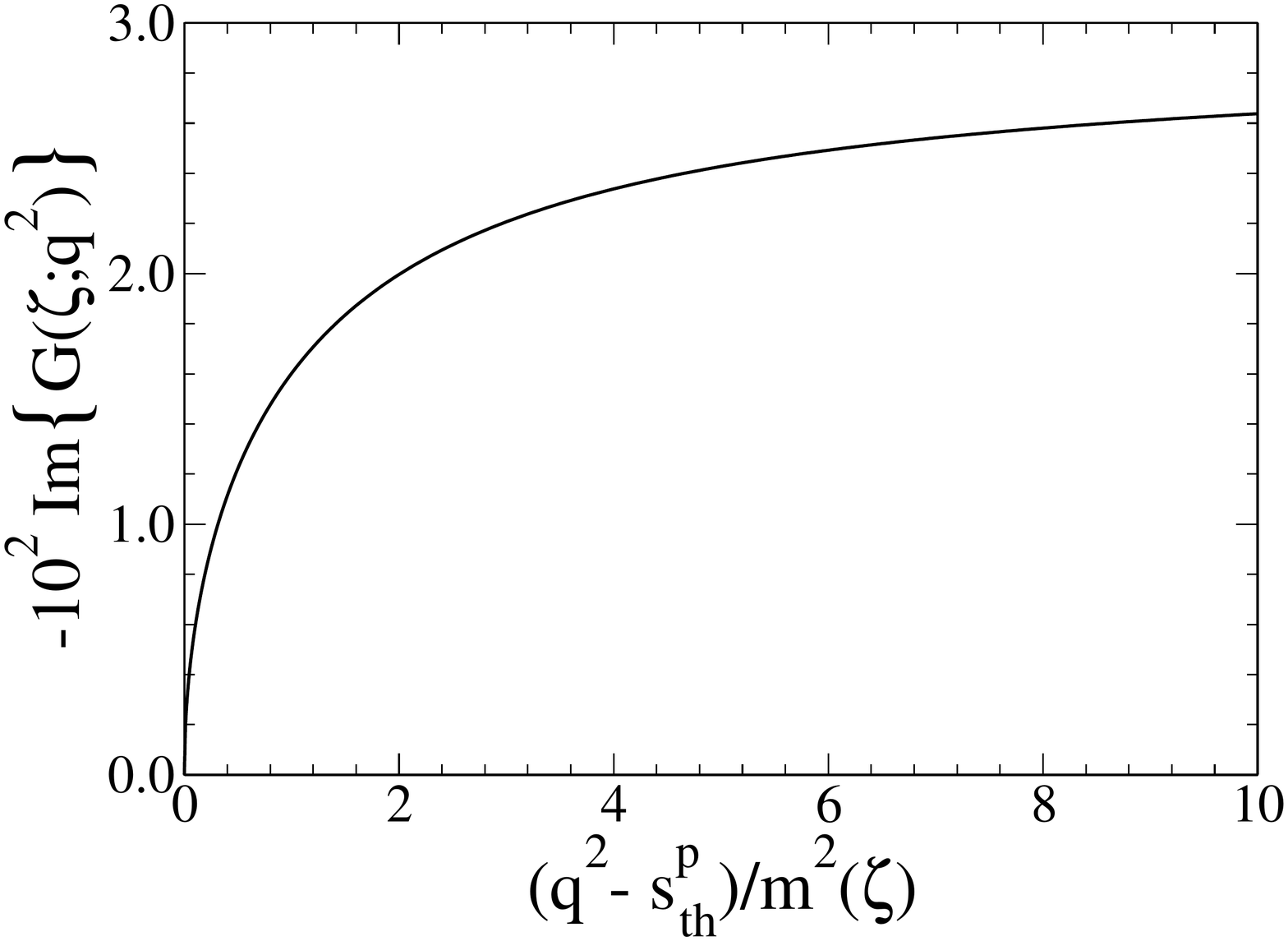}
\vspace*{-0cm}
\caption{ The same as in Fig. \ref{fig_Rchbt}, but for the timelike region.
 Left panel: the real part. Right panel: the imaginary part.}
\label{fig_Rchat}
\end{figure}
In Figs. \ref{fig_Rchbt} and   \ref{fig_Rchat}, the running charge defined in Eq. \eqref{run_ch} is shown for
values below and above the threshold, that in this case holds $s^p_{th}= 4m^2(\zeta)$.
The comparison with the results of Ref. \cite{Sauli:2003zp} 
(where $\Im m\{M(\zeta,p)\}$ has a
negative sign) can be be performed only at the qualitative
level, since the quantitative one is too early for our numerical efforts.
 In any case, one can recognize
quite similar pattern, in particular, for the case of small values 
of the coupling constant $\alpha_{em}$. It
should be pointed out that  the first-order self-energies 
(both fermion and photon ones) depend linearly
upon the coupling constant and therefore, the values of the running 
mass and the {\em relative} running charge
are not affected by such a dependence.

We would like to conclude this Section by shortly anticipating 
  a remark stemming from  the detailed numerical investigations to be presented
elsewhere.
Indeed,  in the quenched
 approximation, one can verify that the integrals relevant for
the
 second iteration are all converging, even the ones including the
  function $\Delta^\prime$ (see  Eq. \eqref{DeltapText}). {Of course, when extending the
  calculations to large values of the coupling constant, extra care on the convergence of the above
  quantities is requested, as one should expect additional singularity issues from Minkowski and Euclidean calculations 
   focusing on the strong-coupling regime, both below and above the critical value (see, e.g.,   
  Refs. \cite{Bashir:2011vg,Sauli:2003zp,Sauli:2020dmx,Hawes:1996ig} for the challenges  beyond the critical value).} 

\section{Conclusions and Perspectives}
\label{sect_CP}
This work belongs to the set of the early attempts (not too much numerous)
 to explore the non-perturbative regime of $QED_{3+1}$ directly in Minkowski space,
  by exploiting the
framework based on  the so-called Nakanishi integral representation
 for describing the self-energies of both fermion and photon. 

The originality of this work, elaborated within the RI'/MOM scheme, 
 lies in the choice of 
the fermion-photon vertex, able to fulfill constraints coming from 
both the Ward-Takashi identity  and the multiplicative renormalizability, that calls for purely
transverse contributions. We have shown that despite the apparent complexity, it is possible to
 derive a well-defined system of equations for the Nakanishi weight functions,
 that we recall are real functions fulfilling a uniqueness theorem, within the Feynman
 diagrammatic framework \cite{nakabook}. 
 In addition, we have presented an initial check
  based on the evaluation 
  of the
 first iteration of the coupled system. In particular, we have  initiated the comparison with
 known results of
   i) the K\"all\'en-Lehmann weights for both fermion and photon, ii)
 the running mass and iii) the charge renormalization function. Beyond this,
 we have also verified that numerical stability remains 
 under control, encouraging toward a more vast numerical investigation.

It has to {be} pointed out that
 the present results readily calls for two natural extensions on a short-time scale.
 First, complete numerical studies should be performed, allowing one to assess  the convergence of the whole approach and to move the
 comparison to a quantitative level, e.g. with the results in Refs. \cite{Jia:2017niz,Sauli:2003zp}.
  Second, the expected residue equal to one at the mass pole should be implemented at the
  level of the NWFs, \emph{i.e.} going from the RI'/MOM scheme to the standard on-shell 
  renormalization scheme.

  On a longer time-scale, the third desirable
  extension would be to move from QED to QCD. 
 An educated reader might object that many ingredients we used are not available or available 
 in a much more complicated way for QCD. For instance, there is no formal
  proof that the propagators of confined particles should have K\"all\'en-Lehmann-like 
  representation (positive definite). Nonetheless, 
  lattice-QCD computations seem to be consistent with a spectral representation
   (although not a positive one)
   \cite{Binosi:2019ecz}. Furthermore,
   the Ward-Takahashi identities must be replaced by the Slavnov-Taylor ones, 
   forcing deep modifications of the {quark-gluon} vertex function, playing an important 
   role in realizing a dynamical breakdown of chiral symmetry.
   Also for this issue, progresses have been recently done in that direction, 
   with the definition 
   of the non-Abelian generalization of the Ball-Chiu vertex 
   \cite{Aguilar:2018epe}. Therefore, despite the technical difficulties 
   to jump from QED to QCD, we believe that such a possibility should deserve a careful
   investigation.
   
    As a final remark, it is appropriate to recall that, for a given
   interacting system,   {our} final goal  is
   to implement and solve  the BSE with dressed
   propagators for both particles and quanta, directly in Minkowski space, i.e. where
   the physical processes take place.
   { In view of this, our present elaboration, that belongs to the phenomenological realm,
     should be considered a
   step forward  for setting  up a 
   viable  tool for the actual investigation of strong-coupling regimes 
   (first below the critical value)},  within a QFT framework  
   where the dynamical ingredients are made transparent and under control.}

\begin{acknowledgements}
We gratefully thank  Massimo Testa  
for very stimulating  discussions, and  Si-xue Qin for illustrating in detail the issues related to the transverse vertex.
C.M. acknowledges discussions with Jose Rodriguez-Quintero and the warm hospitality of INFN Sezione di Roma, where his INFN PostDoc fellowship for the NINPHA project
has been granted. 
\end{acknowledgements}

\appendix
\section{DSE for the fermion self-energy}
\label{app_3}
In this Appendix,  the formal elaboration for obtaining the equation that determines the 
renormalized self-energy $\Sigma_R(\zeta;p)$ is given in details.

\begin{figure}

\centerline{\includegraphics[width=5.0cm]{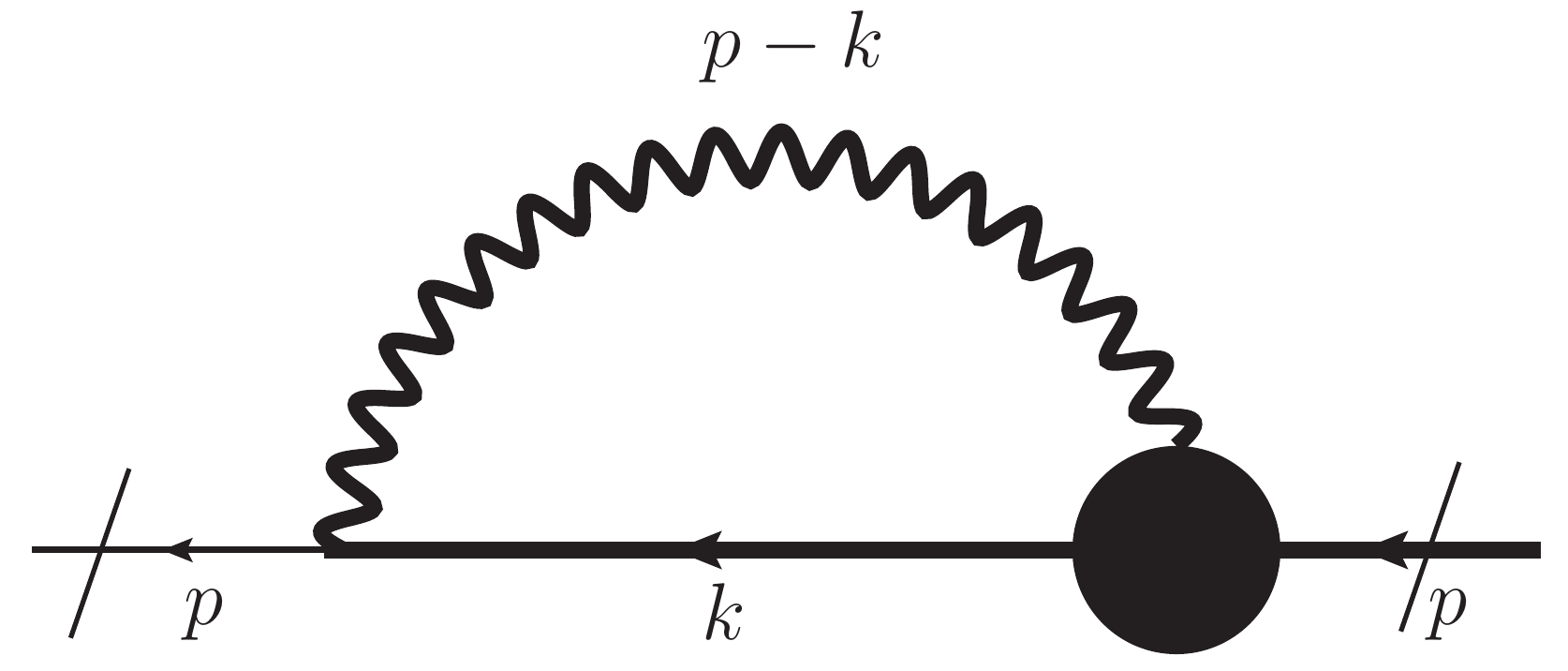}}
\vspace*{-0cm}
\caption{ The pictorial representation of the regularized fermion 
self-energy in
Eq. \eqref{dse2}, with  the external legs amputated. The thick lines are the renormalized propagators of both the fermion
and the photon, respectively, while the thin one is the  free fermion
propagator The full dot represents the renormalized
interaction vertex.}
\label{fig_selF}
\end{figure}

The starting point  is the integral equation fulfilled by the regularized self-energy $\Sigma(\zeta,\Lambda;p)$ 
 (see Itzykson and Zuber \cite{Zuber}, p. 275, for the 
 adopted notations, and also Fig. \ref{fig_selF}), that reads
\be
\Sigma(\zeta,\Lambda;p)=
i~(-ie_0)^2 ~\int_{\Lambda} {d^4k\over
(2\pi)^4}~ \gamma^\beta~   S(\zeta,\Lambda;k)
\nonu \times ~ 
\Gamma^\alpha(\zeta,\Lambda;k,p)~ D_{\alpha\beta}(\zeta,\Lambda; p-k)~~.
\label{gap1}\ee
By introducing in Eq. \eqref{gap1}
the following relation between regularized and renormalized quantities
\be
S(\zeta,\Lambda;k)=Z_2(\zeta,\Lambda)~S_R(\zeta,k)~~, \nonu D_{\alpha\beta}(\zeta,\Lambda; p-k)=Z_3(\zeta,\Lambda)~D^R_{\alpha\beta}(\zeta; p-k)~~,
\nonu
\Gamma^\alpha(\zeta, \Lambda;k,p)={\Gamma^\alpha_R(\zeta;k,p)\over
Z_1(\zeta,\Lambda)}~~, \nonu   e^2_0~Z_3(\zeta,\Lambda)\left[{Z_2(\zeta,\Lambda)\over Z_1(\zeta,\Lambda)}
\right]^2 = e^2_R~~,
\label{renorcos}
 \ee
 one can rewrite
\be
\Sigma_Z(\zeta,\Lambda;p)=-i Z_1(\zeta,\Lambda)~e^2_R ~\int_{\Lambda} {d^4k\over
(2\pi)^4}~ \gamma^\beta~   S_R(\zeta,k)
\nonu~  \Gamma^\alpha_R(\zeta;k,p)~D^R_{\alpha\beta}(\zeta; p-k) 
~~,\label{dse2}
\ee
with 
 $\Sigma_Z(\zeta,\Lambda;p)=Z_2~\Sigma(\zeta,\Lambda;p)$. From Eq.\eqref{self_R1}, one gets the following integral equation for
 the renormalized self-energy 
 \be
 \Sigma_R(\zeta;p)=
 \psla p {\cal A}_R(\zeta;p) + {\cal
 B}_R(\zeta;p)
 \nonu
 =~\Sigma_Z(\zeta,\Lambda;p)-\left.\Sigma_Z(\zeta,\Lambda;p)\right|_{p^2=\zeta^2}
\nonu
= -iZ_1(\zeta,\Lambda)~e^2_R ~\int_{\Lambda} {d^4k\over
  (2\pi)^4}~
\gamma^\beta~ S_R(\zeta,k)
\nonu \times ~\Bigl \{    \Gamma^\alpha_R(\zeta;k,p)~D^R_{\alpha\beta}(\zeta; p-k)
\nonu
-\left[     \Gamma^\alpha_R(\zeta;k,p)~D^R_{\alpha\beta}(\zeta; p-k)~
\right]_{p^2=\zeta^2}\Bigr\}~~.
 \label{app_sigmar}\ee
The scalar functions ${\cal A}_R(\zeta;p)$ and ${\cal B}_R(\zeta;p)$ 
can be obtained by evaluating the following traces
 \be
 {\cal A}_R(\zeta;p)
= {\cal T}_A(\zeta,\Lambda;p) -
\left. {\cal T}_A(\zeta,\Lambda;p)\right|_{p^2=\zeta^2}~~,
\nonu 
{\cal B}_R(\zeta;p)
= {\cal T}_B(\zeta,\Lambda;p) -
\left. {\cal T}_B(\zeta,\Lambda;p)\right|_{p^2=\zeta^2}~~,
\label{AB_TAB}\ee
with
\be
{\cal T}_A(\zeta,\Lambda;p)={1 \over 4p^2} ~\textrm{Tr} 
\Bigl[\psla p~\Sigma_Z(\zeta,\Lambda;p)\Bigr]
~~~, \nonu
{\cal T}_B(\zeta,\Lambda;p)={1 \over 4} ~\textrm{Tr} 
\Bigl[\Sigma_Z(\zeta,\Lambda;p)\Bigr]~~.
\ee
 Since in the Landau gauge the photon is transverse to the momentum transfer,
only the transverse projection $T_{\beta\alpha}\Gamma^\alpha_R(\zeta;k,p)$ is
relevant (see Eq. \eqref{proj} for the definition of $T_{\beta\alpha}$). It is important to notice that the transverse projection of the
Ball-Chiu vertex is not vanishing, i.e. $T_{\beta \alpha}
\Gamma^\alpha_{R;BC}(\zeta;k,p)\ne~0$. In particular, by using Eqs. \eqref{tranT} and
\eqref{eq:BallChiu} one gets
\be
T^\beta_{~\alpha}~\Gamma^\alpha_{R;BC}(\zeta;k,p)
=\frac{\gamma^\beta_T}{2}~F_{{\cal A}_+}(k,p,\zeta)
 \nonu -(\psla p+\psla k) p^\beta_T~ F_{{\cal A}_-}(k,p,\zeta)
 - 2p^\beta_T ~ F_{\cal B}(k,p,\zeta) ~~, 
\nonu\label{dse_Ft}\ee
where the subscript $T$ on a four-vector means 
\be
{\cal V}^\beta_T= {\cal V}^\beta -q^\beta {{\cal V}\cdot q\over q^2}~~,
\ee
with $q=p-k$ (notice that the photon is outcoming), so that   $k^\beta_T=p^\beta_T$.
Moreover,  from Eq. \eqref{FAi} one has 
\be
F_{{\cal A}_+}(k,p,\zeta)= 2-{\cal A}_R(\zeta;k)-{\cal A}_R(\zeta;p)~~,
\nonu
F_{{\cal A}_-}(k,p,\zeta)
   ={{\cal A}_R(\zeta;k)-{\cal A}_R(\zeta;p)\over k^2-p^2}
 \nonu =  -\int_{s_{th}}^\infty ds ~{\rho_A(s,\zeta)\over (k^2-s+i\epsilon)~
   (p^2-s+i\epsilon)}~~,
\nonu
F_{\cal B}(k,p,\zeta)={{\cal B}_R(\zeta;k)-{\cal B}_R(\zeta;p)\over k^2-p^2}
\nonu =-\int_{s_{th}}^\infty ds ~{\rho_B(s,\zeta)\over (k^2-s+i\epsilon)~
   (p^2-s+i\epsilon)}~~.
\ee
For the purely transverse component $ \Gamma^\alpha_{R;T}(\zeta;k,p)$, Eq.
\eqref{vtren_qin}, one
 has
\be
T^\beta_{~\alpha}~\Gamma^\alpha_{R;T}(\zeta;k,p)= 
 -{1\over 2}
\Bigl[(p-k)^2 \gamma^\beta_T
 \nonu +2i  \gamma_5\epsilon^{\beta\mu \nu\rho} 
\gamma_{\mu}p_\nu k_\rho\Bigr]
~ F_{{\cal A}_-}(k,p,\zeta)~~,
\label{dse_Ft1}
\ee
where $(p-k)^\beta_T=0$ has been used.
{Summing the two contributions to $\Gamma^\beta_{R;T}$, one gets:}
\be
\bar\Gamma^\beta_T=\frac{\gamma^\beta_T}{2}~F_{{\cal A}_+}(k,p,\zeta)  
  -\Bigl[(\psla p+\psla k)~p^\beta_T~
  +(p-k)^2 {\gamma^\beta_T\over 2}
  \nonu +i  \gamma_5\epsilon^{\beta\alpha \nu\rho} \gamma_{\alpha}p_\nu k_\rho
\Bigr]~F_{{\cal A}_-}(k,p,\zeta)   
   - 2p^\beta_T ~F_{\cal B}(k,p,\zeta) 
~~.\nonu\ee
After  inserting in Eq. \eqref{dse2}, the expressions of the fermion and
photon propagators in terms of the respective KL representations, i.e. Eqs. \eqref{lehmR} and 
\eqref{lehmRph},   and exploiting Eqs. \eqref{dse_Ft} and \eqref{dse_Ft1},  one 
can obtain 
 the following expressions  for 
 the   traces 
\be
{\cal T}_{A(B)}(\zeta,\Lambda;p)=
-iZ_1(\zeta,\Lambda)~e^2_R ~\int_{0}^\infty 
  d\omega ~\bar\sigma_\gamma(\omega,\zeta,\zeta_p)
  \nonu \times \int_{0}^\infty ds'
~\int_{\Lambda} {d^4k\over(2\pi)^4}~\frac{1}{(p-k)^2-\omega+i\epsilon}
~{1\over  k^2 -s'  +i\epsilon}
\nonu \times~{1\over 4 } \textrm{Tr}\left\{ \left[
   \slashed{k} ~\bar \sigma_V(s',\zeta,s'_{th}) +
  \bar\sigma_S(s',\zeta,s'_{th})\right]
 ~O_{A(B)}\right\}~~,
\nonu\label{app_dse_F1}\ee
where
\be 
O_A={1 \over p^2}~\bar\Gamma^\beta_T~\psla p~\gamma_\beta~~, \quad \quad 
O_B=\bar\Gamma^\beta_T~~\gamma_\beta~~,
\label{eq:OAB}
\ee
and
\be
 \bar\sigma_\gamma(\omega,\zeta,\zeta_p)=\delta\Bigl(\omega-\zeta^2_p\Bigr)
 +\sigma_{\gamma} (\omega,\zeta)~\Theta\Bigl(\omega-\zeta^2_p\Bigr)~~,
\nonu
    \bar{\sigma}_{S(V)}(s',\zeta,s'_{th}) = \delta\Bigl(s'-m^2(\zeta)\Bigr)
   \nonu  +\sigma_{S(V)} (s',\zeta)~\Theta\Bigl(s'-s'_{th}\Bigr)~~.
 \ee

\subsection{Traces evaluation}
\label{subset_cTr}
From Eqs. \eqref{app_dse_F1} and \eqref{eq:OAB}, one gets the
following traces. The one involved in the calculation of ${\cal T}_A$ is 
\be
{1\over 4 } \textrm{Tr}\left\{ \left[
   \slashed{k} ~\bar \sigma_V(s',\zeta,s'_{th}) +
  \bar\sigma_S(s',\zeta,s'_{th})\right]
 ~O_A\right\}
= 
\nonu
=\bar \sigma_V(s',\zeta,s'_{th})~Tr_1 +\bar \sigma_S(s',\zeta,s'_{th})~Tr_2 
~~,\nonu\ee
where 
\be
Tr_1=-{1\over  p^2}~  
 ~
\left\{\left[ { 3\over 2} k\cdot p -{ k^2 p^2-(k\cdot p)^2   \over (p-k)^2}
\right]~F_{{\cal A}_+}(k,p,\zeta)    
  \nonub +\left[\underbrace{\frac{k^2p^2 -(k\cdot p)^2}{(p-k)^2}(k^2+p^2)
  }_{\Gamma_{BC}}
  \nonubb
  -\underbrace{\left({3\over 2}k\cdot p (p-k)^2-k^2p^2+(k\cdot p)^2\right)}_{\Gamma_{T_3}}
   \nonubb  -\underbrace{2(k^2p^2-(k\cdot p)^2)}_{\Gamma_{T_8}} \right]
  ~F_{{\cal A}_-}(k,p,\zeta)\right\}~~,
\label{eq:Tr1}\ee
and 
\be
Tr_2=-{2\over  p^2} 
  ~{k^2 p^2 - (k \cdot p)^2 \over (p-k)^2}  ~F_{\cal B}(k,p,\zeta) ~~.
\nonu
\label{eq:Tr2}\ee
In Eq. \eqref{eq:Tr1}, the underbraces 
 emphasize the contributions   generated by  each term present in  
  the vertex 
  (see Eqs.
\eqref{eq:BallChiu} and \eqref{vtren_qin}). { This is motivated by the needed cooperation for eliminating the contribution produced
by ${\cal A}_R(\zeta;p)$ present in  $F_{{\cal A}+}$. Such a contribution generates a  singular integral  
in ${\cal T}_{A} (\zeta,\Lambda;p^2)$ that cannot be canceled by an analogous term in ${\cal T}_{A} (\zeta,\Lambda;p^2=\zeta^2)$,
since ${\cal A}_R(\zeta;p=\zeta)=0$.
 Also in $F_{{\cal A}-}$ there is ${\cal A}_R(\zeta;p)$, but in a combination with ${\cal
A}_R(\zeta;k)$, such
that it does not plague the further calculation (see below). Notice that also in  ${\cal T}_{B} (\zeta,\Lambda;p^2)$ the same issue will
be met.  In conclusion, all the terms in the vertex
function play an essential  role for properly
  restoring  the multiplicative renormalizability of the self-energy, as expressed in Eq.
  \eqref{app_sigmar}. This result is  expected from the 
  perturbative analysis (see, e.g., \cite{Kizilersu:2009kg}), but it is gratifying  to 
  be achieved  within  a non-perturbative approach. 
}
  
The aforementioned cancellation of ${\cal A}_R(\zeta;p)$ in ${\cal T}_{A}$ can be attained by usefully recasting Eq.
 \eqref{eq:Tr1} as follows
\be
Tr_1=-{1\over  p^2} ~
\left\{\left[\frac{3}{2}k\cdot p-\frac { k^2 p^2-(k\cdot p)^2  }{(p-k)^2}\right]
\nonub \times ~\left[F_{{\cal A}_+}(k,p,\zeta)-(k^2-p^2)F_{{\cal A}_-}(k,p,\zeta)  \right]
 \nonub   
 -\left[2p^2~\left({3\over 2} k\cdot p-\frac{k^2p^2 -(k\cdot p)^2}{(p-k)^2}\right) 
 \nonubb+(k^2p^2-4(k\cdot p)^2) \right]
  ~F_{{\cal A}_-}(k,p,\zeta)\right\}~~,
\label{eq:Final}
\ee
 Then, the problematic ${\cal A}_R(\zeta;p)$  is canceled in the combination 
\be
F_{{\cal A}_+}(k,p,\zeta)-(k^2-p^2)F_{{\cal A}_-}(k,p,\zeta)= 
\nonu =2 \Bigl[1 -
{\cal A}_R(\zeta;k)\Bigr]~~.
\label{app_3_canc}\ee
obtained from the contributions produced by $T_3$ and $T_8$. It must be noticed in Eq. \eqref{eq:Final} that for getting the result
one produces the term 
$$k^2-4(k\cdot p)^2~~,$$ that in principle can generate a singular integral.
Indeed, an other fortunate cancellation takes place  by exploiting  the 4D angular integration and the difference between ${\cal T}_{A}
(\zeta,\Lambda;p^2)$  and ${\cal T}_{A} (\zeta,\Lambda;\zeta^2)$ 
(see also  \ref{subsect_AR}). 
Remarkably the factor of $4$ is essential for obtaining the finite result.

For evaluating the  trace in ${\cal T}_B$ (see Eq. \eqref{app_dse_F1}), 
 one gets  
 \be
 \textrm{Tr}\left\{ \left[
   \slashed{k} ~\bar \sigma_V(s',\zeta,s'_{th}) +
  \bar\sigma_S(s',\zeta,s'_{th})\right]
 ~O_B\right\}
=
\nonu
=\bar \sigma_V(s',\zeta,s'_{th})~Tr_3+\bar \sigma_S(s',\zeta,s'_{th})~Tr_4~~,
 \ee
 where
 \be
 Tr_3= - 2 ~
{k^2p^2-(k\cdot p)^2 \over (p-k)^2}
    ~F_{\cal B}(k,p,\zeta) ~~,
 \ee
 and 
 \be
 Tr_4=\frac{3}{2}~\left[F_{{\cal A}_+}(k,p,\zeta) -
(k^2-p^2) ~F_{{\cal A}_-}(k,p,\zeta) \right]
  \nonu -\left[2 ~{k^2p^2  - (k \cdot p)^2\over (p-k)^2}~ 
+3(p^2-k\cdot p)\right]
  F_{{\cal A}_-}(k,p,\zeta) 
 ~~.\nonu
 \ee
Finally, collecting all the results, one has the following expressions for 
 ${\cal T}_A(\zeta,\Lambda;p)$ and ${\cal T}_B(\zeta,\Lambda;p)$
 \be
 {\cal T}_A(\zeta,\Lambda;p)= iZ_1(\zeta,\Lambda)~e^2_R ~{3\over  p^2}
~\int_{0}^\infty 
  d\omega ~\bar \sigma_\gamma(\omega,\zeta,\zeta_p)
  \nonu \times ~\int_{0}^\infty 
  ds'\int_{\Lambda} {d^4k\over
(2\pi)^4}~ {1\over (p-k)^2-\omega+i\epsilon}~{1\over k^2-s'+i\epsilon}
 \nonu \times~\Bigg\{ \bar \sigma_V(s',\zeta,s'_{th})
\Bigg[ 
\Bigl(k\cdot p  - {2\over 3}{\cal R}(k,p)\Bigr)
\nonu \times~\Bigl( 1- {\cal A}_R(\zeta;k)\Bigr)   
  -\Biggl(p^2\Bigl( k\cdot p-{2\over 3}{\cal R}(k,p)\Bigr) 
  \nonu+{k^2p^2-4(k\cdot p)^2\over 3}\Biggr)
  ~F_{{\cal A}_-}(k,p,\zeta)
\Biggr]
 \nonu + {2\over 3} ~ \bar \sigma_S(s',\zeta,s'_{th})
 ~  {\cal R}(k,p) ~F_{\cal B}(k,p,\zeta) \Bigg\}~~,
 \label{app_3_ar}\ee
 and
 \be
 {\cal T}_B(\zeta,\Lambda;p) =~-iZ_1(\zeta,\Lambda)~e^2_R~3 
~\int_{0}^\infty 
  d\omega ~\bar \sigma_\gamma(\omega,\zeta,\zeta_p)
  \nonu \times~\int_{0}^\infty ds' \int_{\Lambda} {d^4k\over
(2\pi)^4}~ \frac{1}
  {(p-k)^2-\omega+i\epsilon}
~{1\over k^2-s'+i\epsilon}
\nonu \times~\Bigg\{ \bar \sigma_S(s',\zeta,s'_{th}) 
   ~
\Biggl[1
-{\cal A}_R(\zeta;k)-
\nonu\Bigl(p^2
 - 
 k\cdot p+{2\over 3}{\cal R}(k,p)\Bigr)
  F_{{\cal A}_-}(k,p,\zeta) 
\Biggr]
\nonu- \bar \sigma_V(s',\zeta,s'_{th})
~{2 \over 3}{\cal R}(k,p)
    ~F_{\cal B}(k,p,\zeta) \Bigg\} ~~,
\label{app_3_br} \ee
with
\be
{\cal R}(k,p)={k^2p^2  - (k \cdot p)^2\over (p-k)^2}=
{q^2p^2  - (q \cdot p)^2\over q^2}~~.
\ee
with $q=p-k$. 

By using Eq. \eqref{AR_NIR} for ${\cal A}_R(\zeta;k)$ and 
 \eqref{FAi} for $ F_{{\cal A}_-} $ and $ F_{{\cal B}} $, one can write
 ${\cal T}_A$ and  ${\cal T}_B$,  as
 follows
 \be
 {\cal T}_A(\zeta,\Lambda;p)= iZ_1(\zeta,\Lambda)~e^2_R ~{3\over  p^2}
 ~\int_{0}^\infty~d\omega~\bar \sigma_\gamma(\omega,\zeta,\zeta_p)
\nonu \times ~\Biggl\{ 
\int_{0}^\infty~ds'~
\bar\sigma_V(s',\zeta,s'_{th}) ~{\cal I}_1(p,\omega,s')
\nonu+\int_{s_{th}}^\infty~ds~
{\rho_A(s,\zeta)}~
\int_{0}^\infty~ds'~
\bar\sigma_V(s',\zeta,s'_{th})
\nonu \times  ~\Bigl[ {{\cal I}_1(p,\omega,s')\over (\zeta^2-s+i\epsilon)} 
 -{\cal I}_4(p,\omega,s',s) \Bigr]
\nonu
+  \int_{s_{th}}^\infty~ds~
{\rho_A(s,\zeta)\over (p^2-s+i\epsilon)} 
~\int_{s_{th}}^\infty~ds'~
\bar \sigma_V(s',\zeta,s'_{th})~
\nonu \times \Bigl[p^2~ {\cal I}_4(p,\omega,s',s)
+{1\over 3}{\cal I}_5(p,\omega,s',s)\Bigr]
-{2\over 3}
 \int_{s_{th}}^\infty~ds
 \nonu \times
{\rho_B(s,\zeta)\over (p^2-s+i\epsilon)} 
 ~\int_{s_{th}}^\infty~ds'~
\bar \sigma_S(s',\zeta,s'_{th})~{\cal I}_3(p,\omega,s',s)
\Bigg\}
 \nonu\label{app_3_TA}
 \ee
 and
\be{\cal T}_B(\zeta,\Lambda;p) =-i3~Z_1(\zeta,\Lambda)~e^2_R 
 ~\int_{0}^\infty~d\omega~\bar \sigma_\gamma(\omega,\zeta,\zeta_p)
 \nonu \times ~
\Bigg\{  
\int_{0}^\infty~ds'~
\bar \sigma_S(s',\zeta,s'_{th})~{\cal I}_0(p,\omega,s')
\nonu+\int_{s_{th}}^\infty~ds~
{\rho_A(s,\zeta)}
\int_{0}^\infty~ds'~
\bar\sigma_S(s',\zeta,s'_{th})
\nonu \times~\Bigl[{{\cal I}_0(p,\omega,s')\over (\zeta^2-s+i\epsilon)}
-{\cal I}_2(p,\omega,s',s)\Bigr]
\nonu
+\int_{s_{th}}^\infty~ds~
{\rho_A(s,\zeta)\over (p^2-s+i\epsilon)} 
~\int_{0}^\infty~ds'~
\bar \sigma_S(s',\zeta,s'_{th})
\nonu \times ~
\Bigl(p^2{\cal I}_2(p,\omega,s',s)-{\cal I}_4(p,\omega,s',s)
\Bigr)
\nonu+{2\over 3}~\int_{s_{th}}^\infty~ds~
{\rho_B(s,\zeta)\over (p^2-s+i\epsilon)} 
\nonu \times ~\int_{0}^\infty~ds'~
\bar \sigma_V(s',\zeta,s'_{th})~{\cal I}_3(p,\omega,s',s) \Biggr\}
\label{app_3_TB} 
\ee 
where
\be
 {\cal I}_0(p,\omega,s')=\int_{\Lambda} {d^4k\over (2\pi)^4}~{1\over (p-k)^2-\omega+i\epsilon}~
 \nonu \times~{1\over k^2-s'+i\epsilon}
\label{cal_I0}
 \\ & & 
 {\cal I}_1(p,\omega,s')=\int_{\Lambda} {d^4k\over (2\pi)^4}
 ~\left[k\cdot p -{2\over 3}~{\cal R}(k,p)\right]~
 \nonu \times~{1\over (p-k)^2-\omega+i\epsilon}~
 {1\over k^2-s'+i\epsilon}
 \label{cal_I1}
 \\ & &
 {\cal I}_2(p,\omega,s',s)=\int_{\Lambda} {d^4k\over (2\pi)^4}~{1\over (p-k)^2-\omega+i\epsilon}~
 \nonu \times~{1\over k^2-s'+i\epsilon}~{1\over k^2-s+i\epsilon}
 \label{cal_I2}
 \\ & &
{\cal I}_3(p,\omega,s',s)=\int_{\Lambda} {d^4k\over (2\pi)^4}~
 {\cal R}(k,p)~{1\over (p-k)^2-\omega+i\epsilon}
 \nonu \times~
 {1\over k^2-s'+i\epsilon}~{1\over k^2-s+i\epsilon}
\nonu \label{cal_I3}
 \\ & &
 {\cal I}_4(p,\omega,s',s)=\int_{\Lambda} {d^4k\over (2\pi)^4}~\left[k\cdot p -{2\over 3}~
 {\cal R}(k,p)\right]
 \nonu \times~{1\over (p-k)^2-\omega+i\epsilon}~
 {1\over k^2-s'+i\epsilon}~{1\over k^2-s+i\epsilon}
\nonu \label{cal_I4}
 \\ & & 
 {\cal I}_5(p,\omega,s',s)=\int_{\Lambda} {d^4k\over (2\pi)^4}~{k^2p^2  - 
 4(k \cdot p)^2\over (p-k)^2-\omega+i\epsilon}~
\nonu \times~ {1\over k^2-s'+i\epsilon}~{1\over k^2-s+i\epsilon}~
 \label{cal_I5}\ee
 with $\omega,s',s\ge 0$.
 
 It has to  point out that ${\cal I}_0$,
 ${\cal I}_1$, and ${\cal I}_5$ are divergent integrals for $d=4$, and only
 after applying  i) the dimensional regularization and ii) the subtraction of the
 corresponding integrals evaluated at $p^2=\zeta^2$, one gets   finite
 results for ${\cal A}_R$ and ${\cal B}_R$, as it will be  shown in 
  \ref{subsect_BR}  and \ref{subsect_AR}, respectively.

 Differently, the three  integrals ${\cal I}_2$, ${\cal I}_3$ and
  ${\cal I}_4$
 are finite, and  they can be   evaluated by i) applying the Feynman
 parametric formula 
 and  introducing a new variable $q=k-\alpha p$ ($\alpha$ is a proper combination of Feynman
 parameters);  ii) 
  changing  the  variable $q_0\to iq_4$ 
  and iii) eventually  using 
  4D polar coordinates, $q_E\equiv\{q_x,q_y,q_z,q_4\}$,  given by
 \be
 q_E\equiv ~\rho\Bigl \{sin\theta_2~sin\theta_1~
 cos\phi,~sin\theta_2~sin\theta_1 ~sin\phi,
 \nonu
  sin\theta_2~ cos \theta_1, cos \theta_2 \Bigr\}~~.
 \ee
 
 \subsection{Analytic Integrals}
 \label{subsect_Ain}
 The evaluation of the analytic integrals
 in ${\cal T}_A(\zeta,\Lambda;p)$ and ${\cal
 T}_B(\zeta,\Lambda;p)$  represents the most lengthy part of the formal 
 elaboration.  It is helpful to recall that    our goal is to achieve a form of both ${\cal A}_R$ and ${\cal B}_R$
 suitable for exploiting the uniqueness of the NWFs $\rho_A$ and
 $\rho_B$, as suggested by a theorem demonstrated by  Nakanishi for a generic n-leg
 transition amplitude \cite{nakabook}.

To proceed in a simple way, it is very useful to consider spacelike values for the external momentum $p$.
 This choice, as it becomes  immediately clear, simplifies a lot the formal elaboration, and it is not restrictive,
 since the NWFs do
 not depend upon the values of the external momentum, but noteworthy they are used for obtaining  the scalar
 functions ${\cal A}_R$ and ${\cal B}_R$ at any value of $p^2$.  
 
 For the finite integral ${\cal I}_2$  one gets
  \be
 {\cal I}_2(p,\omega,s',s)= \int_0^1 d\xi\int_0^{1-\xi}dv
  \int  {d^4 q\over (2\pi)^4}
  \nonu \times~{2 \over 
 \Bigl[q^2+\xi(1-\xi)p^2 -\xi \omega -v s' -(1-\xi-v) s
 +i\epsilon\Bigr]^3} 
\nonu
 ={i \over (4\pi)^2}\int_0^1 d\xi\int_0^{1-\xi}dv
 \nonu \times~{ 1\over 
 \Bigl[\xi(1-\xi)p^2 -\xi \omega -v s' -(1-\xi-v) s
 +i\epsilon\Bigr]}~~.
 \label{cal_I2b}\ee 
 The last step can be easily carried out without any concern, given the aforementioned choice of
 $p^2< 0$.

The second finite integral, ${\cal I}_3$, becomes
 \be
 {\cal I}_3(p,\omega,s',s)=\int_0^1 d\xi\int_0^{1-\xi}dv\int_0^{1-\xi-v} dw~
 \int {d^4 q\over (2\pi)^4}
 \nonu\times ~{ 6\Bigl[q^2p^2  - (q \cdot p)^2 \Bigr]\over 
\Bigl[q^2+{\cal A}_4(v,w)p^2 -\xi \omega -v s' -ws
+i\epsilon\Bigr]^4}
 \nonu
 ={3\over 2}~p^2~{i\over (4\pi)^2}~\int_0^1 d\xi
 ~\int_0^{1-\xi}dv\int^{1-\xi-v}_0 dw
 \nonu \times~ {1  
\over \Bigl[{\cal A}_4(v,w)p^2 -\xi \omega -v s' -ws
 +i\epsilon\Bigr]}~~,
 \label{cal_I3b}
 \ee
 with
 $
 {\cal A}_4(v,w)=(v+w)(1-v-w)$.

 Finally,  ${\cal I}_4$ can be evaluated by using the result in 
 Eq. \eqref{cal_I3b}, i.e. 
 \be
 {\cal I}_4(p,\omega,s',s)=2~p^2~\int_0^1 d\xi\int_0^{1-\xi}dv
  ~\int  {d^4 q\over (2\pi)^4}
  \nonu \times~{\xi \over 
 \Bigl[q^2+\xi(1-\xi)p^2 -\xi \omega -v  s'-(1-\xi-v) s
 +i\epsilon\Bigr]^3}
\nonu -{2\over 3}~{\cal I}_3(p,\omega,s',s)
\nonu
 =p^2~{i \over (4\pi)^2}\int_0^1 d\xi\int_0^{1-\xi}dv
 \nonu \times ~\left\{{ \xi\over 
 \Bigl[\xi(1-\xi)p^2 -\xi \omega -v s' -(1-\xi-v) s
 +i\epsilon\Bigr]}
 \nonub-\int^{1-\xi-v}_0 dw
 ~ {1  
\over \Bigl[{\cal A}_4(v,w)p^2 -\xi \omega -v s' -ws
 +i\epsilon\Bigr]}\right\}
\nonu\label{cal_I4b} \ee
In the above 4D integration on $q$ the subscript $\Lambda$ 
has been removed since the regularization is not necessary.

\subsection{The ${\cal B}_R$ contribution to the fermion self-energy}
\label{subsect_BR} 
Let us start the evaluation of the contribution  ${\cal B}_R$, since
it contains only one divergent integral, i.e.  ${\cal I}_0$, 
 Eq.\eqref{cal_I0}. To get a finite value for the contribution 
 from ${\cal I}_0$,
 it is compulsory to exploit both the dimensional regularization, that
 legitimates 
 the variable shift  in  the integrand, and the 
 subtraction of the corresponding term in 
   $\left .{\cal
  T}_B(\zeta,\Lambda;p)\right|_{p^2=\zeta^2}$, as shown 
  in Eq. \eqref{AB_TAB}. {With this in mind, we simplify the formal
  elaboration 
    removing  the dependence upon $\Lambda$ in what follows, 
    and directly using the 4D integration.}
  ${\cal B}_R(\zeta;p)$  in 
 Eq. \eqref{AB_TAB}  (see also Eq. \eqref{app_3_TB}) reads 
  \be
 \frac{{\cal B}_R(\zeta;p)}{(\zeta^2-p^2)}
= 3~Z_1(\zeta)~e^2_R 
 ~\int_{0}^\infty d\omega~\bar \sigma_\gamma(\omega,\zeta,\zeta_p)
 ~\int_{0}^\infty ds' 
 \nonu\times~ \Bigg\{  
\bar \sigma_S(s',\zeta,s'_{th})~{\cal D}_0(p,\zeta,\omega,s')    
+\int_{s_{th}}^\infty~ds~
{\rho_A(s,\zeta)}
\nonu \times~
\bar\sigma_S(s',\zeta,s'_{th})
 ~
\Biggl[{{\cal D}_0(p,\zeta,\omega,s')\over \zeta^2-s+i\epsilon}
-{\cal D}_2(p,\zeta,\omega,s',s)
\nonu+ 
{\cal D}_{24}(p,\zeta,\omega,s',s)
\Biggr]
+ \int_{s_{th}}^\infty~ds~
\rho_B(s,\zeta) 
'\nonu \times ~
\bar \sigma_V(s',\zeta,s'_{th})~{\cal D}_3(p,\zeta,\omega,s',s) \Biggr\}~~.
\ee
where $Z_1(\zeta) = Z_1(\zeta,\Lambda\to \infty)$ and
   the  differences of integrals 
   (see Eqs. \eqref{cal_I0}, \eqref{cal_I2b}, \eqref{cal_I3b} and 
    \eqref{cal_I4b}) are defined as follows
   \be
  {\cal D}_0(p,\zeta,\omega,s')=-i~{
  \Bigl[{\cal I}_0(p,\omega,s')-{\cal I}_0(\zeta,\omega,s')\Bigr]\over (\zeta^2-p^2)}
  \label{cal_D0a}
  \\ & &
  {\cal D}_2(p,\zeta,\omega,s',s)=-i~{
  \Bigl[{\cal I}_2(p,\omega,s',s)-{\cal
 I}_2(\zeta,\omega,s',s)\Bigr]\over (\zeta^2-p^2)}
 \nonu\label{cal_D2a}
  \\ & &
  {\cal D}_3(p,\zeta,\omega,s',s)=-i {2\over 3}~{1\over (\zeta^2-p^2)}~
  \nonu \times~\Bigl [ 
 { {\cal I}_3(p,\omega,s',s) \over p^2-s+i\epsilon }
 -{{\cal I}_3(\zeta,\omega,s',s) \over \zeta^2-s+i\epsilon}  \Bigr]
 \label{cal_D3a}
  \\ & &
  {\cal D}_{24}(p,\zeta,\omega,s',s)
 = -i~{1\over (\zeta^2-p^2)}
\nonu \times ~\Bigl[{p^2{\cal I}_2(p,\omega,s',s)
 -{\cal I}_4(p,\omega,s',s) \over p^2-s+i\epsilon}
 \nonu -{\zeta^2{\cal
 I}_2(\zeta,\omega,s',s)-{\cal I}_4(\zeta,\omega,s',s) \over 
 \zeta^2-s+i\epsilon}\Bigr]
 \label{cal_D24a}
  \ee
The actual evaluation of the differences is briefly sketched.
The first one, ${\cal D}_0$, can be obtained from Eq. \eqref{cal_I0} and
recalling the need of the regularization, viz 
\be
  {\cal D}_0(p,\zeta,\omega,s')
  = 
~{ -i \over (\zeta^2-p^2)}~\int_0^1 d\xi~\int  {d^4 q\over (2\pi)^4}~
\nonu \times \bigl[ \chi_0(p,q)-.\chi_0(\zeta,q)\bigr] \nonu
  =
  {1\over (4\pi)^2}
 \int_0^1 d\xi \int_0^1 dv 
 \nonu
 \times~{\xi (1-\xi)  \over 
 \Bigl\{\xi (1-\xi)\Bigl[v\zeta^2+ (1-v)p^2\Bigr] -\xi \omega -(1-\xi) s'
 +i\epsilon\Bigr\}}
 \nonu
  =
  {1\over (4\pi)^2}~\int_0^1 dv
   \int_{0}^\infty dy~
   \int_0^1 {d\xi }~\delta\Bigl[ y - {\xi \omega +(1-\xi) s'
\over \xi (1-\xi)}\Bigr]
  \nonu \times  ~{1
    \over\Bigl[v\zeta^2+ (1-v)p^2 -y +i\epsilon\Bigr]}
  \nonu
  =
-{1\over (4\pi)^2}~\int_0^1 dv
   \int_{0}^\infty dy
  \nonu \times~ \int_0^1 d\xi
    ~{ \Theta \Bigl[\xi(1-\xi) y -\xi \omega -(1-\xi) s'\Bigr]\over 
 \Bigl[v\zeta^2+ (1-v)p^2 -y +i\epsilon\Bigr]^2}~~,
    \label{cal_D0}
\ee
 where  
 \be
 \chi_0(p,q)={1 \over \Bigl[q^2 +\xi (1-\xi) p^2 -\xi \omega -(1-\xi)
 s' +i\epsilon\Bigr]^2}
\nonu \ee and 
 the formal manipulations are  allowed  by  
 the dimensional regularization. 
  Eventually, the last line has been introduced after applying an integration by
  parts for 
 preparing the application 
   of the uniqueness theorem 
 \cite{nakabook} to the
 NWF $\rho_B$. Notice that $\xi \omega +(1-\xi) s'
+\eta~\ge ~0$ and 
   the exchange of the integration on
  $\eta$ and on $y$ has been assumed to be allowed.
 
From Eq. 
 \eqref{cal_I2b}, one gets 
 \be
   \label{cal_D2}
 {\cal D}_2(p,\zeta,\omega,s',s)=
 \frac{1}{(4\pi)^2}
 ~\int_0^1 dv
 \int_{0}^\infty dy~\int_0^1 \frac{d\xi}{\xi (1-\xi)}
 \nonu \times ~\int_0^{1-\xi}dz
 ~ \frac{\delta\Bigl[ y -\frac{\xi \omega +z s' +(1-\xi-z)
 s}{\xi(1-\xi)}\Bigr]}{ \Bigl[v\zeta^2+(1-v)p^2 -y
  +i\epsilon\Bigr]^2}.
\ee
 To evaluate  ${\cal D}_3$, one can introduce the following difference with   
 $A>0$ and $B>0$
 \be
 {\cal D}=~{p^2  
\over \Bigl( p^2-s+i\epsilon \Bigr)~
\Bigl(Ap^2 -B+i\epsilon \Bigr)}
\nonu  -  {\zeta^2  
\over \Bigl (\zeta^2-s+i\epsilon \Bigr)~
  \Bigl (A\zeta^2 -B+i \epsilon \Bigr)} 
 \nonu =
  ~{1\over (sA-B)}~\left[{s  
\over \Bigl( p^2-s+i\epsilon \Bigr)}-{B/A  
\over
\Bigl(p^2 -B/A+i\epsilon \Bigr)}
 \nonub
 -  {s  
\over \Bigl (\zeta^2-s+i\epsilon \Bigr)}+{B/A  
\over
  \Bigl (\zeta^2 -B/A+i \epsilon \Bigr)}\right]
   \nonu
  ={(\zeta^2-p^2)\over (sA-B)}
  \int_0^1dv~
  ~\left\{{s  
\over \Bigl[ v\zeta^2+(1-v)p^2-s+i\epsilon \Bigr]^2}
\nonub-{B/A  
\over
\Bigl[v\zeta^2+(1-v)p^2 -B/A+i\epsilon \Bigr]^2}\right\}
\nonu
 =-(\zeta^2-p^2)~
  \int_0^1dv~\int_0^\infty dy ~y 
  \nonu \times~{
  \Delta'\Bigl[y-s + (s-B/A)\Bigr]
\over \Bigl[ v\zeta^2+(1-v)p^2-y+i\epsilon \Bigr]^2}  
  ~~,
 \label{cal_D}\ee
 where  $\Delta'$ indicates \be
 \Delta'\Bigl[y-s+(sA-B)/A\Bigr]=
 \nonu
 =
 {\delta[y-s + (sA-B)/A]-\delta(y-s)
\over  (sA-B)}
\label{Deltap}
\ee
 From \eqref{cal_I3b} and by suitably modifying Eq. \eqref{cal_D}, 
  one has
  \be
 {\cal D}_3(p,\zeta,\omega,s',s)
 =
 -{1\over (4\pi)^2}~\int_0^1 dv\int_0^\infty dy 
 \nonu \times~{1\over 
\Bigl[\Bigl(v\zeta^2 +(1-v)p^2\Bigr) -y+i\epsilon \Bigr]^2}\int_0^1 d\xi 
 \nonu
 \times ~\int_0^{1-\xi}dt\int^{1-\xi-t}_0 dw~
 y
 \nonu \times ~\Delta' \Bigl[y  -s+{
  (s{\cal A}_4(t,w)-\xi \omega -t s' -ws)/{\cal A}_4(t,w)}\Bigr]
~~,
\nonu\label{cal_D3}\ee
with
$
{\cal A}_4(t,w)=(t+w)(1-t-w)$. 
Finally, from  Eqs. \eqref{cal_I2b}, \eqref{cal_I4b},  \eqref{cal_D} and \eqref{cal_D3},  
 one writes  
 \be
 {\cal D}_{24}(p,\zeta,\omega,s',s)
 = {1 \over  (4\pi)^2~(\zeta^2-p^2)}
\int_0^1 d\xi~(1-\xi)
\nonu
 \times ~\int_0^{1-\xi}dt
 ~\Biggl\{ { p^2\over (p^2-s+i\epsilon)}
 \nonu \times~
 {1\over \Bigl[\xi(1-\xi)p^2 -\xi \omega -t s' -(1-\xi-t) s
 +i\epsilon\Bigr]}
  \nonu - { \zeta^2
  \over (\zeta^2-s+i\epsilon)}
  \nonu \times~{1\over \Bigl[ \xi(1-\xi)\zeta^2-\xi \omega -t s' -(1-\xi-t) s
 +i\epsilon\Bigr]}\Biggr\}
 \nonu+{\cal D}_3(p,\zeta,\omega,s,s')
 \nonu
 =-{1 \over  (4\pi)^2}\int_0^1 dv\int_0^\infty dy
~y \int_0^1 d\xi\int_0^{1-\xi}dt
 \nonu \times ~\Biggl\{(1-\xi)~
 \frac{\Delta' \Bigl[y -s +{ (\xi(1-\xi)s-\xi \omega -t s' -(1-\xi-t) s
 \over \xi (1-\xi)}\Bigr]}{\Bigl[\Bigl(v\zeta^2 +(1-v)p^2\Bigr) -y+i\epsilon \Bigr]^2}
 \nonu
 + \int^{1-\xi-t}_0 \hspace{-0.4cm}{dw}
  ~
  \frac{\Delta' \Bigl[y-s 
 +{({\cal A}_4(t,w)s-\xi \omega -t s' -ws )\over{\cal A}_4(t,w)}\Bigr]}{\Bigl[\Bigl(v\zeta^2 +(1-v)p^2\Bigr) -y+i\epsilon \Bigr]^2}
 \Biggr\}.
\nonu\label{cal_D24}\ee

Inserting Eqs. \eqref{cal_D0}, \eqref{cal_D2}, \eqref{cal_D3},
and \eqref{cal_D24}, one gets
 \be
\frac{{\cal B}_R(\zeta;p)}{(\zeta^2-p^2)}={-3 \over  (4\pi)^2}~Z_1(\zeta)~e^2_R
\int_0^1 dv\int_0^\infty dy ~\int_{0}^\infty~d\omega 
\nonu \times ~\int_0^1 d\xi ~ {\bar \sigma_\gamma(\omega,\zeta,\zeta_p)\over 
\Bigl[\Bigl(v\zeta^2 +(1-v)p^2\Bigr) -y+i\epsilon \Bigr]^2}
\nonu\times \hspace{-0.05cm}
\int_{0}^\infty ds'\,\Bigg\{ 
\bar \sigma_S(s',\zeta,s'_{th})
\Biggl[
 \Theta\Bigl[(1-\xi)(y\xi-s') - \xi \omega \Bigr]
\nonu
+\int_{s_{th}}^\infty\, ds
{\rho_A(s,\zeta)}\,\Bigl(
 {\cal C}^{(0)}_{AS}(\zeta,\omega,s,s',\xi,y)
\nonu
 +y~
 {\cal C}^{(1)}_{AS}(\zeta,\omega,s,s',\xi,y)\Bigr)\Biggr]
 \nonu+y~\bar \sigma_V(s',\zeta,s'_{th})
  ~\int_{s_{th}}^\infty ds\, 
\rho_B(s,\zeta) 
\,
 \int_0^{1-\xi}dt
\int^{1-\xi-t}_0 \hspace{-0.4cm}{dw}~ 
 \nonu\times ~ \Delta' \Bigl[y  -s+
  {{\cal A}_4(t,w)s-\xi \omega -t s' -ws\over{\cal A}_4(t,w)}\Bigr]  
 \Biggr\}~~, 
 \label{app_3_BRf} \ee
with $Z_1(\zeta) = Z_1(\zeta,\Lambda\to \infty)$, 
\be
{\cal C}^{(0)}_{AS}(\zeta,\omega,s,s',\xi,y)=
\frac{ \Theta\Bigl[ y\xi(1-\xi) -\xi \omega -(1-\xi) s'
\Bigr]}{\zeta^2-s+i\epsilon}~ 
 \nonu
 + {1\over ~\xi (1-\xi)}\int_0^{1-\xi} dz
 ~\delta\Bigl[ y -{\xi \omega +z s' +(1-\xi-z)
   s\over \xi(1-\xi)}\Bigr]
 ~~,
\nonu\ee
and
\be
{\cal C}^{(1)}_{AS}(\zeta,\omega,s,s',\xi,y)=
\int_0^{1-\xi}dt
 \Biggl\{ (1-\xi)
 \nonu \times~
  \Delta' \Bigl[y  -s+ { \xi (1-\xi)s-\xi \omega -t s' -(1-\xi-t) s
   \over \xi (1-\xi)}\Bigr]
  \nonu
 +
 \int^{1-\xi-t}_0 \hspace{-0.4cm}{dw} ~\Delta' \Bigl[y -s
 +{{\cal A}_4(t,w)s-\xi \omega -t s' -ws\over{\cal A}_4(t,w)}\Bigr]
   \Biggr\}.
\nonu\ee

\subsection{The ${\cal A}_R$ contribution }
\label{subsect_AR}
In order to evaluate ${\cal A}_R$, {(recall $p^2<0$, but without loss of generality
 on the final result for the NWFs)} one has to face with  the divergent behavior 
of 
 ${\cal I}_1$, Eq. \eqref{cal_I1}, and ${\cal I}_5$, Eq. \eqref{cal_I5}. 
 The strategy is exactly the same we have applied to ${\cal I}_0$ in the subsec.
 \ref{subsect_BR}, combining the dimensional regularization 
 for shifting the integration variable and then exploiting the subtraction. 

One can write from 
Eq. \eqref{AB_TAB}
\be
\frac{{\cal A}_R(\zeta;p)}{\zeta^2-p^2}= -3 Z_1(\zeta,\Lambda\to\infty)~e^2_R 
 ~\int_{0}^\infty~d\omega~\bar \sigma_\gamma(\omega,\zeta,\zeta_p)
\nonu \times~\int_{0}^\infty~ds'
 ~\Biggl\{ 
\bar\sigma_V(s',\zeta,s'_{th}) ~{\cal D}_1(p,\zeta,\omega,s')
\nonu
+\int_{s_{th}}^\infty~ds~
{\rho_A(s,\zeta)}~
\bar\sigma_V(s',\zeta,s'_{th})
 ~\Biggl[ {{\cal D}_1(p,\zeta,\omega,s')\over (\zeta^2-s+i\epsilon)}
 \nonu
 -{\cal D}_4(p,\zeta,\omega,s',s) 
\nonu + {\cal D}_4^\prime(p,\zeta,\omega,s',s)
+{\cal D}_5(p,\zeta,\omega,s',s)\Biggr]
-
 \int_{s_{th}}^\infty~ds
 \nonu \times ~
\rho_B(s,\zeta) 
\bar \sigma_S(s',\zeta,s'_{th})~{\cal D}_3(p,\zeta,\omega,s',s)
\Bigg\}~~.
 \ee
where 
\be
{\cal D}_1(p,\zeta,\omega,s')= -i~{1\over (\zeta^2-p^2)}
\nonu \times~\Bigl[{{\cal I}_1(p,\omega,s')\over p^2}-{{\cal
 I}_1(\zeta,\omega,s')\over \zeta^2}\Bigr]
\label{cal_D1a}
\\ & &
{\cal D}_4(p,\zeta,\omega,s',s)= -i{1\over (\zeta^2-p^2)}
\nonu \times~\Bigl[{{\cal I}_4(p,\omega,s',s)
 \over p^2}-{{\cal I}_4(\zeta,\omega,s',s)\over \zeta^2}\Bigr]
 \label{cal_D4a}
\\  & &
 {\cal D}_4^\prime(p,\zeta,\omega,s',s)=-i{1\over (\zeta^2-p^2)}
 \nonu \times ~
\Bigl[{{\cal I}_4(p,\omega,s',s)
 \over (p^2-s+i\epsilon)}-{{\cal I}_4(\zeta,\omega,s',s)\over 
 (\zeta^2-s+i\epsilon)}\Bigr]
\label{cal_D4pa}
\\  & &
{\cal D}_5(p,\zeta,\omega,s',s)=-i~{1\over 3(\zeta^2-p^2)}
\nonu \times~\Bigl[
{{\cal I}_5(p,\omega,s',s)\over p^2~(p^2-s+i\epsilon)}-{{\cal I}_5(\zeta,\omega,s',s)\over 
\zeta^2~~(\zeta^2-s+i\epsilon)}\Bigr]
\label{cal_D5a}
\ee
 By
  exploiting Eq. \eqref{cal_D0}, ${\cal D}_1$ can be evaluated as follows
\be
 {\cal D}_1(p,\zeta,\omega,s')
 =-i{1\over (\zeta^2-p^2)}\int_0^1 d\xi~\int  {d^4 q\over (2\pi)^4}
 ~
\nonu \times
 \Biggl\{{\xi \over \Bigl[q^2 +\xi (1-\xi)p^2 -\xi \omega -(1-\xi)
 s' +i\epsilon\Bigr]^2}
 \nonu
 -{\xi \over  \Bigl[q^2 +\xi (1-\xi)\zeta^2 -\xi \omega -(1-\xi)
 s' +i\epsilon\Bigr]^2}
\nonu
  -{4\over 3} \int_0^{1-\xi}dt \Biggl[
 {q^2  - (q\cdot p)^2/p^2 \over~\Bigl[q^2 +t(1-t)p^2 -\xi \omega 
 -ts'+i\epsilon\Bigr]^3}
 \nonu 
 -{q^2  - (q\cdot \zeta)^2/\zeta^2 \over \Bigl[q^2 +t(1-t)\zeta^2 
 -\xi \omega 
 -ts'+i\epsilon\Bigr]^3}\Biggr]\Biggr\}=
 \nonu
=-{1\over (4\pi)^2}\int_0^1 dv \int_{0}^\infty dy 
~{1\over \Bigl[v\zeta^2+ (1-v)p^2 -y
 +i\epsilon\Bigr]^2}
 ~
 \nonu
 \times \int_0^1 d\xi \Biggl\{\xi~
 \Theta\Bigl[y \xi(1-\xi)- \xi \omega -(1-\xi) s' \Bigr]
 \nonu
 - \int_0^{1-\xi}dt~
 \Theta\Bigl[ yt(1-t) - \xi \omega -t s' \Bigr]
 \Biggr\}
\label{cal_D1}\ee
  From Eq. \eqref{cal_I4b}, one gets 
 \be
 {\cal D}_4(p,\zeta,\omega,s',s)
={1\over (4\pi)^2}
 \int_0^1 dv
 \int_{0}^\infty dy
 \nonu \times ~{1\over 
 \Bigl\{v\zeta^2+(1-v)p^2 -y
  +i\epsilon\Bigr\}^2} \int_0^1 {d\xi }\int_0^{1-\xi}dt
\nonu\times  \Biggl\{{1\over (1-\xi)}
~
  \delta\Bigl[y -{\xi \omega +t s' +(1-\xi-t) s\over \xi(1-\xi)}\Bigr] 
 \nonu 
 -\int^{1-\xi-t}_0 {dw\over {\cal A}_4(t,w)}
 \delta\Bigl[ y -{\xi \omega +t s' +w s\over{\cal A}_4(t,w)}\Bigr]
 \Biggr\} ~~,
 \nonu \label{cal_D4}\ee
 and
 \be
 {\cal D}_4^\prime(p,\zeta,\omega,s',s)= 
 -{1 \over (4\pi)^2}\int_0^1 dv \int_0^\infty dy
 \nonu \times ~
 \int_0^1 d\xi \int_0^{1-\xi}dt~{y\over \Bigl[v\zeta^2+(1-v)p^2 -y
 +i\epsilon\Bigr]^2}
\nonu
\times 
 \Biggl\{{\xi}
 \Delta'\Bigl[y -s + {s\xi(1-\xi)-\xi \omega -t s' -(1-\xi-t) s
 \over \xi(1-\xi)}\Bigr]
 \nonu
 -\int^{1-\xi-t}_0 \hspace{-0.4cm}{dw} 
 ~ \Delta'\Bigl[y-s +{s{\cal A}_4(t,w)-\xi \omega -t s' -ws \over \mathcal{A}_4(t,w)} 
\Bigr]
\Biggr\}~.
\nonu \label{cal_D4p}\ee

Recalling that one has first to apply the dimensional regularization to ${\cal I}_5$, 
from Eq. \eqref{cal_I5} and using Eq. \eqref{cal_D}, and introducing
one has 
 \be
{\cal D}_5(p,\zeta,\omega,s',s)= 
-i~{2\over 3(\zeta^2-p^2)}
~\int_0^1 d\xi\int_0^{1-\xi}dt
\nonu \times ~
 ~\int  {d^4 q\over (2\pi)^4} \left[\chi_5(q,p) - \chi_5(q,\zeta) \right]
 \nonu
 =-~{1\over (4\pi)^2}\int_0^1 dv
  \int_0^\infty dy
  \nonu \times~
 {y\over \Bigl[\Bigl(v\zeta^2 +(1-v)p^2\Bigr) -y+i\epsilon \Bigr]^2}
 \int_0^1 d\xi~{\xi^2} ~\int_0^{1-\xi}dt
\nonu\times~\Delta' \Bigl[y  -s+{s\xi(1-\xi)-\xi\omega -t s'-(1-\xi-t)s \over \xi (1-\xi)}
\Bigr]
~~.
\nonu\label{cal_D5}
  \ee
  with 
  \be
  \label{eq:dummyfunction}
  \chi_5(q,p) = {1 \over p^2~(p^2-s+i\epsilon)}
  \nonu \times  \frac{4\Bigl[q^2p^2  - (q \cdot p)^2\Bigr] 
  -3p^2q^2- 3p^4 \xi^2}{ \Bigl[q^2+\xi (1-\xi)p^2 -\xi\omega -t s'-(1-\xi-t)s
 +i\epsilon\Bigr]^3}~.
\nonu\ee
By using Eqs.  \eqref{cal_D3}, \eqref{cal_D1}, \eqref{cal_D4}, 
 \eqref{cal_D4p}  and \eqref{cal_D5}, one has
 \be
 \frac{{\cal A}_R(\zeta;p)}{\zeta^2-p^2}= {3e^2_R Z_1(\zeta) \over (4\pi)^2}
\int_0^1 dv\int_0^\infty dy \int_{0}^\infty d\omega \int_0^1 d\xi 
\nonu \times ~\int_{0}^\infty ds'
 { \bar \sigma_\gamma(\omega,\zeta,\zeta_p) \over \Bigl[\Bigl(v\zeta^2 +(1-v)p^2\Bigr) -y+i\epsilon \Bigr]^2}
 \nonu
 \times 
~\Biggl\{ 
\bar\sigma_V(s',\zeta,s'_{th}) 
\Biggl[
~\Bigl(\xi~
 \Theta\Bigl[y\xi (1-\xi) - \xi \omega -(1-\xi) s'\Bigr]
\nonu - \int_0^{1-\xi}dt~
 \Theta\Bigl[ yt(1-t) - \xi \omega -t s'\Bigr] \Bigr)
 \nonu
 +\int_{s_{th}}^\infty~ds~
{\rho_A(s,\zeta)}~
 \Bigl({\cal C}^{(0)}_{AV}(\zeta,\omega,s,s',\xi,y)
\nonu +y~{\cal C}^{(1)}_{AV}(\zeta,\omega,s,s',\xi,y)\Bigr)\Biggr]
- \bar \sigma_S(s',\zeta,s'_{th}) \nonu \times~
 \int_{s_{th}}^\infty~ds~
\rho_B(s,\zeta) 
~
 \int_0^{1-\xi}dt\int^{1-\xi-t}_0 {dw}~
  y
  \nonu \times ~\Delta' \Bigl[y  -s+
  {s{\cal A}_4(t,w)-\xi \omega -t s' -ws\over{\cal A}_4(t,w)}\Bigr]\Biggr\}
~~,\nonu
 \label{app_3_ARf}\ee
 with  
 $Z_1(\zeta) = Z_1(\zeta,\Lambda\to \infty)$,
 \be
 {\cal C}^{(0)}_{AV}(\zeta,\omega,s,s',\xi,y)= 
 \frac{\xi
 \Theta\Bigl[y\xi (1-\xi) - \xi \omega -(1-\xi) s'\Bigr]}{\zeta^2-s+i\epsilon}
 \nonu
 - \int_0^{1-\xi}dt~
\frac{\Theta\Bigl[ yt(1-t) - \xi \omega -t s'\Bigr]}{\zeta^2-s+i\epsilon}
 \nonu
 + \int_0^{1-\xi}dt
 \Biggl\{{1\over (1-\xi)}
 ~\delta\Bigl[y -{\xi \omega +t s' +(1-\xi-t) s\over \xi(1-\xi)}\Bigr] 
\nonu -\int^{1-\xi-t}_0 \hspace{-0.4cm}{dw\over {\cal A}_4(t,w)}
 \delta\Bigl[ y -{\xi \omega +t s' +w s\over{\cal A}_4(t,w)}\Bigr]
 \Biggr\}
~~,
 \ee
 and
\be
 {\cal C}^{(1)}_{AV}(\zeta,\omega,s,s',\xi,y)= 
\int_0^{1-\xi}dt
 ~\Biggl\{ (1+\xi)\xi
 \nonu \times ~
 \Delta'\Bigl[y -s +{s\xi(1-\xi)-\xi \omega -t s' -(1-\xi-t) s\over \xi(1-\xi)}
\Bigr]
 \nonu
 -\int^{1-\xi-t}_0 \hspace{-0.4cm}{dw}
 ~ \Delta'\Bigl[y -s+{s{\cal A}_4(t,w) -\xi \omega -t s' -ws 
 \over{\cal A}_4(t,w)} \Bigr]  
\Biggr\} 
~.
\nonu \ee
\section{DSE for the photon self-energy}
\label{app_4}
\begin{figure}
\centerline{\includegraphics[width=5.0cm]{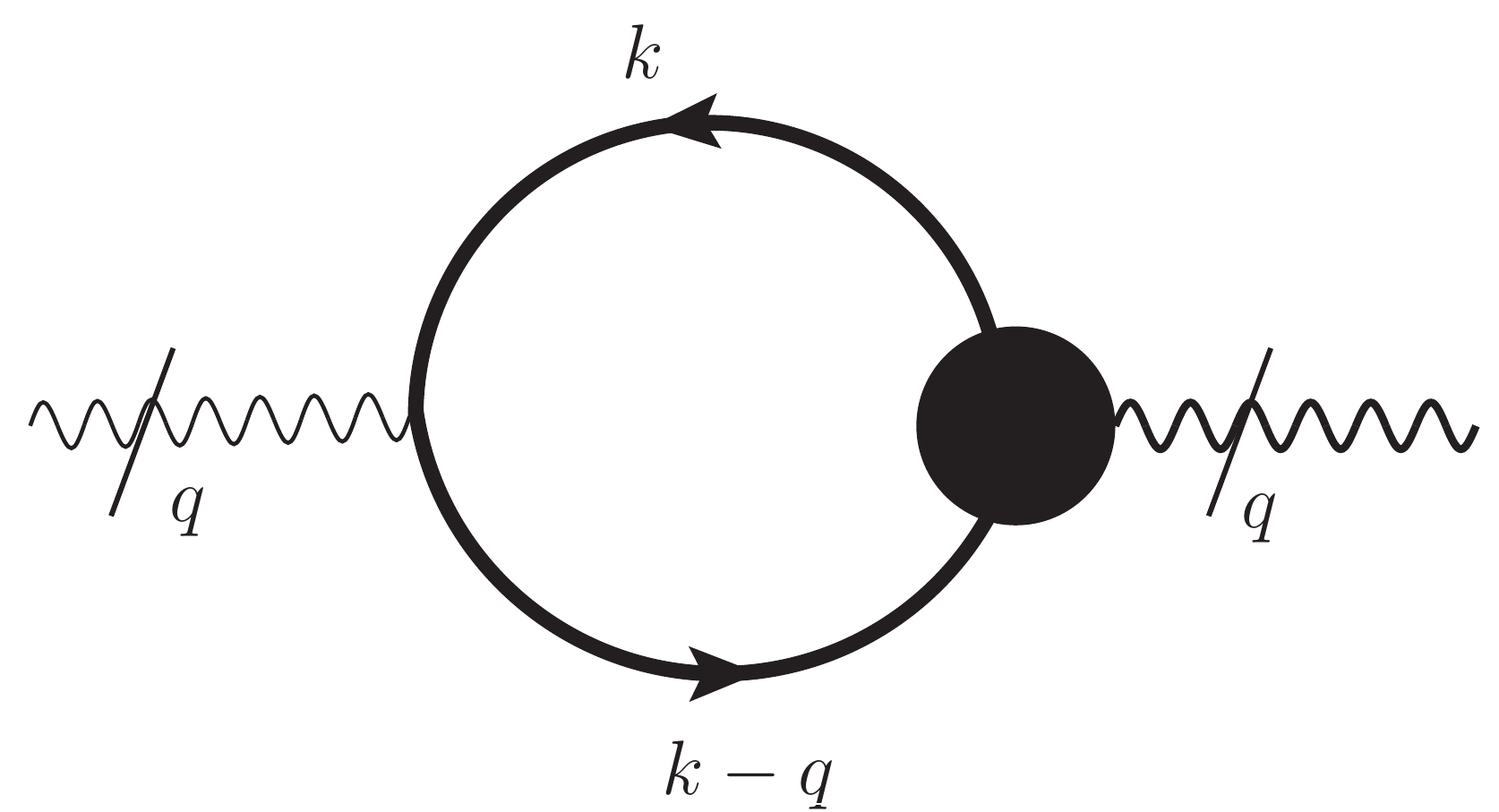}}
\vspace*{-0cm}\caption{ The pictorial representation of the regularized photon 
self-energy in
Eq. \eqref{self_regPh}, with the external legs amputated. The thick lines are the renormalized propagators 
of i) the fermion and antifermion pair
and ii) the incoming photon, while the thin one is the  free photon 
propagator. The full dot represents the renormalized
interaction vertex.}
\label{fig_selfph}
\end{figure}

This Appendix is devoted to obtain the integral equation determining the renormalized photon
self-energy. Eq. \eqref{self_Rph1}.  The initial step is given by the DSE for the regularized
 polarization tensor, Eq. \eqref{ten_pol}, that we rewrite here for convenience,
 (the kinematical quantities are shown in Fig. \ref{fig_selfph}) 
\be
  \label{app_4_vt}
  \Pi^{\mu\nu} (\zeta,\Lambda;q)= -q^2~T^{\mu\nu}~\Pi(\zeta,\Lambda;q)
  = -i  {Z_1(\zeta,\Lambda)\over Z_3(\zeta,\Lambda)}~e_R^2
  \nonu \times 
    \int_{\Lambda}  \frac{d^4k}{(2\pi)^4} \textrm{Tr}\Bigl\{\gamma^\mu
    S_R(\zeta,k)\Gamma^\nu_{R}(\zeta;k,k-q)S_R(\zeta,k-q)\Bigr\}
\nonu
\ee
where Eq. \eqref{renorcos} has been used for the renormalized quantities.
From Eq. \eqref{app_4_vt} and  the properties \eqref{projN}, it follows that
the renormalized photon self-energy,  Eq. \eqref{self_Rph1} reads
\be
 \Pi_R(\zeta;q)=~\Bigl[{\cal T}_P(\zeta,\Lambda;q)-
 \left.{\cal T}_P(\zeta,\Lambda;q)\right|_{q^2=\zeta^2_p}\Bigr]
\label{app_4_Pir}\ee
with
\be
{\cal T}_P(\zeta,\Lambda;q)= 
  ~-i 
     Z_1(\zeta,\Lambda)~{4\over 3}~{e_R^2\over q^2}~\int_{s_{th}} ds
    \nonu \times~ \int_{s'_{th}} ds'
    \int_{\Lambda}  \frac{d^4k}{(2\pi)^4} ~{1 \over k^2-s +i\epsilon} 
    \nonu \times~{1 \over (k-q)^2-s' +i\epsilon} 
    {{\cal P}_{\mu\nu}\over 4}
    ~\textrm{Tr}\Biggl\{~\gamma^\mu
    ~\left[
    \psla k ~\bar \sigma_V(s,\zeta,s_{th})
    \nonub+\bar \sigma_S(s,\zeta,s_{th})\right]~\Gamma^\nu_R(\zeta;k,k-q)
  \nonu \times~\left[
   (\psla k-\psla q)~\bar \sigma_V(s',\zeta,s'_{th}) + \bar \sigma_S(s',\zeta,s'_{th})\right]
  \Biggr\}
  \label{calTP}
  \ee
where we have used: i) the KL representation of the fermion propagator, Eq. \eqref{lehmR}, 
ii)  the definitions of $\bar \sigma_{V(S)}$ in Eq.\eqref{def_sig}. From the vertex contributions Eqs. \eqref{eq:BallChiu}, \eqref{vtren_qin} and the relation $\gamma_5\epsilon^{\mu\alpha \nu\rho} \gamma_{\alpha}(k-q)_\nu k_\rho=\gamma_5\epsilon^{\mu\alpha \nu\rho} \gamma_{\alpha}k_\nu q_\rho$, one can define the relevant trace:
\be
Sp^{\mu\nu} =  \bar \sigma_V(s',\zeta,s'_{th})  \bar \sigma_V(s,\zeta,s_{th}) 
Tr^{\mu\nu}_1 
\nonu+
\bar \sigma_S(s',\zeta,s'_{th})  \bar \sigma_S(s,\zeta,s_{th})Tr^{\mu\nu}_2 
\nonu
   + \bar \sigma_V(s',\zeta,s'_{th})\bar \sigma_S(s,\zeta,s_{th}) Tr^{\mu\nu}_3
  \nonu +\bar \sigma_S(s',\zeta,s'_{th})\bar \sigma_V(s,\zeta,s_{th})
Tr^{\mu\nu}_4
  \label{trace_TP}
\ee
where
\be
Tr^{\mu\nu}_i=~
{1\over 4}\textrm{Tr}\Biggl\{ O^\mu_i~
\Biggl[\frac{\gamma^\nu}{2}~
  F_{{\cal A}_+}(k,k-q,\zeta)
  \nonu -\frac{(2\psla k-\psla q) (2k-q)^\nu}{2}~ F_{{\cal A}_-}(k,k-q,\zeta)
 \nonu - (2k-q)^\nu~ F_{\cal B}(k,k-q,\zeta) 
   -{1 \over 2}
\left[q^2 \gamma^\nu
\nonub- q^\nu \psla q+2i  \gamma^5\epsilon^{\nu\alpha \beta\rho} 
\gamma_{\alpha}k_\beta q_\rho\right]~F_{{\cal A}_-}(k,k-q,\zeta)     \Biggr]~
  \Biggr\} \nonu
\ee
with
\be
O^\mu_1=(\psla k-\psla q)\gamma^\mu ~\psla k~~,
\quad \quad
O^\mu_2=\gamma^\mu~~, \nonu
O^\mu_3=(\psla k-\psla q)\gamma^\mu ~~,
\quad \quad
O^\mu_4= \gamma^\mu ~\psla k~~.
\ee
Performing the  traces  (recall that $\epsilon^{0123}=1$), one has for the first trace
\be
Tr^{\mu\nu}_1=~
{\tau^{\mu\nu}_1\over 2}
F_{{\cal A}_+}(k,k_p,\zeta)
\nonu
  -{(k+k_p)^\nu\over 2}\Bigl[k_p^\mu~k\cdot ( k+k_p)  - ( k+k_p)^\mu k_p \cdot k
  \nonu+ k^\mu k_p\cdot ( k+k_p)
 \Bigr]~ F_{{\cal A}_-}(k,k_p,\zeta)
  -
\Bigl[{1\over 2}~\underbrace{(q^2 ~\tau^{\mu\nu}_1 - \tau^{\mu\nu}_2)}_{T_3}
\nonu \underbrace{-g^{\mu\nu}~\Bigl(k^2 q^2- (k\cdot q)^2\Bigr)
-  \tau^{\mu\nu}_3 ~k\cdot q
 +\tau^{\mu\nu}_4}_{T8}
\Bigr]
\nonu \times ~F_{{\cal A}_-}(k,k_p,\zeta) ~~,
\ee
where the following notation has been introduced for getting a more compact expression
\be
k_p=k-q
\nonu
\tau^{\mu\nu}_1=k_p^\mu ~ k^\nu+ k_p^\nu ~ k^\mu -
  g^{\mu\nu} k_p\cdot k ~~,
\nonu
\tau^{\mu\nu}_2=q^\nu \Bigl( k_p^\mu~k\cdot  q
  - q^\mu k_p \cdot k + k^\mu k_p\cdot q \Bigr)~~,
\nonu
\tau^{\mu\nu}_3= q^\mu k^\nu +q^\nu k^\mu~~,
\nonu
\tau^{\mu\nu}_4=q^\mu q^\nu k^2 +k^\mu k^\nu  q^2~~.
\nonumber \ee
Moreover, the  contributions from the transverse vertexes, i.e. $T_3$ and $T_8$,
 have been properly emphasized.
 
After introducing
 \be 
 {\cal K}^{\mu\nu}= k^\mu k^\nu -{1 \over 2}( q^\mu  k^\nu+ q^\nu  k^\mu)
 ~~~,
 \ee
 one can obtain:
\be
Tr^{\mu\nu}_1 
 = \Bigl[{\cal K}^{\mu\nu} -
 {g^{\mu\nu} \over 2}  (k^2-q\cdot k)\Bigr]
~\Bigl[F_{{\cal A}_+}(k,k-q,\zeta)
\nonu
 +(2k\cdot q-q^2) F_{{\cal A}_-}(k,k-q,\zeta)\Bigr]
\nonu
 -2\Bigl[ \tau_4^{\mu\nu}
+k^2~{\cal K}^{\mu\nu}
-k\cdot q~\tau_3^{\mu\nu}
 \Bigr]
 ~F_{{\cal A}_-}(k,k-q,\zeta)
 \nonu
 +  g^{\mu\nu}~\Bigl(k^2 q^2- (k\cdot q)^2\Bigr)
~F_{{\cal A}_-}(k,k-q,\zeta)  ~~,  
\ee
The remaining traces are given by 
\be
Tr^{\mu\nu}_2
  = \frac{g^{ \mu \nu}}{2}~
  F_{{\cal A}_+}(k,k-q,\zeta)
  -\Biggl[(2 k- q)^\mu (2k-q)^\nu 
  \nonu+ \underbrace{\left(q^2 g^{ \mu \nu} -q^\mu q^\nu \right)}_{T_3}\Biggr]~ \frac{F_{{\cal A}_-}(k,k-q,\zeta) }{2}
\nonu
={g^{ \mu \nu}\over 2}~~
  \Bigl[F_{{\cal A}_+}(k,k-q,\zeta)-q^2~F_{{\cal A}_-}(k,k-q,\zeta)\Bigr]
\nonu  - 2{\cal K}^{\mu\nu}~ F_{{\cal A}_-}(k,k-q,\zeta) ~~,
\nonu
\nonu
Tr^{\mu\nu}_3=~
 -(k-q)^\mu~(2k-q)^\nu~  F_{\cal B}(k,k-q,\zeta)~~,
\nonu
 =-\Bigl[2{\cal K}^{\mu\nu} +q^\mu
 (q^\nu - k^\nu )\Bigr]~F_{\cal B}(k,k-q,\zeta) ~~,
 \nonu \nonu
Tr^{\mu\nu}_4
  =- k^\mu~(2k-q)^\nu~ F_{\cal B}(k,k-q,\zeta)
  \nonu
  =
  -\Bigl[2 {\cal K}^{\mu\nu}
 + k^\nu q^\mu\Bigr]~F_{\cal B}(k,k-q,\zeta) ~~.\ee
Saturating the tensor $Sp^{\mu\nu}$ with ${\cal P}^{\mu\nu}$ one gets 
for $T_P$ (recall that $g_{\mu\nu}{\cal P}^{\mu\nu}=0$)
\be
{\cal T}_P(\zeta,\Lambda;q)= 
  -i 
     Z_1(\zeta,\Lambda)~{4\over 3}~{e_R^2\over q^2}~\int_{s_{th}} ds\int_{s'_{th}} ds'
\nonu \times~    \int_{\Lambda}  \frac{d^4k}{(2\pi)^4} ~{1 \over k^2-s +i\epsilon} 
    ~{1 \over (k-q)^2-s' +i\epsilon}  
 \nonu \times ~  \Biggl\{ \bar \sigma_V(s',\zeta,s'_{th})  \bar \sigma_V(s,\zeta,s_{th})~
 \Biggl[\Bigl({\cal R}_1(k,q) +3k\cdot q 
 \Bigr)
\nonu \times ~2\Bigl (1-{\cal A}_R(\zeta,k)\Bigr)
-2\Bigl[ (k-q)^2~\Bigl({\cal R}_1(k,q)+3 k\cdot q\Bigr)  
\nonu -2 k^2q^2  
+2 (k\cdot q)^2 
 \Bigr]
~F_{{\cal A}_-}(k,k-q,\zeta)
 \Biggr] 
-2\bar \sigma_S(s',\zeta,s'_{th}) 
\nonu \times  \bar \sigma_S(s,\zeta,s_{th})~\Bigl(
 {\cal R}_1(k,q) +3 k\cdot q 
   \Bigr)~F_{{\cal A}_-}(k,k-q,\zeta)  
 \nonu - \bar \sigma_S(s',\zeta,s'_{th})  \bar \sigma_V(s,\zeta,s_{th})~
 \Bigl[2 {\cal R}_1(k,q) 
 +3k\cdot q 
 \nonu+3 \Bigl(2(k\cdot q)-q^2\Bigr)\Bigr]~F_{\cal B}(k,k-q,\zeta)  
  -\bar \sigma_V(s',\zeta,s'_{th})  
  \nonu \times ~\bar \sigma_S(s,\zeta,s_{th})  ~\Bigl[2 {\cal R}_1(k,q) +3k\cdot q\Bigr]~ F_{\cal B}(k,k-q,\zeta)\Biggr\} 
 \nonu\label{calTP_1}\ee
with
\be
{\cal R}_1(k,q)=k^2-4{(k\cdot q)^2\over q^2}
\ee
and  (see Eq. \eqref{app_3_canc}, with $p\to k-q$ )
\be
2\Bigl( 1-{\cal A}_R(\zeta,k)\Bigr)=F_{{\cal A}_+}(k,k-q,\zeta)
\nonu 
-(2 k\cdot q -q^2)~ F_{{\cal A}_-}(k,k-q,\zeta)
\ee
As in the case of the fermion self-energy (see  \ref{app_3}), 
inserting the expressions of ${\cal F}_{A_+}$, ${\cal F}_{A_-}$ and ${\cal F}_B$
in terms of the NWFS (see Eq. \eqref{FAi})  one can
write
\be
{\cal T}_P(\zeta,\Lambda;q)=
 -i  Z_1(\zeta,\Lambda)~{4\over 3}~{e_R^2\over q^2}~\int_{s_{th}}^\infty ds~
     \int_{0}^\infty ds'
     \nonu
     \times~
    \Biggl\{ \bar\sigma_V(s',\zeta,s'_{th}) ~ \bar\sigma_V(s,\zeta,s_{th})
 \Biggl[ 2~{\cal I}_6(q,s,s') 
 \nonu \times~\Biggl(1
 + \int_{s_{th}}^\infty d\omega ~
{\rho_A(\omega,\zeta)\over 
(\zeta^2-\omega +i\epsilon)}\Biggr)  
+2\int_{s_{th}}^\infty d\omega 
\nonu \times 
   ~\rho_A(\omega,\zeta)\Bigl(
   \omega ~{\cal I}_7(q,s,s',\omega)-2 ~{\cal I}_8(q,s,s',\omega)
 \Bigr)  \Biggr] 
\nonu
+2\bar\sigma_S(s',\zeta,s'_{th}) ~ \bar\sigma_S(s,\zeta,s_{th})
\nonu \times~
\int_{s_{th}}^\infty d\omega 
   ~\rho_A(\omega,\zeta)~{\cal I}_7(q,s,s',\omega)
  \nonu +\Bigl[ \bar\sigma_S(s',\zeta,s'_{th}) ~ \bar\sigma_V(s,\zeta,s_{th})
 \nonu  +\bar\sigma_V(s',\zeta,s'_{th}) ~ ~\bar\sigma_S(s,\zeta,s_{th})
   \Bigr]~
  \int_{s_{th}}^\infty d\omega 
   ~\rho_B(\omega,\zeta)
   \nonu \times ~
   \Bigl(2 {\cal I}_7(q,s,s',\omega)   -3{\cal I}_9(q,s,s',\omega) \Bigr)~ 
 \nonu + 3 \bar\sigma_S(s',\zeta,s'_{th}) ~ \bar\sigma_V(s,\zeta,s_{th})~\int_{s_{th}}^\infty d\omega 
   ~\rho_B(\omega,\zeta) 
   \nonu \times~      
 \Bigl(2{\cal I}_{9}(q,s,s',\omega)-q^2{\cal I}_{10}(q,s,s',\omega)\Bigr)~  
 \Biggr\}
\label{app_4_Tpq}\ee
where the integrals ${\cal I}_i$, are defined as follows
\be
{\cal I}_6(q,s,s')=\int_{\Lambda} {d^4k\over (2\pi)^4}~
 {\left[k^2-4{(k\cdot q)^2\over q^2}+3k\cdot q\right]\over k^2-s+i\epsilon}
 \nonu \times~{1\over (k-q)^2-s'+i\epsilon} 
\label{cal_I6a}
\\ & &
 {\cal I}_7(q,s,s',\omega)=\int_{\Lambda}\frac{d^4k}{(2\pi)^4}
\frac{ \left[k^2-4{(k\cdot q)^2\over q^2}+3k\cdot q\right]}{k^2-s+i\epsilon} 
\nonu \times {1 \over \left[(k-q)^2-s'+i\epsilon\right]\left[ k^2-\omega+i\epsilon\right]\left[(k-q)^2-\omega+i\epsilon \right]}
\nonu\label{cal_I7a}
\\ & &
{\cal I}_8(q,s,s',\omega)=\int_{\Lambda}\frac{d^4k}{(2\pi)^4}
\frac{\left[k^2q^2-(k\cdot q)^2\right]}{k^2-s+i\epsilon}
\nonu\times ~{1\over\left[(k-q)^2-s'+i\epsilon\right]\left[ k^2-\omega+i\epsilon\right]\left[(k-q)^2-\omega+i\epsilon \right]}
\nonu\label{cal_I8a}
\\ & &
{\cal I}_9(q,s,s',\omega)=\int_{\Lambda} {d^4k\over (2\pi)^4}
 ~{(k\cdot q)\over k^2-s+i\epsilon}
 \nonu \times~{1\over [(k-q)^2-s'+i\epsilon]~
 [ k^2-\omega+i\epsilon]~[ (k-q)^2-\omega+i\epsilon]}
\nonu\label{cal_I9a}
\\ & &
{\cal I}_{10}(q,s,s',\omega)=\int_{\Lambda} {d^4k\over (2\pi)^4}
 ~{1\over k^2-s+i\epsilon}
\nonu \times  ~{1\over [(k-q)^2-s'+i\epsilon]~
 [ k^2-\omega+i\epsilon]~[ (k-q)^2-\omega+i\epsilon]}
\nonu\label{cal_I10a}\ee
with $s,s',\omega\ge 0$. {Recall that the external momentum $q^2$ is 
chosen spacelike, for the sake of simplicity in the formal 
 elaboration.} 

Notice that  ${\cal I}_6$ {presents an apparent quadratically divergence}, 
as expected. {Therefore, we exploit} dimensional regularization, like  the integral ${\cal
I}_0$ (see Eq. \eqref{cal_D0}), and 
 obtain
 \be
 {\cal I}_6(q,s,s')
 =i {3q^2\over (4\pi)^2}\int_0^1 d\xi ~
  \int_0^\infty  d x
  \nonu \times~
 {\xi (1-\xi)~x \over \Bigl[-x + \xi (1-\xi)q^2 -\xi s'-(1-\xi)s +i\epsilon\Bigr]^2}
  \label{cal_I6p}\ee
  that has a logarithmic divergence, harmless once we subtract the corresponding
  integral evaluated at $\zeta^2_p$ (see Eq. \eqref{app_4_Pir}).
  
   The other  integrals are convergent and after applying the Feynman
   parametrization and the change of variable $p_0\to ip_4$, one gets
   %
   %
   \be
     {\cal I}_7(q,s,s',\omega)
      =i{3\over (4\pi)^2}~ q^2~ \int_0^1 d\xi 
 ~ \int_0^{1-\xi} dv
  \nonu\times~  \int_0^{1-\xi-v} dw
  ~{{\cal A}_4(v,w)\over \Bigl[D^a_{ph}(q,s',s,v,w,\xi,\omega)
 +
 i\epsilon\Bigr]^2}~,
 \label{cal_I7p}
   \ee
   %
   %
 \be
 {\cal I}_8(q,s,s',\omega)
 =i {3~q^2\over 2(4\pi)^2}\int_0^1 d\xi 
  ~ \int_0^{1-\xi} dv
  \nonu\times~ \int_0^{1-\xi-v}dw
  ~ {1\over \Bigl[D^a_{ph}(q,s',s,v,w,\xi,\omega)
  +
 i\epsilon\Bigr]}~,
 \label{cal_I8p}
\ee
%
%
 \be 
 {\cal I}_{9}(q,s,s',\omega)
 =i {q^2\over (4\pi)^2}\int_0^1 d\xi 
  ~ \int_0^{1-\xi} dv
\nonu\times~  \int_0^{1-\xi-v}dw
  ~{(v+w)\over \Bigl[D^a_{ph}(q,s',s,v,w,\xi,\omega)
  +
 i\epsilon\Bigr]^2}~,
 \label{cal_9p}
\ee
   %
   %
\be
 {\cal I}_{10}(q,s,s',\omega)
 =i {1\over (4\pi)^2}~\int_0^1 d\xi 
  ~ \int_0^{1-\xi} dv
  \nonu \times ~\int_0^{1-\xi-v} dw~
  {\over \Bigl[D^a_{ph}(q,s',s,v,w,\xi,\omega)+
 i\epsilon\Bigr]^2}~,
 \label{cal_10p}
 \ee
 where
 \be
 D^a_{ph}(q,s',s,v,w,\xi,\omega)={\cal A}_4(v,w)q^2 -v s'-(\xi+w) \omega 
 \nonu -(1-v-\xi-w)s~.\nonumber \ee
By using  Eq. \eqref{app_4_Tpq} and 
the definition in Eq. \eqref{app_4_Pir},  
the photon self-energy reads (see  Eq. \eqref{NIRph})
\be
\Pi_R(\zeta;q)
= Z_1(\zeta,\Lambda)~{4\over 3}~e_R^2 ~(\zeta^2_p-q^2)~
  ~\int_{0}^\infty ds~
     \int_{0}^\infty ds'~
  \nonu \times~  \Biggl\{ \bar\sigma_V(s',\zeta,s'_{th}) ~ \bar\sigma_V(s,\zeta,s_{th})
 \Biggl[ 2~{\cal D}_6(q,\zeta_p,s,s') 
 \nonu \times ~\Biggl(1
 + \int_{s_{th}}^\infty d\omega ~
{\rho_A(\omega,\zeta )\over 
(\zeta^2-\omega +i\epsilon)}\Biggr)  
+2\int_{s_{th}}^\infty d\omega 
   ~\rho_A(\omega,\zeta)
   \nonu \times~\Bigl(
   \omega ~{\cal D}_7(q,\zeta_p,s,s',\omega)-2 ~{\cal D}_8(q,\zeta_p,s,s',\omega)
 \Bigr)  \Biggr] 
\nonu
+2\bar\sigma_S(s',\zeta,s'_{th})  ~\bar\sigma_S(s,\zeta,s_{th})
\nonu \times ~
\int_{s_{th}}^\infty d\omega 
   ~\rho_A(\omega,\zeta)~{\cal D}_7(q,\zeta_p,s,s',\omega)
  \nonu + \bar\sigma_S(s',\zeta,s'_{th}) ~ \bar\sigma_V(s,\zeta,s_{th})
  \nonu \times ~
  \int_{s_{th}}^\infty d\omega 
   ~\rho_B(\omega,\zeta)~
    {\cal D}_{7,9}(q,\zeta_p,s,s',\omega)   
 \nonu +  \bar\sigma_V(s',\zeta,s'_{th}) ~ ~\bar\sigma_S(s,\zeta,s_{th})~\int_{s_{th}}^\infty d\omega 
   ~\rho_B(\omega,\zeta)
   \nonu \times ~       
 \Bigl( {\cal D}_{7,9}(q,\zeta_p,s,s',\omega) +
 {\cal D}_{9,10}(q,\zeta_p,s,s',\omega)) \Bigr)
 \Biggr\}
 \nonu \label{app_4_selfR}\ee
 where
\be
{\cal D}_n(q,\zeta_p,s,s')= -i{1\over (\zeta^2_p-q^2)}~
\nonu \times ~\left[{{\cal I}_n(q,s,s')\over q^2}-
{{\cal I}_n(\zeta_p,s,s')\over \zeta^2_p}\right]~~,
\label{cal_D6_10}
 \ee
 with $n=6,7,8,9,10$,
 and
 \be
 {\cal D}_{7,9}(q,\zeta_p,s,s',\omega)= 
2{\cal D}_{7}(q,\zeta_p,s,s',\omega)
\nonu-3{\cal D}_{9}(q,\zeta_p,s,s',\omega)~~,
\label{cal_D79a}
\\ & &
{\cal D}_{9,10}(q,\zeta_p,s,s',\omega)= 
6{\cal D}_{9}(q,\zeta_p,s,s',\omega)
\nonu-3{\cal D}_{10}(q,\zeta_p,s,s',\omega)~~.
\label{cal_D910a}
\ee

The explicit expressions of ${\cal D}_6$, ${\cal D}_7$ and ${\cal D}_8$, are  
 \be
 {\cal D}_6(q,\zeta_p,s,s')= 
 {6\over (4\pi)^2}~ \int_0^1 d\xi ~
 \xi^2 (1-\xi)^2 
 \int_0^1 dv
\nonu \times ~\int_0^\infty  d x
 {x \over \Bigl[D^b_{ph}(q,s',s,v,x,\xi,\omega,\zeta_p)
 +i\epsilon\Bigr]^3
 }
\nonu
= {3\over (4\pi)^2}~ \int_0^1 d\xi ~
 \xi (1-\xi) \int_0^1 dv \int_0^\infty dy
\nonu \times ~ {\delta\Bigl[ y - (\xi s'+(1-\xi)s)/(\xi(1-\xi)) \Bigr] \over \Bigl[ (1-v)q^2 
 +v\zeta^2_p-y+i\epsilon\Bigr]}
\nonu
= -{3\over (4\pi)^2}~ \int_0^1 d\xi ~
 \xi (1-\xi) \int_0^1 dv \int_0^\infty dy
\nonu \times ~ {\Theta\Bigl[ y\xi(1-\xi) - \xi s'-(1-\xi)s \Bigr] \over \Bigl[ (1-v)q^2 
 +v\zeta^2_p-y+i\epsilon\Bigr]^2}~~,
\label{cal_D6} 
\ee
with
\be
D^b_{ph}(q,s',s,v,x,\xi,\omega,\zeta_p)=-x + \xi (1-\xi)q^2 
\nonu-\xi s'-(1-\xi)s 
 +v\xi(1-\xi)(\zeta^2_p-q^2)
\nonumber 
\ee
 \be
 {\cal D}_7(q,\zeta_p,s,s',\omega)= 
 {6\over (4\pi)^2}~  \int_0^1 d\xi 
  ~ \int_0^{1-\xi} dv\int_0^{1-\xi-v} \hspace{-0.4cm}{dw}
  \nonu \times ~\int_0^1 dt~
 {{\cal A}^2_4(v,w)\over \Bigl[D^c_{ph}(q,s',s,v,w,t,\xi,\omega,\zeta_p)
 + i\epsilon\Bigr]^3
}
 \nonu 
 =-{3\over (4\pi)^2}~  \int_0^1 d\xi 
  ~ \int_0^{1-\xi} dv \int_0^{1-\xi-v} {dw\over {\cal A}_4(v,w)}
 \nonu \times ~\int_0^1 dt\int_0^\infty dy  
~
 {{\partial\over \partial y}\delta \Bigl[ y-{\cal A}_7(s,s',\omega,v,\xi,w)
 \Bigr]\over \Bigl[(1-t)q^2 + t \zeta^2_p
 -y
+ i\epsilon\Bigr]^2}~,
\label{cal_D7}
\ee
with
\be
D^c_{ph}(q,s',s,v,w,t,\xi,\omega,\zeta_p)=D^a_{ph}(q,s',s,v,w,\xi,\omega)
\nonu+ tv(1-v) (\zeta^2_p-q^2)~~,
\nonumber\ee
 \be
 {\cal D}_8(q,\zeta_p,s,s',\omega)= 
 {3\over 2(4\pi)^2}~\int_0^1 d\xi 
  ~ \int_0^{1-\xi} dv
  \nonu \times ~ \int_0^{1-\xi-v} dw 
   ~{\cal A}_4(v,w)\int_0^1dt~
 \nonu \times {1\over \Bigl[D^c_{ph}(q,s',s,v,w,\xi,\omega) +i\epsilon\Bigr]^2}
 \nonu
 ={3\over 2(4\pi)^2}~\int_0^1 d\xi 
  ~ \int_0^{1-\xi} dv\int_0^{1-\xi-v} {dw \over  {\cal A}_4(v,w)}
\nonu \times~\int_0^1dt \int_0^\infty dy {\delta\Bigl[y -{\cal A}_7(s,s',\omega,v,\xi,w)
 \Bigr]\over \Bigl[(1-t)q^2 +t\zeta^2_p-y
+i\epsilon\Bigr]^2}~~,
\label{cal_D8}
\ee
In view of the application of the uniqueness theorem for extracting $\rho_\gamma$, it is useful to apply an integration by part while
evaluating  $ {\cal D}_{9}$ and $ {\cal D}_{10}$.  One gets
\be
 {\cal D}_9(q,\zeta_p,s,s',\omega)
={2\over (4\pi)^2}~\int_0^1  d\xi
  ~ \int_0^{1-\xi} dv
  \nonu \times ~\int_0^{1-\xi-v} dw~{(v+w)\over {\cal A}^2_4(v,w)} 
  \nonu
  \times~\int_0^1dt
 \int_0^\infty dy ~{\delta\Bigl[y -{\cal A}_7(s,s',\omega,v,\xi,w)
 \Bigr]\over
   \Bigl[(1-t)q^2 +
   t\zeta^2_p-y + i\epsilon\Bigr]^3}=
 \nonu
=-{1\over (4\pi)^2}~\int_0^1 d\xi 
  ~ \int_0^{1-\xi} dv \int_0^{1-\xi-v} dw~{(v+w)\over {\cal A}^2_4(v,w)}
  \nonu \times
   \int_0^1 dt
 \int_0^\infty dy ~{{\partial\over \partial y}\delta\Bigl[y -{\cal A}_7(s,s',\omega,v,\xi,w)
 \Bigr]\over
   \Bigl[(1-t)q^2 +
t\zeta^2_p-y + i\epsilon\Bigr]^2}~,
\label{cal_D9}
\ee
and
\be
 {\cal D}_{10}(q,\zeta_p,s,s',\omega)
={2\over (4\pi)^2}~\int_0^1 d\xi 
  ~ \int_0^{1-\xi} dv
  \nonu \times ~\int_0^{1-\xi-v} {dw\over {\cal A}^2_4(v,w)} 
  \int_0^1dt
  \int_0^\infty dy  \nonu
   \times~
  {\delta\Bigl[y-{\cal A}_7(s,s',\omega,v,\xi,w)
  \Bigr]
  \over \Bigl[(1-t)q^2 +
  t\zeta^2_p-y + i\epsilon\Bigr]^3}=
\nonu
=-{1\over (4\pi)^2}~\int_0^1 d\xi 
  ~ \int_0^{1-\xi} dv\int_0^{1-\xi-v} {dw\over {\cal A}^2_4(v,w)}
  \nonu \times
  \int_0^1dt
 \int_0^\infty dy  ~
  {{\partial\over \partial y}\delta \Bigl[y-{\cal A}_7(s,s',\omega,v,\xi,w)
 \Bigr]
  \over \Bigl[(1-t)q^2 +
t\zeta^2_p-y + i\epsilon\Bigr]^2}~,
\label{cal_D10}\ee
with
\be
{\cal A}_7(s,s',\omega,v,\xi,w)=
\nonu
={v s'+(\xi+w) \omega  +(1-\xi-v-w)s\over {\cal A}_4(v,w)}
~~.\nonumber \ee
Finally, one has
\be
{\cal D}_{7,9}(q,\zeta_p,s,s',\omega)= 
-{3\over (4\pi)^2}~\int_0^1 d\xi 
  ~ \int_0^{1-\xi} dv
  \nonu\times\int_0^{1-\xi-v} dw ~{v+w\over {\cal A}^2_4(v,w)}~
  \Bigl[1 -2 (v+w)\Bigr]\int_0^1 dt
 \nonu\times \int_0^\infty dy ~{{\partial\over \partial y}\delta \Bigl[y -{\cal A}_7(s,s',\omega,v,\xi,w)
 \Bigr]\over
   \Bigl[(1-t)q^2 +
t\zeta^2_p-y + i\epsilon\Bigr]^2}~~,
\ee 
and
\be
{\cal D}_{9,10}(q,\zeta_p,s,s',\omega)=
{3\over (4\pi)^2}~\int_0^1 d\xi 
  ~ \int_0^{1-\xi} dv
  \nonu \times\int_0^{1-\xi-v} {dw \over {\cal A}^2_4(v,w)}~
  \Bigl[1 - 2(v+w)\Bigr] \int_0^1 dt
\nonu \times  \int_0^\infty dy ~{{\partial\over \partial y}\delta \Bigl[y -{\cal A}_7(s,s',\omega,v,\xi,w)
 \Bigr]\over
   \Bigl[(1-t)q^2 +
t\zeta^2_p-y + i\epsilon\Bigr]^2}
\nonu\ee
 Before obtaining the integral equation fulfilled by the NWF $\rho_\gamma$, it is useful to show 
  that the  last two terms in Eq. \eqref{app_4_selfR} are equal. As a matter of fact, one can recast the following term in a different form by
reintroducing the fourth Feynman parameter, i.e.
\be 
\int_{s_{th}}^\infty ds~
     \int_{0}^\infty ds'\bar\sigma_S(s',\zeta,s'_{th}) ~ \bar\sigma_V(s,\zeta,s_{th})
      \int_0^1 d\xi
      \nonu \times
  ~ \int_0^{1} dv\int_0^{1} dw \int_0^1 du~(v+w)
  \nonu \times ~\delta(1-u-w-v-\xi)~{1 -2 (v+w)\over (v+w)^2 (1-v-w)^2}
   \nonu \times
   ~{\partial\over \partial y}\delta\Bigl[y -{vs'+(\xi+w)\omega) +us \over (v+w) (1-v-w) }\Bigr] 
  \ee
 Then,  re-naming the following variables:
 i) $s\to s'$, ii) $v\to u$ and iii) $\xi\to w$
 and iv) by exploiting the delta function, one has $u+\xi=1-v-w$ and can write
\be
\int_{0}^\infty ds~
     \int_{0}^\infty ds'~\bar\sigma_S(s,\zeta,s_{th}) ~ \bar\sigma_V(s',\zeta,s'_{th}) 
     \int_0^1 d\xi
     \nonu \times 
  ~ \int_0^{1} dv\int_0^{1} dw \int_0^1 du~(1-v-w)
  \nonu \times ~\delta(1-v-\xi -u-w)~{1 -2 (1-v-w)\over (1-v-w)^2 (v+w)^2}
  \nonu \times  ~{\partial\over \partial y}
  \delta\Bigl[y -{us+(w+\xi)\omega) +vs' \over (1-v-w) (v+w) }\Bigr]=
   \nonu
   =-\int_{0}^\infty ds~
     \int_{0}^\infty ds'~~\bar\sigma_S(s,\zeta,s_{th}) ~ \bar\sigma_V(s',\zeta,s'_{th})
      \int_0^1 d\xi
 \nonu\times ~ \int_0^{1} dv\int_0^{1} dw \int_0^1 du~(1-v-w)
  \nonu \times ~\delta(1-v-\xi -u-w)~{1 -2 (v+w)\over (1-v-w)^2 (v+w)^2}
  \nonu\times  ~{\partial\over \partial y}\delta\Bigl[y -{vs'+(w+\xi)\omega) + us\over (1-v-w) (v+w) }\Bigr]
   ~~~~~Q.E.D.
\nonu\ee
After introducing the uniqueness theorem in Eq. \eqref{app_4_selfR}, 
one gets for  $\rho_\gamma$ 
(see Eq. \eqref{NIRph})
\be
{ \Theta(y-s^p_{th})~\rho_\gamma(y,\zeta)=} -
\lim_{\Lambda \to \infty}Z_1(\zeta,\Lambda)~{e_R^2\over (2\pi)^2} 
\int_{0}^\infty ds
\nonu \times ~
     \int_{0}^\infty ds'~ \int_0^1 d\xi
~  \Biggl\{ \bar\sigma_V(s',\zeta,s'_{th}) ~ \bar\sigma_V(s,\zeta,s_{th})
\nonu \times
 \Biggl[2 
 \xi (1-\xi)  
 \Theta\Bigl[ y \xi(1-\xi)- \xi s'-(1-\xi)s \Bigr] 
 \nonu\times ~\Biggl(1
 + \int_{s_{th}}^\infty d\omega ~
{\rho_A(\omega,\zeta, \Lambda)\over 
(\zeta^2-\omega +i\epsilon)}\Biggr) 
 +2\int_{s_{th}}^\infty d\omega 
   ~\rho_A(\omega,\zeta) 
   \nonu \times
  ~ \int_0^{1-\xi} dv\int_0^{1-\xi-v} {dw\over {\cal A}_4(v,w)}
    \nonu\times~\Biggl(\omega~ 
~{\partial\over \partial y}\delta \Bigl[ y-{\cal A}_7(s,s',\omega,v,\xi,w)\Bigr]   
  \nonu +
   \delta\Bigl[y -{\cal A}_7(s,s',\omega,v,\xi,w)\Bigr]\Biggr)
     \Biggr] 
+~\bar\sigma_S(s',s'_{th}\zeta) 
\nonu\times~ \bar\sigma_S(s,\zeta,s_{th})~
\int_{s_{th}}^\infty d\omega 
   ~\rho_A(\omega,\zeta)~  
  \int_0^{1-\xi} dv 
  \nonu \times~\int_0^{1-\xi-v} {dw\over {\cal A}_4(v,w)}
~{\partial\over \partial y}\delta \Bigl[ y-{\cal A}_7(s,s',\omega,v,\xi,w)\Bigr] 
  \nonu 
  +
  2\bar\sigma_S(s',\zeta,s'_{th}) ~ \bar\sigma_V(s,\zeta,s_{th})~
  \int_{s_{th}}^\infty d\omega 
   ~\rho_B(\omega,\zeta)
 \nonu \times ~ \int_0^{1-\xi} dv \int_0^{1-\xi-v} dw ~(v+w)~{1 -2 (v+w)\over {\cal A}_4^2(v,w)}
  \nonu \times ~{\partial\over \partial y}\delta\Bigl[y -{\cal A}_7(s,s',\omega,v,\xi,w) \Bigr]   
 \Biggr\}~~,
 \nonu \label{app_4_rhogf}
\ee
with  \be
Z_1(\zeta),=\lim_{\Lambda \to \infty}Z_1(\zeta,\Lambda)~,
\nonu  {\cal A}_4=(t+w)~(1-t-w)\nonu
{\cal A}_7(s,s',\omega,v,\xi,w)=
\nonu
={\Bigl[v s'+(\xi+w) \omega  +(1-\xi-v-w)s\Bigr]\over {\cal A}_4(v,w)}
~.\nonumber \ee


\section{ First iteration}
\label{app_5}
This Appendix is devoted to present a first analytic result obtained by iterating one time the coupled system we have obtained.

The inputs are given by the zeroth-order NWFs $\rho_A$,  $\rho_B$ and $\rho_\gamma$
, i.e.
\be
  \label{eq:initialpoint}
  \rho_A^{(0)}(s,\zeta)  =  \rho_B^{(0)}(s,\zeta)=\rho^{(0)}_\gamma(s,\zeta)= 0~~,
 \ee
 Hence, the KL weights of the fermion and photon propagators (see Eqs.
 \eqref{klwf},  \eqref{sigmag} and \eqref{def_sig}) read 
 \be
  \bar{\sigma}^{(0)}_V(s,\zeta)  =   \delta(s-m^2(\zeta)) ~~,\nonu
  \bar{\sigma}^{(0)}_S(s,\zeta)  =   m(\zeta)~ \delta(s-m^2(\zeta))~~, 
 \nonu \bar{\sigma}^{(0)}_\gamma(s,\zeta,\zeta_p)  =   \delta(s-\zeta^2_p)~~,
\ee
and  the renormalization constants become (see Eq. \eqref{Z2_NIR})
\be
 Z_2^{(0)}  =  1 +\int_{s_{th}}^\infty ds~ {\rho_A^{(0)}(s,\zeta)\over \zeta^2-s+i\epsilon}=1=Z_1^{(0)} 
\ee
By inserting   the tree-level expressions, Eq. \eqref{eq:initialpoint}, in Eq. \eqref{rhoA} 
one obtains the first iteration for $\rho_A$, viz
\be
\Theta\Bigl(y-s_{th}\Bigr)~ \rho^{(1)}_A(y,\zeta)
  = 
 {3\over (4\pi)^2} ~e^2_R 
 ~\int_{0}^\infty~d\omega~\delta(\omega-\zeta^2_p)
\nonu \times~\int_0^1 d\xi~\int_{0}^\infty~ds'~\delta(s'-m^2(\zeta)) 
 \nonu
 \times
 ~\Biggl[\xi~
 \Theta\Bigl[y\xi(1-\xi) - \xi \omega -(1-\xi) s'\Bigr]
 \nonu - \int_0^{1-\xi}dt~
 \Theta\Bigl[ y t(1-t)- \xi \omega -t s'\Bigr] \Biggr]
 \nonu 
=   \frac{3e_R^2 }{(4\pi)^2}~  
\int_0^1 d\xi
 \Biggl\{\xi~\Theta\Bigl[y\xi(1-\xi)- (1-\xi)m^2(\zeta)  -\xi \zeta_p^2\Bigr] 
\nonu
- \int _0^{1-\xi}dt~\Theta\Bigl[yt(1-t)-tm^2(\zeta)-\xi \zeta_p^2 \Bigr] 
\Biggr\}
\nonu
 = \frac{3e_R^2 }{(4\pi)^2}  
	       \Biggl\{\int_0^1 d\xi~\xi~\Theta\Bigl[\xi(1-\xi)y- (1-\xi)m^2(\zeta)  -\xi \zeta_p^2\Bigr] 
	   \nonu
	       - \int_0^1 d\xi\int_0^1dt'~\Theta(t'-\xi)
 \nonu\times ~\Theta\Bigl[t'(1-t')y-(1-t') m^2(\zeta)-\xi \zeta_p^2 \Bigr] \Biggr\}~~.	       
\label{rhoA1_1}
\ee	 
Notice that the two theta functions imply also
\be
\xi y-m^2(\zeta)\ge 0 ~~, \quad \quad  t' y-m^2(\zeta)\ge 0  ~~.     
\ee
The second integral is usefully manipulated as follows 
\be
\int_0^1 d\xi\int _0^{1}dt'~\Theta(t'-\xi)
\nonu\times~\Theta\Bigl[t'(1-t')y-(1-t')m^2(\zeta)-\xi \zeta_p^2 \Bigr]
\nonu
=\int_0^1 dt' ~\Theta[t'y-m^2(\zeta)]~\int_0^1 d\xi~
\nonu\times~\Biggl\{\Theta (t'-\xi)~
\Theta\left[t'(1-t')y-(1-t')m^2(\zeta)-t' \zeta_p^2 \right]
\nonu
+ \Theta\left[t'(1-t')y-(1-t')m^2(\zeta)-\xi \zeta_p^2 \right]~
\nonu \times
~\Theta\Bigl\{t'\zeta_p^2 -[t'(1-t')y-(1-t')m^2(\zeta)]\Bigr\}~\Biggr\}
\nonu
=\int_0^1 dt' ~\Theta[t'y-m^2(\zeta)]
\nonu \times~\Biggl\{t'~
\Theta\left[t'(1-t')y-(1-t')m^2(\zeta)-t' \zeta_p^2 \right]
\nonu
+ {t'(1-t')y-(1-t')m^2(\zeta)\over \zeta_p^2} 
\nonu \times
~\Theta\Bigl[t'\zeta_p^2 -\Bigl(t'(1-t')y-(1-t')m^2(\zeta)\Bigr)\Bigr]~\Biggr\}~,
\label{rhoA1_2}
\ee
where the two contributions are obtained  by exploiting  the two sets of inequalities
\be
{t'(1-t')y-(1-t')m^2(\zeta)\over \zeta^2_p} \ge t' \ge \xi
~~, \nonu
t' \ge {t'(1-t')y-(1-t') m^2(\zeta)\over  \zeta^2_p} \ge \xi~~.
\ee
Both sets are generated by the constraints on the variable $\xi$ in the first line of \eqref{rhoA1_2}. 

Recollecting the above results one obtains the following  expression of $\rho^{(1)}_A$ 
\be
\Theta\Bigl(y-s_{th}\Bigr)~ \rho^{(1)}_A(y,\zeta)
  = 
 ~{3\over (4\pi)^2} ~e^2_R ~\Theta(y-m^2(\zeta))
 \nonu \times ~\int_{{m^2(\zeta)/ y}}^1 dt'  
~ {t^{\prime 2} y -t'(y+m^2(\zeta))  +m^2(\zeta)\over \zeta_p^2} 
\nonu \times
~\Theta\Bigl[ t^{\prime 2} y -t'(y+m^2(\zeta)-\zeta_p^2)  +m^2(\zeta)\Bigr]
\label{rhoA1_3}\ee
The constraints imposed by the theta function on $t'$ can be obtained from 
 the solutions of the 
second-order equation, that read
\be
t'_\pm={1 \over 2y}~\Bigl[y+m^2(\zeta)-\zeta^2_p
\nonu\pm
 \sqrt{[y-m^2(\zeta)-\zeta_p^2]^2-4m^2(\zeta)\zeta_p^2}~\Bigr]
\label{xi_ex}\ee
It is important to notice that both real and complex-conjugated solutions 
are allowed, due to the presence of IR-regulator $\zeta^2_p$ in the
discriminant. The complex-conjugated solutions  lead to an 
IR-dependent contribution in $\rho^{(1)}_A$ 
that properly vanishes in the limit $\zeta_p\to0$
 matching the  constraint
 expected from the lhs of
Eq. \eqref{rhoA1_3}. Notably, this term guarantees the
continuity of $\rho_A$ and therefore of the K\"all\'en-Lehman weights, when
approaching the physical threshold $y=m^2(\zeta)$. 

The real positive solutions $t'_\pm$ are obtained when
\be
y\ge y_+=[m(\zeta)+ \zeta_p]^2~~~{\rm or}~~~ y_-=[m(\zeta)- \zeta_p]^2\ge y~~,
\nonu\ee
with also
$
y-\zeta^2_p+m^2(\zeta)\ge 0$.
The constraint $y\ge m^2(\zeta)$ in Eq. \eqref{rhoA1_3}
 excludes $y_-$  and one remains with
$
\Theta\Bigl[y-(m+ \zeta_p)^2\Bigr]$.
 In this case, one can easily show that the real solutions fulfill 
\be 
 1\ge t^{r}_\pm \ge {m^2(\zeta)\over y}~~.
\ee
Therefore { $$t'\in ~ \Bigl[{m^2(\zeta)\over y}, t^r_-\Bigr]\cup 
\Bigl[t^r_+,1\Bigr]~~~.$$}

{ The discriminant is negative,  when
\be
[m(\zeta)+ \zeta_p]^2\ge y\ge [m(\zeta)- \zeta_p]^2~~,
\ee and   $t'$ does not have any constraint, i.e. $t'\in [m^2(\zeta)/y,1]$. }
Moreover,  taking  into account $\Theta[y-m(\zeta)]$ one remains with
 $[m(\zeta)+ \zeta_p]^2\ge y\ge m^2(\zeta)$, that generates 
  an IR-dependent term with no impact for
  $\zeta^2_p \to 0$.

In conclusion, $\rho^{(1)}_A$ is given by
\be
 \Theta\Bigl(y-s_{th}\Bigr)~ \rho^{(1)}_A(y,\zeta)
  = 
  -{3e^2_R\over (4\pi)^2 \zeta^2_p}\Theta(y-m^2(\zeta))
  \nonu \times
  \Biggr\{\Theta\Bigl[[m(\zeta)+ \zeta_p]^2- y\Bigr]~ 
  {\Bigl(y-m^2(\zeta)\Bigr)^3 \over 6y^2}
 \nonu
 + \Theta\Bigl[y-[m(\zeta)+ \zeta_p]^2\Bigr]~ 
 \Biggl[ {\Bigl(y-m^2(\zeta)\Bigr)^3 \over 6y^2} 
 \nonu
 + 
 \left.{2 t^{\prime 3} y -3t^{\prime 2} 
 (y+m^2(\zeta)) +6m^2(\zeta) t' \over 6}\right|_{t'_-}^{t'_+}\Biggr]\Biggr\}=
 \nonu
 =-{e^2_R\over 2(4\pi)^2}~{1\over \zeta^2_py^2}~\Theta(y-m^2(\zeta))
 \Biggr\{\Theta\Bigl[y-[m(\zeta)+ \zeta_p]^2\Bigr]
 \nonu \times ~\Bigl(y-m^2(\zeta)\Bigr)^3
  ~ 
 \Bigl[1- f(y,\zeta,\zeta_p^2)\Bigr]
 \nonu +\Theta\Bigl[[m(\zeta)+ \zeta_p]^2- y\Bigr]~\Bigl(y-m^2(\zeta)\Bigr)^3
  \Biggr\} 
 \label{app5_rhoA}\ee 
 where 
 \be
 f(y,\zeta,\zeta^2_p)=\sqrt{1 -\zeta^2_p~{2y+2m^2(\zeta)-\zeta^2_p\over (y-m^2(\zeta))^2}}~
 \nonu\times ~\Biggl[ 1+\zeta^2_p~{y+m^2(\zeta)-2 \zeta^2_p\over \Bigl(y-m^2(\zeta)\Bigr)^2}
   \Biggr]~~,
 \ee
 with $y>~[m(\zeta)+ \zeta_p]^2$.
  To complete our analysis, let us  consider $\Theta\Bigl(y-s_{th}\Bigr)~ \rho^{(1)}_A(y,\zeta)$ for $\zeta_p\to
  0$. In particular, one remains with  the following limit
\be
 \lim_{\zeta_p\to 0}~ 
 {1\over \zeta^2_p}~\Bigl[1 -f(y,\zeta,\zeta^2_p)\Bigr]=0
 \ee
 since
$$ f(y,\zeta,\zeta^2_p) \sim 1 +{\zeta^4_p\over 2}~f''(0,y)$$

 Therefore one gets
 \be
 \lim_{\zeta_p\to 0}~ \Theta\Bigl[y-(m(\zeta)+\zeta_p)^2\Bigr]~\rho^{(1)}_A(y,\zeta)=
 0
 \nonu
 \lim_{y\to \infty}~\Theta\Bigl[y-(m(\zeta)+\zeta_p)^2\Bigr]~\rho^{(1)}_A(y,\zeta)=0
 \label{app5_rhoAl}\ee

   Starting from Eq. \eqref{rhoB} and  repeating analogous steps one has  for $\rho^{(1)}_B$
\be
\Theta\Bigl(y-s_{th}\Bigr)~\rho^{(1)}_B(y,\zeta)
 =
 -~{3 \over  (4\pi)^2}~e^2_R~ m(\zeta)
\nonu \times ~\int_{0}^\infty~d\omega~\delta(\omega-\zeta^2_p)
~\int_0^1 d\xi  \int_{0}^\infty~ds'~ \delta(s'-m^2(\zeta))
\nonu \times~
 \Theta\Bigl[ y\xi(1-\xi) - \xi \omega -(1-\xi) s'\Bigr]
=-~\frac{3}{(4\pi)^2} ~e_R^2~m(\zeta) 
\nonu \times ~ \int_0^1 d\xi ~\Theta\Bigl[y \xi(1-\xi)-\xi (\zeta_p^2-m^2(\zeta))-m^2(\zeta)   \Bigr] \nonu 
                = -~ \frac{3}{(4\pi)^2} ~e_R^2~m(\zeta) ~ \Theta\Bigl[y-[m(\zeta)+\zeta_p]^2\Bigr]~(\xi_+ - \xi_-)
		  \nonu 
 = -~ \frac{3e_R^2}{(4\pi)^2}~ m(\zeta) ~  \Theta\Bigl[y-[m(\zeta)+\zeta_p]^2\Bigr]~ 
	\nonu \times~	\frac{1}{y}\sqrt{[y-(m^2(\zeta)+\zeta_p^2)]^2-4m^2(\zeta)\zeta_p^2}
\label{app5_rhoB}\ee
with  $\xi_\pm$ given in Eq. \eqref{xi_ex} after changing $t'\to \xi$. 
 Differently from  $\rho^{(1)}_A(y,\zeta)$, this time no IR-dependent 
issue is present.  For
completeness, the following relevant limits have to considered.

\be
\lim_{\zeta_p\to 0}~\rho^{(1)}_B(y,\zeta)=
\nonu =-~\frac{3}{(4\pi)^2} ~e_R^2~ m(\zeta) ~  \Theta\Bigl[y-m(\zeta)^2\Bigr]~
{y-m^2(\zeta)\over y}
\nonu
\lim_{y\to \infty}~\rho^{(1)}_B(y,\zeta)=-~\frac{3}{(4\pi)^2} ~e_R^2~m(\zeta)
\label{app5_rhoBl}\ee
It should be emphasized that the regularized ${\cal
B}_Z(\zeta,\Lambda;p)$,
Eq. \eqref{self_reg},
 obtained
from the above $\rho^{(1)}_B$ and by
taking into account the limits in Eq. \eqref{app5_rhoBl}, shows
the expected singular behavior in both  IR and UV regions.

For $\rho_\gamma$ in Eq.  \eqref{rhog}, one gets the following first iteration
\be
\Theta(y-s^p_{th})~\rho^{(1)}_\gamma(y,\zeta)= -{e_R^2\over (2\pi)^2} 
\int_{0}^\infty ds~\delta(s-m^2(\zeta))
\nonu \times~     \int_{0}^\infty ds'~\delta(s'-m^2(\zeta)) \int_0^1 d\xi~2 \xi (1-\xi)
\nonu\times~  \Theta\Bigl[ y\xi(1-\xi) - \xi s'-(1-\xi)s \Bigr] 
= ~-{2\over (2\pi)^2}~e_R^2 
\nonu\times~\int_0^1 d\xi~ \xi (1-\xi) 
 \Theta\Bigl[ y \xi(1-\xi)- m^2(\zeta) \Bigr]
=
\nonu
 =  ~- \frac{e_R^2}{3(2\pi)^2}\Theta(y)\Theta(y-4m^2(\zeta))~
 \Bigl(1+2{m^2(\zeta) \over y}\Bigr)
 \nonu \times~\sqrt{1-4\frac{m^2(\zeta)}{y}}
\label{app5_rhog}\ee
with
\be
\xi_\pm = \frac{1}{2}\left(1\pm\sqrt{1-4\frac{m^2(\zeta)}{y}}\right) 
\ee
Let us recall that $s^p_{th}=4 m^2(\zeta)$. Moreover 
\be
\lim_{y\to \infty}~\rho^{(1)}_\gamma(y,\zeta)=\frac{e_R^2}{3(2\pi)^2}
\ee
and therefore $Z_3^{(1)}$ is logarithmically divergent, (see Eq. \eqref{Z3_NIR}).
The first-order photon self-energy, Eq. \eqref{NIRph}, is given  by
\be
 \Pi^{(1)}_R(\zeta,q^2) 
 =
 \int_{s^p_{th}}^\infty ds~\frac{(\zeta^2_p-q^2)~\rho^{(1)}_\gamma(s,\zeta)}
 {(q^2-s+i\epsilon)~(\zeta^2_p-s+i\epsilon)}
 \nonu
 =-\frac{e_R^2}{3 (2\pi)^2}~\int_{s^p_{th}}^\infty ds~{
 (\zeta^2_p-q^2)~\Bigl[s+2m^2(\zeta)\Bigr]\over s~(q^2-s+i\epsilon)~(\zeta^2_p-s+i\epsilon)}
\nonu \times  \sqrt{1-4\frac{m^2(\zeta)}{s}}
 \label{app5_PiR}
\ee
{Notice that the imaginary part of $\Pi^{(1)}_R(\zeta,q^2)$, when $q^2> 4m^2(\zeta)$ coincides with the result
that can be found in}
Ref \cite{Zuber}.
\section{ Formulas Summary}
\label{app_6}
For the sake of a quick focus on the main formal results have been obtained in the paper, in
 this Appendix we list the initial 
expressions useful for 
 for a numerical calculations  of KL weights in terms of NWFs.

The three NWFs $\rho_A$, $\rho_B$ and $\rho_\gamma$ fulfill the following integral equations
\be
\Theta\Bigl(y-s_{th}\Bigr)~ \rho_A(y,\zeta)
  = 
 {3\over (4\pi)^2}~e^2_R \lim_{\Lambda \to \infty}~Z_1(\zeta,\Lambda)
\nonu \times 
 ~\int_{0}^\infty~d\omega~\bar \sigma_\gamma(\omega,\zeta,\zeta_p,\Lambda)
~\int_0^1 d\xi~\int_{0}^\infty~ds'
\nonu~\Biggl\{ 
\bar\sigma_V(s',\zeta,s'_{th},\Lambda) ~
 \Biggl[\xi~
 \Theta\Bigl(y\xi (1-\xi) - \xi \omega -(1-\xi) s'\Bigr)
 \nonu- \int_0^{1-\xi}dt
\Theta\Bigl( yt (1-t) - \xi \omega -t s'\Bigr) \Biggr]
+  \bar\sigma_V(s',\zeta,s'_{th},\Lambda)
\nonu \times
\Biggl[~\int_{s_{th}}^\infty~ds~
 ~\rho_A(s,\zeta,\Lambda)~{\cal C}^{(0)}_{AV}(\zeta,\omega,s,s',\xi,y)
\nonu 
+  y~\int_{s_{th}}^\infty~ds~
 ~\rho_A(s,\zeta,\Lambda)~{\cal C}^{(1)}_{AV}(\zeta,\omega,s,s',\xi,y)\Biggr]
\nonu-  y\bar \sigma_S(s',\zeta,s'_{th},\Lambda)
 \int_0^{1-\xi}dt\int^{1-\xi-t}_0 {dw} 
 \nonu \times~\int_{s_{th}}^\infty ds
~\rho_B(s,\zeta,\Lambda)  
\nonu \times ~  \Delta' \Bigl[y  -s+
  {s {\cal A}_4(t,w)-\xi \omega -t s' -ws\over{\cal A}_4(t,w)}\Bigr]\Biggr\}
~~,
\ee
\be
\Theta\Bigl(y-s_{th}\Bigr)~\rho_B(y,\zeta)
 =
 -~{3 \over  (4\pi)^2}~e^2_R~\lim_{\Lambda \to \infty}~
 Z_1(\zeta,\Lambda )
\nonu \times~\int_{0}^\infty~d\omega~\bar \sigma_\gamma(\omega,\zeta,\zeta_p,\Lambda)
~\int_0^1 d\xi \int_{0}^\infty~ds'
\nonu \times~\Bigg\{  
\bar \sigma_S(s',\zeta,s'_{th},\Lambda)~ 
 ~
 \Theta\Bigl[ y\xi (1-\xi) - \xi \omega -(1-\xi) s'\Bigr]
\nonu+
\bar\sigma_S(s',\zeta,s'_{th},\Lambda)~
 \nonu \times \Biggl[~\int_{s_{th}}^\infty~ds~
{\rho_A(s,\zeta,\Lambda)}~{\cal C}^{(0)}_{AS}(\zeta,\omega,s,s',\xi,y)
\nonu+y~
 ~\int_{s_{th}}^\infty~ds~
{\rho_A(s,\zeta,\Lambda)}~{\cal C}^{(1)}_{AS}(\zeta,\omega,s,s',\xi,y)\Biggr]
\nonu+y~
\bar \sigma_V(s',\zeta,s'_{th},\Lambda)~
 \int_0^{1-\xi}dt\int^{1-\xi-t}_0 {dw}
\nonu \times ~ \int_{s_{th}}^\infty~ds~
\rho_B(s,\zeta,\Lambda) 
~\nonu \times 
  ~\Delta' \Bigl[y  -s
  +{s {\cal A}_4(t,w) -\xi \omega -t s' -ws\over{\cal A}_4(t,w)}\Bigr]
 \Biggr\}
~~,
\ee
\be
{ \Theta(y-s^p_{th})~\rho_\gamma(y,\zeta)=} -{e_R^2\over (2\pi)^2}~
\lim_{\Lambda\to\infty}~Z_1(\zeta,\Lambda)\int_{0}^\infty ds 
\nonu\times ~
     \int_{0}^\infty ds'~ \int_0^1 d\xi
~  \Biggl\{ \bar\sigma_V(s',\zeta,s'_{th},\Lambda) ~ 
\bar\sigma_V(s,\zeta,s_{th},\Lambda)
\nonu \times ~
 2 
 \xi (1-\xi)  
 \Theta\Bigl[ y \xi(1-\xi)- \xi s'-(1-\xi)s) \Bigr] 
 \nonu\times ~\left( 1
 + \int_{s_{th}}^\infty d\omega ~
{\rho_A(\omega,\zeta ,\Lambda)\over 
(\zeta^2-\omega +i\epsilon)}  \right) 
 \nonu +\int_{s_{th}}^\infty d\omega 
   ~\rho_A(\omega,\zeta,\Lambda)~{\cal C}_\gamma(s,s',\xi,\omega) 
  \nonu+
  2\bar\sigma_S(s',\zeta,s'_{th},\Lambda) ~ \bar\sigma_V(s,\zeta,s_{th},\Lambda)~
  \nonu \times~\int_{s_{th}}^\infty d\omega 
   ~\rho_B(\omega,\zeta,\Lambda)~
  ~ \int_0^{1-\xi} dv\int_0^{1-\xi-v} dw ~(v+w)\nonu \times
  ~{1 -2 (v+w)\over {\cal A}_4^2(v,w)}
   ~{\partial\over \partial y}\delta \Bigl[y -{\cal A}_7(s,s',\omega,v,\xi,w) \Bigr]   
 \Biggr\}
~~,
\nonu \ee
with $\Delta'$, ${\cal C}^{(i)}_{AV}$, ${\cal C}^{(i)}_{AS}$ and ${\cal C}_{\gamma}$ 
from Eqs. \eqref{DeltapText},\eqref{cal_AV0}, \eqref{cal_AV1},
 \eqref{cal_AS0}, \eqref{cal_AS1} and \eqref{cal_gamma}, and  the following equations that relate the NWFs to the KL weights 
 (see Eq. \eqref{klwf})
\be
   \sigma_V(\omega,\zeta) =  \frac{D_I\Bigl[1-(\zeta^2-\omega)\pva\Bigr]-\rho_A(\omega,\zeta)D_R}{D_R^2+\pi ^2 D_I^2} 
  \nonu
   \sigma_S(\omega,\zeta) =  \frac{D_I\Bigl[m(\zeta)+(\zeta^2-\omega)\pvb\Bigr]+\rho_B(\omega,\zeta)D_R}{D_R^2+\pi ^2 D_I^2} \nonu
  \ee
  where $\omega \ge s_{th}= (m(\zeta)+\zeta_p)^2$,  the notation $\langle \rho_{A,B} \rangle $ means
  \be
  \langle \rho_{A,B} \rangle = \textrm{P.V.}\int_{s_{th}}^\infty ds \frac{\rho_{A,B}(s,\zeta)}{(p^2-s)(\zeta^2-s+i\epsilon)}
  \ee
  and 
  \be
  D_R=\omega\Bigl[(1-(\zeta^2-\omega)\pva)^2-\pi^2 \rho_A^2(\omega,\zeta) \Bigr] 
  \nonu- 
  \Bigl[(m(\zeta)+(\zeta^2-\omega)\pvb )^2-\pi^2 \rho_B^2(\omega,\zeta)\Bigr] ~~,
  \nonu
  D_I=2 \omega\rho_A(\omega,\zeta)\Bigl[1-(\zeta^2-\omega)\pva\Bigr]
  \nonu+ 2\rho_B(\omega,\zeta)\Bigl[m(\zeta)+
  (\zeta^2-\omega)\pvb\Bigr]~~.
  \ee 
  Moreover, one has (see Eq. \eqref{sigmag})
  \be
 \sigma_\gamma(\omega,\zeta)=~-~{1 \over  (\omega-\zeta^2_p)}
 \nonu \times ~{\rho_\gamma(\omega,\zeta)\over\Bigl[(1+(\zeta^2_p-\omega)\pvg)^2+\pi^2
 \rho^2_\gamma(\omega,\zeta)\Bigr]}~~,
 \label{sigmagf}
\ee
with $\omega\ge~ \zeta^2_p$.
  
  The fundamental  renormalization constant $Z_2=Z_1$ 
  is given by
  \be
 \lim_{\Lambda \to \infty} Z_2(\zeta,\Lambda)= 1 + \lim_{\Lambda \to \infty}\int_{s_{th}}^\infty ds 
   ~\frac{\rho_A(s,\zeta,\Lambda )}{\zeta^2-s+i\epsilon}~.
   \ee
   
\end{document}